\documentclass[11pt,letterpaper,leqno]{article}

\usepackage[utf8]{inputenc}
\usepackage[english]{babel}
\usepackage{palatino}

\usepackage[margin=.9in]{geometry}
\usepackage{multicol}
\usepackage{ragged2e}
\usepackage{enumitem}
\usepackage{epigraph}
\usepackage{chngcntr}

\renewcommand{\baselinestretch}{1.19}

\usepackage{amsmath}
\usepackage{amssymb}
\usepackage{amsfonts}
\usepackage{amscd}
\usepackage{mathtools}
\usepackage{amsthm}
\usepackage{thmtools}
\usepackage{bm}
\usepackage{bbm}
\usepackage{xfrac}  
\DeclareMathAlphabet{\mathstandardcal}{OMS}{cmsy}{m}{n}
\newcommand{\Lagr}{\mathstandardcal{L}}
\newcommand{\Man}{\mathstandardcal{M}}
\usepackage{mathrsfs}
\usepackage[cal=boondox]{mathalfa}

\usepackage[usenames,dvipsnames,svgnames,table]{xcolor}

\definecolor{customblue}{rgb}{0.2235, 0.4157, 0.6941}
\definecolor{customred}{rgb}{0.8000, 0.1451, 0.1608}
\definecolor{ao}{rgb}{0.0, 0.5, 0.0}
\definecolor{afuchs}{rgb}{0.57, 0.36, 0.51}
\definecolor{amber}{rgb}{1, 0.75, 0}
\definecolor{air}{rgb}{0.36, 0.54, 0.66}

\usepackage{graphicx}
\usepackage{epsfig} 
\graphicspath{ {images/} }
\usepackage{float}
\usepackage{placeins}
\usepackage{caption}
\usepackage{subcaption}
\usepackage{multirow}
\usepackage{tabularx}
\usepackage{booktabs}
\usepackage{siunitx} 

\usepackage{tikz}
\usetikzlibrary{calc, intersections, through, backgrounds, shadings, shapes.geometric, arrows.meta, patterns}
\usepackage{pgfplots}
\usepgfplotslibrary{fillbetween}

\usepackage{etoolbox}
\usepackage{apptools}

\usepackage[bottom,multiple]{footmisc}
\usepackage{natbib}
\setcitestyle{authoryear,round,longnamesfirst=false,maxnames=2}

\usepackage{xr-hyper}

\usepackage[hyperfootnotes=false]{hyperref} 
\hypersetup{
    pdfborder=0 0 0,
    colorlinks=true, 
    urlcolor=MidnightBlue, 
    citecolor=MidnightBlue, 
    linkcolor=MidnightBlue,
    filecolor=MidnightBlue
}

\usepackage{nameref}
\usepackage{cleveref}

\theoremstyle{plain}

\newtheorem{prop}{Proposition}
\newtheorem{lem}{Lemma}
\newtheorem{rmk}{Remark}
\newtheorem{cl}{Claim}

\newtheorem{assu}{Assumption}
\newtheorem{mydef}{Definition}

\AtAppendix{\counterwithin{prop}{subsection}}
\AtAppendix{\counterwithin{lem}{subsection}}
\AtAppendix{\counterwithin{cor}{subsection}}
\AtAppendix{\counterwithin{cl}{subsection}}

\newcommand\nc{\newcommand}
\nc\on{\operatorname}

\newrobustcmd*{\citefirstlastauthor}{\AtNextCite{\DeclareNameAlias{labelname}{given-family}}\citeauthor}

\makeatletter
\def\@footnotecolor{gray!70!black}
\define@key{Hyp}{footnotecolor}{%
 \HyColor@HyperrefColor{#1}\@footnotecolor%
}
\patchcmd{\@footnotemark}{\hyper@linkstart{link}}{\hyper@linkstart{footnote}}{}{}

\renewcommand\@biblabel[1]{}

\makeatother

\hypersetup{footnotecolor=black}

\begin{document}
\title{Automation, AI, and the \\ Intergenerational Transmission of Knowledge}

\author{Enrique Ide\thanks{Department of Economics, IESE Business School, Carrer d'Arn\'{u}s i de Gar\'{\i} 3-7, 08034 Barcelona, Spain (eide@iese.edu). I am grateful for the valuable comments and suggestions of David Autor, Heski Bar-Isaac, Matthew Beane, David De la Croix, Andrea Galeotti, Luis Garicano, Jie Gong, Dan Gross, Chad Jones, Anton Korinek, Jin Li, Suraj Malladi, V\'{i}ctor Mart\'{i}nez de Alb\'{e}niz, Crist\'{o}bal Otero, Sebasti\'{a}n Otero, Esteban Rossi-Hansberg, Raffaella Sadun, Eduard Talam\`{a}s, Chris Tonetti, and seminar participants at Chicago FED, IESE, IMF, Kellogg Strategy, Luohan Academy, and NBER Org. Econ. I also acknowledge the financial support of IESE through the High Impact Initiative-course 2024/2025. I declare I have no relevant or material financial interests that relate to the research described in this paper.}}
\date{\today}
\maketitle 

\begin{abstract} Motivated by concerns that AI-driven entry-level automation may disrupt early-career learning, this paper examines how technological change affects the intergenerational transmission of tacit knowledge---practical, hard-to-codify skills acquired through workplace interaction. I develop a task-based overlapping-generations model in which novices acquire tacit knowledge by working alongside experts. Knowledge-transfer contracts are incomplete because tacit knowledge is embodied and non-verifiable. In equilibrium, endogenous growth arises because only the most knowledgeable experts manage production and transmit their expertise to multiple novices, diffusing best practices. I show that improvements in entry-level automation increase output upon adoption but can reduce growth and welfare, even without reducing entry-level employment. This occurs when such improvements reallocate novices away from the most productive experts, slowing the diffusion of best practices. By contrast, technological improvements that increase the number of novices learning from the most productive experts strengthen knowledge transmission and raise growth.\end{abstract} 
 
\newpage 

\section{Introduction} 

Recent advances in Artificial Intelligence (AI) have fueled optimistic forecasts of dramatic increases in productivity and economic growth. Unlike earlier technological innovations---such as robots or conventional computers---AI systems can learn and adapt from examples, enabling them to perform work once considered inherently human \citep{autor2024applying, brynjolfsson2023generative, AIKE}. As a result, many technologists foresee an economic transformation that could rival or even surpass that of the Industrial Revolution \citep{amodei,hassabis,altman}.

However, the widespread adoption of AI may carry unintended long-term costs. By allowing senior workers to accomplish more tasks independently, AI-driven automation may reduce entry-level opportunities and threaten the implicit contract in which younger workers exchange their labor for training and the prospect of future promotion (\citeauthor{roose}, {\em \color{MidnightBlue} The New York Times}, \citeyear{roose}; \citeauthor{WSJ}, {\em \color{MidnightBlue} The Wall Street Journal}, \citeyear{WSJ}; \citeauthor{bloomberg}, {\em \color{MidnightBlue} Bloomberg.com}, \citeyear{bloomberg}). Such arrangements have historically been crucial for transmitting tacit knowledge---practical insights that cannot be fully articulated yet remain indispensable for carrying out complex work---raising concerns about how future generations will acquire essential expertise \citep{beaneWSJ,beane2024skillcode,garicanoAIbecker}.

Motivated by these concerns, this paper develops a framework for analyzing how different forms of technological change---automation, task creation, and labor- and capital-augmenting technologies---reshape the intergenerational transmission of tacit knowledge and, through that channel, long-run growth. The central premise is that technological change affects current and future productivity not only by changing how tasks are performed, but also by changing who works with whom and, therefore, who learns from whom.

Existing research on automation and AI has mainly focused on how new technologies directly affect productivity, labor-market outcomes, and the distribution of income and wealth \citep[e.g.][]{autor2003, acemoglu2011chapter, moll2022uneven, AKR, autor2025expertise, AIKE}. However, much less attention has been paid to how these innovations reshape the interpersonal interactions through which valuable workplace skills are transmitted. This paper addresses that gap by embedding technological change within a growth model of knowledge diffusion and learning \citep[e.g.,][]{lucas2009ideas,lucas2014knowledge,perla2014equilibrium,de2018clans}. In doing so, it provides a unified framework for analyzing how new technologies jointly affect current output, knowledge transmission, and long-run growth.

The baseline model features overlapping generations, with individuals living for two periods: first as novices, then as experts. Novices are identical at birth, have no wealth, and supply one unit of labor inelastically. Experts, by contrast, differ in skill. This skill level is observable, but the knowledge underlying it is tacit. Outsiders can therefore assess an expert’s productivity, but the insights and processes through which it is achieved are accessible only through close interaction.

Production combines expert skill with routine tasks that can be performed either by novices or by machines. Machines are more productive than novices, but can perform only a subset of tasks; the rest must be carried out by novices. Because the non-automated portion of production requires multiple units of novice labor, production gives rise to many-to-one matching: each novice works for exactly one expert, while any expert who produces hires multiple novices. By working alongside an expert, a novice gains intimate access to that expert’s tacit knowledge and enters the next period with the same skill level as her mentor.\footnote{The baseline assumes that all novice-performed routine tasks provide uniform access to the expert’s tacit knowledge. Section \ref{sec:robust} and Online Appendix \ref{sec:hetero} relax this assumption by allowing some routine tasks to have no learning value.}

Since future earnings rise with mentor skill, novices are willing to accept lower wages to work alongside more skilled experts. Those higher future earnings reflect both the greater productivity novices will attain as experts and the lower wages their own novices will accept in exchange for access to more valuable expertise. However, because expert knowledge is tacit, it cannot be fully articulated and is therefore non-verifiable. Moreover, because it is tied to personal experience and thus embodied in the individual, the resulting human capital cannot be seized or repossessed and therefore cannot serve as collateral. Consequently, novices cannot finance access to better mentors by borrowing against the future returns to acquired expertise. Lacking initial wealth, they face a zero lower bound on wages. Concurrent labor thus becomes the only way to compensate experts for the opportunity to learn, giving rise to an apprenticeship-like arrangement.

In equilibrium, all automatable tasks are assigned to machines, and each expert cohort is partitioned endogenously by skill. The lowest-skilled experts are inactive because hiring novices at equilibrium wages is unprofitable. Above them lies an active region, which itself splits into two: a \textit{price-cleared} tier in which experts hire novices at strictly positive wages, and an \textit{oversubscribed} tier at the top, where the zero lower bound binds and wages can no longer clear the market for those positions.\footnote{The oversubscribed region is consistent with the severe rationing of entry-level positions---and the corresponding willingness of junior associates to accept unusually long hours relative to pay---often observed at top professional services firms (see Sections \ref{sec:rosen} and \ref{sec:sch} for details).} This endogenous partition then governs the evolution of expertise over time: the next generation’s skill distribution is obtained by truncating the current distribution at the marginal active expert and reweighting the surviving upper tail by experts’ span of control---the number of novices each active expert employs.

When the initial distribution of expert skill is fat-tailed,\footnote{Although assuming an unbounded, fat-tailed initial skill distribution may appear strong at first sight, it is better interpreted as a reduced-form representation of an economy in which agents with bounded skills draw ideas from an unbounded latent distribution with fat tails \citep[``Plato’s realm of Forms'' in the language of][]{buera}.} this process generates endogenous asymptotic growth. The most productive experts replicate their skills by passing them on to multiple novices, while the fat upper tail prevents the pool of mentors from becoming too sparse as the economy develops. The asymptotic growth rate of output is therefore determined by experts’ span of control and the tail thickness of the initial skill distribution. Furthermore, once skills are normalized relative to the least-skilled expert in each cohort, the cross-sectional distribution of skills converges asymptotically to a Pareto law.

I then use the framework to study several technological shocks, beginning with automation. I first show that expanding the set of automatable tasks raises output upon adoption, but can also reduce subsequent growth. The immediate output gain reflects two margins: routine tasks are reallocated from novices to more productive machines, and the novices released by high-skill experts allow a new tier of lower-skill experts to produce. The subsequent growth slowdown, in turn, reflects the fact that this same reallocation of novice labor worsens mentor composition. Some novices are matched with worse experts, acquire lower-quality tacit knowledge, and later become worse mentors themselves. The best practices of the leading experts thus diffuse more slowly, imposing a persistent drag on growth. Importantly, this slowdown does not require an improvement in automation to reduce entry-level employment or deprive novices of hands-on practice; a compositional deterioration in the pool of mentors is sufficient.

Through this displacement channel, the automation shock can reduce aggregate welfare, defined as the discounted sum of the income of current and future generations, relative to a no-adoption benchmark. This welfare loss stems from the zero lower bound on wages. In a frictionless environment, novices would be able to borrow against future earnings to preserve access to valuable training positions---especially since the same automation shock makes expertise more valuable. Instead, novices cannot directly pay oversubscribed experts for access, so those experts fail to fully internalize the future income generated by the mentorship positions that automation displaces. Put differently, the zero lower bound forces novices to finance elite mentorship solely through their labor, the very currency that an improvement in entry-level automation devalues.

Displacement, however, operates alongside a countervailing force. Improvements in automation can also expand experts’ scale of operation by reducing the supervision time required to complete a given production process. If this scale effect dominates, the total number of novices working with the very best experts rises, even if each production process requires fewer novices. The automation shock then accelerates the diffusion of best practices and raises growth. Even so, at least in these early stages of adoption, the preliminary evidence on generative AI discussed below seems more consistent with displacement than with offsetting scale effects. This may be because current AI automation is still ``so-so'' \citep{AcemogluRestrepoNBER}, generating limited time savings, or because expert supervision is constrained by hard cognitive or attentional limits (see Section \ref{sec:scale} for details).

Beyond automation, the framework yields analogous implications for other forms of technological change. Technologies that create new labor-intensive tasks raise output on impact and increase long-run growth, even without scale effects. They do so by inducing the most skilled experts to hire more novices, thereby accelerating the diffusion of best practices. By contrast, technologies that raise the productivity of novices or machines in the tasks they already perform increase long-run growth only if they expand experts’ scale of operation.

Taken together, these results imply that the effects of new technologies on knowledge transmission can be ambiguous. The same technology can automate some tasks, create new ones, and augment labor or capital in existing tasks. Yet these forces operate through a common margin: they reallocate junior workers across experts, thereby changing whom those workers learn from. Although direct evidence on this technology-induced reallocation remains scarce, evidence from other shocks supports its relevance: weak labor-market conditions at entry and idiosyncratic shifts in large-firm hiring can redirect young workers across initial placements, with persistent consequences for earnings and skill development \citep{oyer2006initial,arellano2022effects,arellano2024career}. Future empirical work on how technological change shapes expertise formation should therefore look beyond aggregate entry-level employment and track how junior workers are reallocated across firms and mentors.

While applicable to many technologies, these results are especially relevant to the recent advent of generative AI. Early evidence comparing employment changes across age groups and occupations differentially exposed to AI suggests that junior workers have experienced weaker relative employment outcomes in more AI-exposed occupations \citep{berger2,lichtinger2025,canaries2025,klein2025generative,atkinson2026young}. This pattern is consistent with some displacement of junior labor occurring in such environments. Although the available evidence does not yet track affected workers' subsequent placements, anecdotal reports suggest that some graduates may be taking positions at smaller, mid-tier firms that previously struggled to attract such talent \citep{ellis2026,lahart2026}.

In this sense, the evidence so far suggests that current iterations of generative AI may be a form of ``seniority-biased technological change'' \citep{lichtinger2025}: a technology that substitutes for entry-level execution while complementing expert judgment. Holding expertise fixed, such a change is unambiguously beneficial. But expertise is endogenous, and its intergenerational transmission is governed by incomplete contracts. As a result, to the extent that such technologies reduce junior workers' access to the most productive mentors and firms, seniority-biased technological change can erode the future supply of the very expertise it complements.

\subsection{Related Literature}

This paper bridges three strands of literature. It connects the task-based literature on technology and labor demand with the literature on idea flows and economic growth by examining how new technologies reshape the workplace interactions through which tacit skills are transmitted across generations. It also contributes to the literature on the economics of apprenticeships by showing how early-career learning links technology adoption to aggregate productivity over time.

The task-based approach to technology and labor demand was developed by \citet{zeira1998workers}, \citet{autor2003}, and \citet{acemoglu2011chapter}. In this framework, production requires the completion of a range of tasks allocated across factors of production according to comparative advantage. Technological change therefore affects labor demand, wages, inequality, and productivity by altering who performs which tasks and at what cost \citep{acemoglu2018race,AcemogluRestrepoNBER,acemoglu2022tasks,AKR}. Building on this approach, \citet{autor2025expertise} model occupations as bundles of tasks and show that automation can be expertise-leveling, reducing occupational expertise requirements.\footnote{Other important contributions that use the task-based approach to study automation, AI, and technological change include \citet{aghion2017artificial}, \citet{moll2022uneven}, \citet{korinek}, \citet{jones2024framework}, \citet{freund2025job}, \citet{jones2025past}, \citet{acemoglu2022automation}, \citet{demirer2026chaining}, and \cite{althoff2026task}. For analyses of AI’s effects on output and growth outside the task-based framework, see \citet{gansG} and \citet{beraja2026value}.}

More recently, \citet{ide2024turing,AIKE,GKW} build on the knowledge-hierarchies literature of \citet{garicano2000hierarchies} and \citet{garicano2004inequality,garicano2006organization} to study AI’s distinctive implications. Knowledge hierarchies are a specialized version of the task-based framework in which task linkages emerge endogenously from organizational choices aimed at using tacit knowledge efficiently \citep{garicano2015knowledge}. Using this framework, \cite{ide2024turing,AIKE,GKW} highlight the importance of AI autonomy: nonautonomous AI disproportionately benefits the least knowledgeable---resembling the expertise-leveling logic of \citet{autor2025expertise}---whereas autonomous AI is expertise-amplifying, primarily benefiting the most knowledgeable.

I contribute to the task-based literature by studying how technological change affects the intergenerational transmission of tacit knowledge and, through that channel, the evolution of productivity. My model builds on the canonical task-based framework. At the same time, it also draws inspiration from the knowledge-hierarchies literature in emphasizing the economic importance of tacit knowledge and the role of organizations in determining who works with whom.

The literature on idea flows and economic growth emphasizes how knowledge diffuses through interactions among individuals and firms, shaping productivity and growth \citep{kortum1997research,eaton1999international,eaton2002technology,lucas2009ideas,lucas2014knowledge}.\footnote{Other important contributions include \cite{buera2020global}, \cite{jarosch2021learning}, \cite{perla2021equilibrium}, and \cite{benhabib}. For a survey on the literature, see \cite{buera}.} My paper contributes to this literature by studying a margin largely absent from existing models: how technological change reshapes who learns from whom in production, with consequences for the diffusion of best practices.

In the context of idea flows and economic growth, the papers closest to mine are \cite{perla2014equilibrium}, \cite{de2018clans}, and \cite{caicedo2019learning}. In both my paper and \cite{perla2014equilibrium}, long-run growth is tied to the endogenous evolution of a fat-tailed productivity distribution as better practices diffuse through the economy. In \cite{perla2014equilibrium}, this evolution is driven by firm imitation: low-productivity firms upgrade by drawing a new productivity level from the distribution of producing firms. In my model, by contrast, diffusion is intergenerational and workplace-based: the most productive experts replicate their expertise by transmitting it to the novices who work with them and later become experts themselves. This distinction introduces a scarce-mentor bottleneck and gives technological change a central role in reshaping who learns from whom.

As in \cite{caicedo2019learning}, my paper embeds organizations within a model of idea flows and knowledge diffusion. In their framework, however, agents learn through stochastic encounters, while organizational assignment determines production roles, output, and wages. In my model, by contrast, workplace assignment is itself the learning technology: novices learn from the experts with whom they work. \cite{de2018clans} is closer on intergenerational transmission: it also studies an overlapping-generations economy in which tacit knowledge passes from experts to novices. But \cite{de2018clans} assumes random matching and focuses on how preindustrial apprenticeship institutions helped mitigate masters’ incentives to shirk on training effort. By contrast, my paper assumes that novices compete to work with the best experts and studies how technological change affects the transmission of tacit knowledge.

This workplace-based learning structure also connects my paper to the literature on the economics of apprenticeships: novices acquire tacit knowledge by working alongside experts, effectively exchanging current labor for access to expertise. As \citet{mokyr2019} emphasizes, apprenticeship contracts are inherently incomplete because they govern the transfer of tacit knowledge. That incompleteness can distort training and lead to inefficient knowledge transmission \citep{garicano2017relational,de2018clans,fudenberg2019training,fudenberg2021working}.

I contribute to this literature by studying how technological change reshapes apprenticeship-like arrangements and, through them, the intergenerational transmission of tacit knowledge. Closest is the complementary work of \cite{GR2025AI}, which extends \cite{garicano2017relational} to analyze how AI affects apprenticeship viability. Their analysis, however, focuses on the dynamics of a single expert–novice relationship. By contrast, my paper studies apprenticeship-like arrangements in general equilibrium, where technology changes the allocation of novices across experts and, through that allocation, the evolution of expertise and long-run growth.

Finally, my paper is closely related to several contemporaneous contributions that examine how new technologies affect learning, skill formation, and internal labor market structure. \citet{acemoglu2026collapse} show how agentic AI can substitute for human learning effort and erode the stock of public knowledge. \citet{afrouzi2026automation} study how automation and task creation shape learning-by-doing and worker-to-manager transitions, while \citet{friebel2026pyramids} study how AI shocks reshape firms’ junior-to-senior span, generating hiring freezes and transitional oscillations. Relative to these papers, I emphasize a different margin: who learns from whom. In my framework, technological change shapes skill formation by reallocating novices across experts and thereby altering the expertise transmitted to future generations, even without changing aggregate entry-level employment.

\subsection{Roadmap}

The remainder of the paper is organized as follows. Section \ref{sec:rosen} lays out the conceptual foundations of the analysis. Section \ref{sec:baseline} presents the baseline model and defines competitive equilibrium. Section \ref{sec:longrun} characterizes long-run equilibrium outcomes. Section \ref{sec:advances} studies the effects of technological change. Section \ref{sec:robust} extends the framework to incorporate additional margins that shape technology's effects on tacit knowledge transmission. Section \ref{sec:final} concludes.

\section{Conceptual Foundations} \label{sec:rosen}

This section outlines the conceptual foundations of the analysis: the tacit dimension of expertise, the inherent constraints on its transfer, and its transmission across generations.

\vspace{3mm}

\noindent \textit{The Tacit Dimension of Expertise}.--- Codifiable knowledge refers to explicit rules and procedures that can be easily transmitted via manuals, books, and databases. Tacit knowledge, in contrast, encompasses intuitive skills and insights that are difficult to articulate precisely \citep{polanyi,foray2004economics}. These two forms of knowledge are interdependent: codified instructions usually require tacit knowledge to be properly interpreted and adapted to real-world complexities. As \citet[][pp. 14-15]{mokyr2002gifts} explains:

\begin{quote} Tacit knowledge is needed to obtain inexpensive, reliable access to codified instructions […] no set of instructions [...] can ever be complete. It would be too expensive to write a complete set of instructions for every technique. Judgement, dexterity, experience, and other forms of tacit knowledge inevitably come into play when technique is executed. \end{quote}

While professional capability relies on both forms of knowledge, codifiable elements can be standardized and scaled. This paper therefore focuses on the tacit dimension. Hereafter, I use ``expertise'' and ``tacit knowledge'' interchangeably to refer to the unwritten insights required to execute and orchestrate well-defined tasks. In investment banking, for instance, populating financial models and verifying due diligence documents are largely standardized procedures. However, transforming their completion into financial advice requires the judgment to select appropriate models, identify anomalies that standard procedures might overlook, and adjust methods when unforeseen contingencies arise.

Because tacit knowledge is tied to personal experience, it is embodied in the individual \citep{polanyi,garicano2000hierarchies}. Thus, while tacit insights can be transferred, their transmission requires the expert’s direct involvement. This makes the expert's time a key bottleneck to the use and diffusion of this knowledge \citep{foray2004economics,garicano2015knowledge,AIKE}.

\vspace{3mm}

\noindent \textit{Labor as a Mechanism and Currency for Acquiring Tacit Knowledge}.--- Lacking a codifiable structure, tacit knowledge must be acquired experientially through active execution and collaboration with more experienced individuals. As \citet[][p. 55]{polanyi1962personal} observes:

\begin{quote} An art which cannot be specified in detail cannot be transmitted by prescription [...] It can be passed on only by example from master to apprentice [...] By watching the master and emulating his efforts, the apprentice unconsciously picks up the rules of the art, including those which are not explicitly known to the master himself. \end{quote}

Thus, labor supplied alongside an expert serves a dual role: it is both a productive input and a mechanism for assimilating tacit knowledge \citep{lave1991situated,brown1991organizational,beane2019shadow}. This apprenticeship model is pervasive in expertise-intensive professions (e.g., medicine, law, finance), where novices perform routine tasks under expert supervision, observing how experts apply judgment to integrate these inputs into a cohesive whole \citep{garicano2017relational,beane2024skillcode}.

Independent trial-and-error can also help acquire tacit skills, but it remains an inefficient substitute for mentorship. For instance, exploiting office closures and subsequent return-to-office mandates at a Fortune 500 firm, \citet{emanuel2026power} show that younger engineers receive more feedback and later write higher-quality code when co-located with more experienced colleagues. They also find that these engineers are subsequently more likely to be poached for better jobs, consistent with co-location fostering the accumulation of valuable human capital. The value of mentorship is also visible in entry-level labor markets in professional services, such as finance and law, where novices routinely accept demanding hours and low effective pay to work alongside top practitioners.\footnote{For example, junior investment bankers routinely work 100- to 110-hour weeks, systematically underreporting to evade 80-hour institutional limits \citep{glickmanSaeedy2025,saeedy2024a}. In extreme cases, the physical toll of these conditions has proved fatal, as seen in the exhaustion-related death of an intern in 2013 after consecutive overnight shifts and the 2024 death of an associate following repeated 100-hour workweeks \citep{saeedy2024b}.}$^{,}$\footnote{While this phenomenon could also reflect other mechanisms---such as novices trying to acquire a prestigious signal of high ability---evidence on the long-term effects of early-career placements suggests that skill development is an important part of the explanation \citep{oyer2006initial,arellano2022effects,arellano2024career}. I return to this evidence in Section \ref{sec:scale}.}

The contracts governing the transfer of tacit knowledge, however, are inherently incomplete \citep{mokyr2019}. Because such expertise resists precise articulation, it is non-verifiable; because it is embodied in the individual, it cannot be seized or repossessed, and therefore cannot be pledged as collateral. These features make it difficult to finance training through claims on the future returns to acquired expertise \citep{becker1964human, garicano2017relational}. Concurrent labor thus becomes the natural means of payment in this environment.

\vspace{3mm}

These observations motivate the structure of the baseline model. The model represents early-career workers as novices and more experienced practitioners as experts. Expertise is embodied in experts and transmitted through joint production. The assignment of novices to experts therefore determines the expertise novices carry into the future and, in turn, the quality of mentors available to subsequent cohorts. Because access to expertise cannot be financed through enforceable claims on future returns, novices rely on concurrent labor to pay for access to mentorship. Technological change can therefore affect productivity not only by changing how tasks are performed, but also by changing which experts novices learn from and the value of novice labor as a means of financing mentorship. Section \ref{sec:baseline} formalizes these margins.

\section{The Baseline Model} \label{sec:baseline}

Section \ref{sec:environment} introduces the environment and maps its key assumptions to the conceptual foundations of Section \ref{sec:rosen}. Section \ref{sec:formal} then describes agents’ problems, defines competitive equilibrium, and establishes existence and uniqueness. Doing so upfront allows the analysis to focus directly on long-run outcomes and the effects of technological shocks. To help readers navigate the notation used throughout the analysis, Appendix \ref{app:glossary} provides a glossary of the core mathematical symbols.

\subsection{The Environment and Its Interpretation} \label{sec:environment}

\noindent \textit{A Broad Overview of the Environment}.--- Time is discrete and infinite, indexed by $t = 0, 1, 2, \dots$. The economy consists of overlapping generations of risk-neutral individuals who live for two periods: first as novices, then as experts. Each cohort has unit mass, so the total population remains constant at two. All agents discount future payoffs by the common discount factor $\beta \in (0,1)$.

Novices are identical at birth, lack wealth, and inelastically supply one unit of labor.  Experts, in contrast, differ in their skill $q$. This skill level is observable, but the knowledge behind it is tacit. Consequently, external observers can perfectly assess an expert's productivity but cannot discern the insights or processes that drive those results without closer interaction.

Let $F_t(q)$ denote the cumulative distribution function (CDF) of expert skills at time $t$, with $x_t \equiv \min \operatorname{support} \{F_t\}$. This distribution is the aggregate state of the economy and evolves endogenously as novices work alongside experts and assimilate their expertise in the manner described below. The initial distribution $F_0$ is exogenously given, has a finite mean, and admits a continuous and strictly positive probability density function $f_0(q)$ on its support $[x_0, +\infty)$. 

There is a single final good, which serves as the numeraire. An expert produces output by combining her skill with a continuum of routine tasks in the $[0, N]$ interval. Tasks are not scalable: in each expert’s production process, every task must be completed exactly once. A unit measure of tasks can be completed using one unit of novice labor or, when feasible, one machine. For simplicity, experts do not supply routine labor themselves. 

Novices can perform all routine tasks with productivity $h$, while machines can perform only tasks in $[0,I] \subseteq [0,N]$, but do so with higher productivity $m>h$. The output of an expert is the product of her skill and a Cobb-Douglas aggregate of routine-task contributions. Thus, if an expert with skill $q$ decides to produce (is ``active'') and assigns a feasible measure $k \in [0, I]$ of automatable tasks to machines---and the remaining measure $N-k$ to novice labor---her output is given by:\footnote{See Appendix \ref{sec:production} for the relationship between this specification and standard task-based models \citep[e.g.][]{acemoglu2011chapter,acemoglu2018race,AcemogluRestrepoNBER}. Unlike those settings, tasks in this framework are non-scalable.}$$q m^{k} h^{N-k} $$

I restrict attention to the case in which experts require multiple novices to produce. As I formally show in Section \ref{sec:output}, this assumption is necessary for the economy to exhibit long-run growth:

\begin{assu}[Experts Need Multiple Novices] \label{assu:M1}
The measure of non-automatable tasks exceeds one:
\[
N - I > 1
\]
\end{assu}

Following \cite{AKR}, each machine is produced instantaneously using $\rho>0$ units of the final good and fully depreciates after one period, implying a rental rate of $\rho$. On the labor side, matching is many-to-one: each novice works for exactly one expert and uses her single unit of labor to perform a unit measure of routine tasks, while each active expert hires multiple novices to carry out all tasks not assigned to machines (a measure strictly greater than one by Assumption \ref{assu:M1}).

By working alongside an expert, a novice observes and assimilates the tacit insights behind the expert's skill. Hence, a novice matched with an expert of skill $q$ becomes an expert of identical skill in the subsequent period. In this sense, the employment relationship is also the channel through which tacit knowledge is transmitted.

However, the non-verifiability of tacit knowledge and the inalienability of the resulting human capital prevent novices from pledging future earnings as collateral. Lacking initial wealth, they face a liquidity constraint that forces wages to be non-negative ($w_t(q) \ge 0$). This leaves concurrent labor as the only means of compensating an expert for the opportunity to learn, giving rise to an apprenticeship-like arrangement.

The economic environment is thus parsimoniously described by a small set of primitives: the discount factor $\beta$, the initial distribution of skills $F_0$, the rental rate of machines $\rho$, the parameters of the production technology $(N, I, h, m)$, and the zero lower bound on wages.

\vspace{3mm}

\noindent \textit{Interpretation}.--- Having described the economic environment, I now relate the model's core assumptions to the conceptual framework of Section \ref{sec:rosen}.

First, for tractability, the baseline model imposes a strict division of labor between execution and judgment: novices and machines perform routine tasks, while experts orchestrate production. This structure mirrors knowledge-intensive industries, where early-career professionals (and, more recently, AI) execute well-structured tasks, while senior practitioners integrate those inputs into a cohesive whole. Section \ref{sec:robust} returns to this simplification by discussing an extension in which each expert chooses whether to act as an orchestrator or to supply one unit of routine labor. The section argues that allowing this occupational choice preserves the paper’s main insights.

Second, the assumption that tasks are non-scalable reflects experts’ limited time and attention: each expert can oversee only the measure $N$ of tasks required for a single production process. This formalizes a core principle of Section \ref{sec:rosen}: because expertise is embodied in the individual, its application and transmission cannot be easily scaled. In investment banking, for instance, a senior partner can oversee only a finite number of concurrent transactions, regardless of skill. Admittedly, this fixed-capacity assumption is stark: it rules out scale-of-operation effects. Section \ref{sec:robust} relaxes this restriction and examines the implications of the resulting scale effects.

Third, the model assumes that joint production is necessary for skill acquisition. By executing routine tasks, a novice gains direct insight into how the expert transforms these inputs into output, unconsciously learning ``the rules of the art.'' Online Appendices \ref{sec:plato} and \ref{sec:nu} relax this assumption by allowing skill development through independent innovation as well. Within joint production, I impose two further simplifications: routine tasks provide uniform access to the expert's methods, and skill transmission is exact, so novices always inherit their mentor's skill. Online Appendices \ref{sec:hetero} and \ref{sec:imperfect_transmission} study extensions that relax these two simplifications, and Section \ref{sec:further} discusses their implications.

Fourth, because novices lack initial wealth and cannot borrow, labor is their only means to pay for training ($w_t(q) \ge 0$). Rather than explicitly microfounding the contractual frictions that preclude such borrowing, I impose this non-negativity constraint as a reduced-form proxy for the contract incompleteness inherent in the transfer of tacit knowledge, as discussed in Section \ref{sec:rosen}.\footnote{Thus, novices’ inability to borrow should be interpreted broadly. One interpretation is an external credit constraint: because the tacit knowledge acquired through mentorship is non-verifiable and embodied in the novice, it cannot be pledged as collateral, so financial institutions may be unable to lend against its future returns. A related interpretation is a Becker-style appropriability constraint: these same features may prevent the expert or firm providing mentorship from obtaining an enforceable claim, whether in money or in kind, on the novice’s future returns to that expertise once she becomes an expert herself \citep{becker1964human, garicano2017relational, GR2025AI}. \label{foot:zlb}} A zero wage can be interpreted literally as unpaid labor (e.g., culinary \textit{stages}), or more broadly as novices accepting demanding hours relative to nominal pay, resulting in low effective compensation (e.g., medical residencies, or junior associates in law and finance).\footnote{Note that the model abstracts from the intensive margin of labor supply; this is purely an out-of-model interpretation.}

Fifth, the model treats experts’ skill levels as observable, while the tacit insights behind them remain hidden from outsiders. This should be interpreted not as literal perfect observability, but as a tractable polar case of directed sorting, in which novices rank potential mentors using observable markers of expertise, such as professional standing, past performance, institutional affiliation, or training lineage. Such directed sorting is consistent with the observation that novices in white-collar professions often sacrifice current compensation to work alongside top practitioners (see Section \ref{sec:rosen}).

Finally, the initial skill distribution, $F_0$, is assumed to have unbounded support. Following \citet{buera}, this assumption serves as a reduced-form approximation for an economy where an initial population with bounded skills draws ideas from an unbounded latent distribution (``Plato's realm of Forms''). Online Appendix \ref{sec:plato} formalizes this extension.

\subsection{Optimization, Matching, and Equilibrium} \label{sec:formal}

\noindent \textit{Experts' Problem}.---Given the machine rental rate $\rho$ and the wage schedule $w_t$, an expert with skill $q$ at time $t$ maximizes final-period income. Specifically, the expert chooses whether to produce or remain inactive. If active, the expert assigns a measure $k \in [0, I]$ of automatable tasks to machines. The expert's optimal income is therefore:
 $$\Pi_t(q) = \max \left\{ 0, \, \max_{k \in [0,I]} \left[ qm^{k} h^{N-k} - \rho k - w_t(q) (N-k) \right] \right\}$$

The following lemma establishes a central result for the subsequent analysis.

\begin{lem} \label{lem:cut} Even if novice labor is free ($w_t(q)=0$), full automation ($k=I$) is strictly optimal for all active experts with skill: $$ q >  \frac{\rho I}{h^N[ (m/h)^I-1]}$$ \end{lem}

\begin{proof} See Appendix \ref{app:thresh}. \end{proof}

Intuitively, high-skill experts benefit more from the productivity premium of machines $(m > h)$, while the direct cost advantage of using novice labor is bounded by the zero lower bound on wages. Thus, even when novice labor is free, $w_t(q)=0$, the productivity premium of machines dominates for sufficiently high $q$. Full automation is therefore strictly optimal for sufficiently skilled experts.

I then impose the following assumption:

\begin{assu}[Full Adoption of Automation Technologies] \label{assu:full} 
The lower bound of the initial skill distribution, $x_0$, satisfies: $$x_0 >  \frac{\rho I}{h^N[ (m/h)^I-1]}$$ Since all experts in subsequent periods possess skills $q \ge x_0$ (as shown below), all active experts at every date, regardless of their skill, optimally assign $I$ tasks to machines and hire $N - I$ novices.
\end{assu}

\noindent Online Appendix \ref{sec:full} shows that, for the equilibrium class analyzed here, this restriction affects only a finite initial transition. If the initial minimum skill lies below the full-automation threshold, some active experts may initially choose less than full automation. Skill accumulation, however, raises the economy’s minimum skill above the threshold in finite time, after which all active experts fully automate.

Given Assumption \ref{assu:full}, the optimal income of expert $q$ at time $t$ reduces to:\begin{equation} \label{eq:inc}
\Pi_t(q) \equiv \max\left\{0,\; S(q)-w_t(q)(N-I) \right\}
\end{equation}where $S(q) \equiv q m^{I} h^{N-I} - \rho I$ is the surplus generated by an active expert with skill $q$ before novice compensation. Expert $q$ at time $t$ weakly prefers producing to remaining inactive if and only if  $S(q) \ge w_t(q)(N-I)$. 

I restrict attention to equilibria in which the expert income schedule, $\Pi_t(q)$, is continuous and non-decreasing in $q$. The continuity restriction merely streamlines the exposition; one can show that any equilibrium featuring a non-decreasing income schedule is necessarily continuous. The monotonicity requirement, in turn, selects outcomes robust to the free disposal of skills---where an expert can costlessly mimic a less-skilled peer. Under this selection, and up to zero-profit production indifferences, the set of active experts at time $t$, denoted by $A_t$, can be represented as an upper interval: \[
A_t \equiv  [a_t, \infty), \ \text{for some $a_t \ge x_t$ }
\]where, by convention, the cutoff expert $a_t$ is included in $A_t$.

\vspace{3mm}

\noindent \textit{Novices' Problem}.--- Given the current wage schedule $w_t$ and next period's expert income schedule $\Pi_{t+1}$, novices at time $t$ evaluate potential matches by the lifetime income they deliver. Matching with an expert of skill $q$ yields: $$w_t(q) + \beta \Pi_{t+1}(q)$$
where $\Pi_{t+1}(q)$ is the optimal income of an expert with skill $q$ in period $t+1$, as defined by (\ref{eq:inc}). Throughout, whenever this continuation schedule is evaluated below the support of $F_{t+1}$, I use the convention $\Pi_{t+1}(q)=0$; equivalently, $\Pi_{t+1}(q)=0$ for $q<x_{t+1}$.

\vspace{3mm}

\noindent \textit{Within-Period Competitive Matching}.--- I now characterize the equilibrium outcome at time $t$, taking $F_t$ and $\Pi_{t+1}$ as given. This requires determining the wage schedule and matching arrangement that prevails in that period.

I model within-period matching as a competitive assignment subject to experts’ fixed capacity and the zero lower bound on wages. Apart from these constraints, there are no search frictions, informational asymmetries, or coordination failures. Expert-novice matching at time $t$ is therefore represented as an assignment from novices to the set of active experts $A_t$, with inactive experts receiving no novices. Because novices are ex ante identical, this assignment is summarized by a finite Borel measure $\Lagr_t$ on $A_t$, where $\Lagr_t(E)$ denotes the mass of novices assigned to experts in the measurable set $E\subseteq A_t$.

A feasible assignment must satisfy two conditions. First, each active expert optimally hires exactly $N-I$ novices, so:
\begin{equation}\label{eq:feas}
\Lagr_t(E)=(N-I)\int_E dF_t(q), \ \ \text{for any measurable $E\subseteq A_t$}
\end{equation}
Second, novices have unit mass and supply labor inelastically, so all novices must be assigned: $\Lagr_t(A_t)=1$. Since $A_t=[a_t,\infty)$, these two feasibility conditions together imply the aggregate market-clearing condition:\begin{equation}\label{eq:mktclearing}
1=(N-I)\bigl[1-F_t(a_t)\bigr].
\end{equation}

The next definition adapts the canonical static, frictionless assignment framework \citep[e.g.,][Section 2.1]{becker1973theory,eeckhout2011identifying} to an environment with a zero lower bound on wages:

\begin{mydef}[Within-Period Competitive Matching] \label{def:within} Given $F_t$ and $\Pi_{t+1}$, a within-period competitive matching consists of a set of active experts $A_t =[a_t,\infty)$, an assignment $\Lagr_t$ on $A_t$, and a non-negative wage schedule $w_t(q)$ such that:\vspace{-2mm} \begin{enumerate}[leftmargin=*,itemsep=0.05em]
\item \textbf{Feasibility:} The assignment satisfies (\ref{eq:feas}) and (\ref{eq:mktclearing}).
\item \textbf{Expert Participation:} The expert income schedule is given by:  $$\Pi_t(q)= \left\{\begin{array}{cl}  0 & \text{if $q \in [x_t, a_t)$} \\ 
S(q) - (N-I)w_t(q) & \text{if $q \in [a_t, \infty)$} \end{array} \right.$$with $S(q)-(N-I)w_t(q)\le 0$ for all $q\in[x_t,a_t)$, and $\Pi_t(q) \ge 0$ for all $q \in [a_t, \infty)$.
\item \textbf{Stability:} There do not exist an expert $q' \in [x_t, \infty)$, a wage $w' \ge 0$, and an active expert $q \in [a_t, \infty)$ such that:$$S(q') - (N-I)w' > \Pi_t(q')  \quad \text{and} \quad w' + \beta\Pi_{t+1}(q') > w_t(q) + \beta\Pi_{t+1}(q). $$  \end{enumerate} \end{mydef}

Feasibility and expert participation restate, respectively, the conditions defining a feasible assignment and the income implications for active and inactive experts. The new condition in Definition \ref{def:within} is stability. It rules out any expert---active or inactive---offering a nonnegative wage that both raises her income and gives novices a strictly higher lifetime payoff than they receive from some active match. Thus, although the zero lower bound may prevent wages from fully clearing access to the most desirable experts, the realized matching remains competitively stable: no mutually profitable poaching opportunity remains available at non-negative wages.

Online Appendix \ref{sec:within} then shows that an (essentially) unique within-period competitive matching exists whenever $\Pi_{t+1}$ is continuous and non-decreasing, and $F_t$ admits a continuous and strictly positive density on its support $[x_t, \infty) \subseteq [x_0,\infty)$.\footnote{The qualification ``essentially'' reflects the fact that the equilibrium pins down only a lower bound on the wages of inactive experts, not a unique wage schedule for them. Specifically, for $q<a_t$, equilibrium requires $w_t(q)\geq S(q)/(N-I)$, so that production is unprofitable. The exact wage schedule for inactive experts is therefore indeterminate. This indeterminacy is immaterial because inactive experts neither produce nor hire novices, nor do they affect payoffs or allocations.}

In equilibrium, the aggregate market-clearing condition (\ref{eq:mktclearing}) uniquely determines the active set $A_t =[a_t,\infty)$, and the assignment $\Lagr_t$ follows immediately from (\ref{eq:feas}). Moreover, because $N-I>1$ (Assumption \ref{assu:M1}), only a strict subset of experts is active in each period, i.e., $a_t>x_t$. Stability then requires the marginal active expert to earn zero income, $\Pi_t(a_t)=0$, implying that $w_t(a_t)=S(a_t)/(N-I)>0$.\footnote{That $w_t(a_t)>0$ follows from the strict monotonicity of $S(\cdot)$ and the fact that Assumption \ref{assu:full} also implies that $S(x_0)>0$. Together, these imply $S(a_t)>S(x_t)\ge S(x_0)>0$.} The lifetime income of a novice (``novice utility'') matched with expert $a_t$ at time $t$ is then:\begin{equation} \label{eq:u} U_t \equiv w_t(a_t) + \beta \Pi_{t+1}(a_t) = \frac{S(a_t)}{N-I} + \beta \Pi_{t+1}(a_t) \end{equation}

Online Appendix \ref{sec:within} also shows that stability of the competitive matching further implies that the wage schedule can be written as:\begin{equation}\label{eq:wages} w_t(q) = \left\{ \begin{array}{cl} S(a_t)/(N-I) & \text{if $q \in [x_t,a_t)$} \\
 \max\{0,U_t-\beta\Pi_{t+1}(q)\}  & \text{if $q \in [a_t, \infty)$}  \end{array} \right. \end{equation}For inactive experts ($q < a_t$), this wage schedule ensures that production is unprofitable. For active experts ($q \ge a_t$), the schedule implies that those with $\beta \Pi_{t+1}(q) < U_t$ pay strictly positive wages and deliver novices the baseline utility $U_t$. By contrast, experts with $\beta \Pi_{t+1}(q) \ge U_t$ pay zero wages and deliver lifetime income weakly above $U_t$. Consequently, $U_t$ is the minimum lifetime income any novice can secure at time $t$.

Although the framework abstracts from explicit ``queuing'' or ``rationing'' by specifying only a realized assignment rather than an explicit matching protocol, these two cases admit a natural interpretation. Positions that deliver exactly $U_t$ are \emph{price-cleared}: competition drives wages down to offset the future value of superior training, leaving novices indifferent among these positions. In contrast, positions delivering strictly more than $U_t$ are \emph{oversubscribed}: the zero lower bound prevents further adjustment, so novices strictly prefer these matches. Access to such capacity-limited positions is therefore determined by the realized assignment, generating latent ``excess demand.''

\vspace{3mm}

\noindent \textit{Intertemporal Linkages}.--- Having characterized the equilibrium at time $t$ given $F_t$ and $\Pi_{t+1}$, I now turn to the intertemporal linkages.

The first linkage runs from future expert incomes to current expert incomes through current wages. Because novices inherit their mentor’s skill, experts extract the value of mentorship through lower wages, subject to the zero lower bound. Substituting the active-expert branch of the wage schedule in (\ref{eq:wages}) into (\ref{eq:inc}) makes this dependence explicit: $$\Pi_t(q) = \max\{0, S(q) - (N-I)\overbrace{\max\{0, U_t - \beta\Pi_{t+1}(q)\}}^{\text{$=w_t(q)$ for $q \ge a_t$} }\}$$

The second linkage is the law of motion for the distribution of experts’ skills. Because each novice inherits her mentor's skill, and active experts each hire $N-I$ novices, the skill distribution evolves according to:\begin{equation} \label{eq:evolution} F_{t+1}(q) = \left\{ \begin{array}{cl} 0 & \text{if $q  \le a_t$} \\ (N-I) \left[ F_t(q)-F_t(a_t)\right]  & \text{if $q>a_t$} \end{array} \right. \end{equation}

\begin{figure}[t!]
\centering
   \includegraphics[scale=0.49]{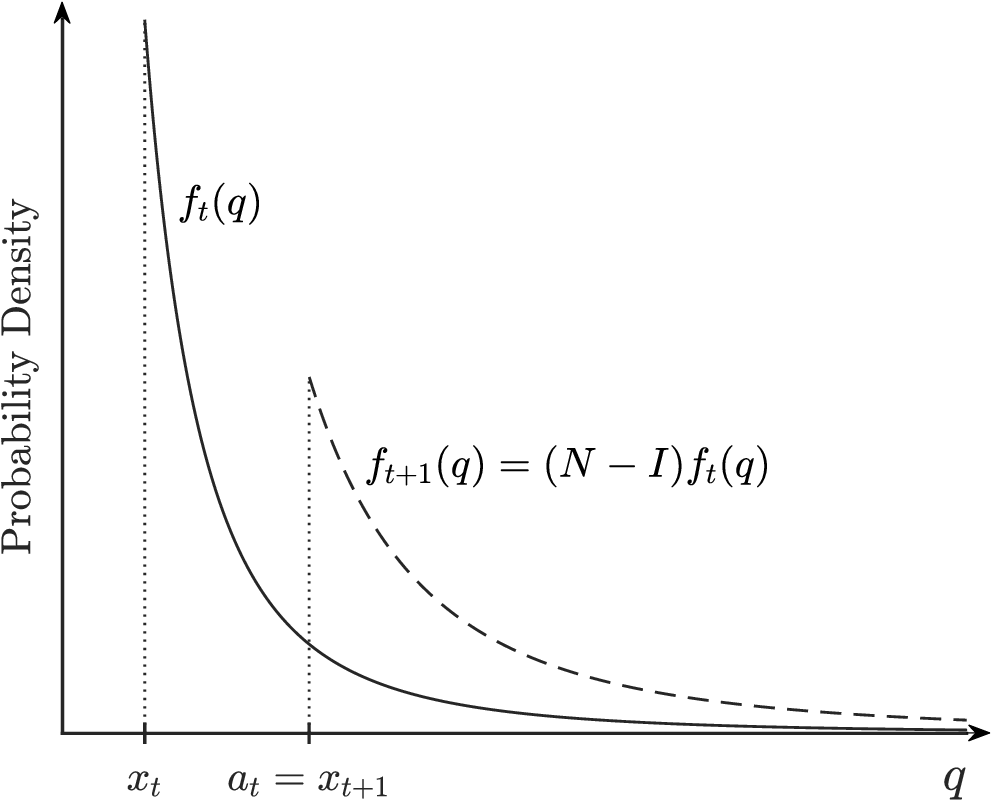}
  \captionsetup{justification=centering}
\caption{The Evolution of the Distribution of Experts' Skills over Time} \label{fig:evolution}
\end{figure}

Equation (\ref{eq:evolution}) defines a valid CDF, with the aggregate market-clearing condition (\ref{eq:mktclearing}) ensuring that $\lim_{q \to \infty} F_{t+1}(q)=1$. Figure \ref{fig:evolution} illustrates how the density $f_{t+1}(q) \equiv dF_{t+1}(q)/dq$ is obtained by truncating $f_t(q) \equiv dF_t(q)/dq$ at the marginal active expert $a_t$ and rescaling the surviving upper tail by experts’ \textit{span of control}, $N-I$. Hence, the support of $F_{t+1}$ is $[x_{t+1},\infty)=[a_t,\infty) \subseteq [x_t,\infty) \subseteq [x_0,\infty)$. Moreover, because $F_0$ admits a continuous and strictly positive density on $[x_0,\infty)$, equation (\ref{eq:evolution}) implies that each $F_t$ admits a continuous and strictly positive density on its support for all $t>0$. Thus, the regularity conditions on $F_t$ needed for the existence and uniqueness of within-period competitive matching hold at every date.

\vspace{3mm}

\noindent \textit{Dynamic Competitive Equilibrium}.--- Given the current skill distribution $F_t$ and a continuation income schedule $\Pi_{t+1}$, within-period matching determines the active set $A_t=[a_t,\infty)$, baseline novice utility $U_t$, wage and income schedules $w_t$ and $\Pi_t$, and assignment $\Lagr_t$ in period $t$. The active set then determines the next period’s skill distribution $F_{t+1}$, which in turn governs the subsequent matching market and continuation income schedule $\Pi_{t+1}$. A dynamic competitive equilibrium requires these objects to be mutually consistent for all $t \ge 0$, given the initial distribution $F_0$. 

\begin{mydef}[Dynamic Competitive Equilibrium] \label{def:equilibrium} A dynamic competitive equilibrium consists of the following sequences.\begin{itemize}[leftmargin=*,noitemsep]
\item Skill distributions $\{F_t\}_{t \ge 0}$, evolving according to the law of motion (\ref{eq:evolution}) with $F_0$ given.
\item Marginal active experts $\{a_t\}_{t \ge 0}$, satisfying the aggregate market-clearing condition (\ref{eq:mktclearing}).
\item Assignments $\{\Lagr_t\}_{t \ge 0}$, constructed using (\ref{eq:feas}).
\item Baseline novice utilities $\{U_t\}_{t \ge 0}$ alongside wage and income schedules $\{w_t, \Pi_t\}_{t \ge 0}$, jointly determined by (\ref{eq:inc}), (\ref{eq:u}), and (\ref{eq:wages}).
\end{itemize}
 \end{mydef}
 
The equilibrium features a block-recursive structure.  Given the current distribution $F_t$, the aggregate market-clearing condition (\ref{eq:mktclearing}) uniquely determines the active cutoff $a_t$. Because $F_t$ and $a_t$ fully characterize $F_{t+1}$ via equation (\ref{eq:evolution}), the dynamical system of allocations $\{a_t, F_t\}_{t \ge 0}$ evolves autonomously forward in time. In contrast, equilibrium prices---utilities, wages, and incomes---are forward-looking. Because they depend on future states, they cannot be determined from the current state alone. The intuition for how this system is solved is developed in Section \ref{sec:sch}.

For now, I simply state the existence and uniqueness of equilibrium. With these technical foundations secured, Section \ref{sec:longrun} proceeds directly to characterize the economy’s long-run outcomes.

\begin{prop}[Existence and Uniqueness of a Dynamic Competitive Equilibrium] \label{prop:existence} For any initial distribution $F_0$ satisfying the model's assumptions, there exists a unique dynamic competitive equilibrium within the class where expert income $\Pi_t(q)$ is continuous and non-decreasing in $q$. This equilibrium is block-recursive: the sequence of allocations $\{a_t, F_t\}_{t \ge 0}$ is determined independently of the sequence of prices $\{U_t, w_t, \Pi_t\}_{t \ge 0}$. Moreover, for all $t \ge 0$, the wage schedule $w_t(q)$ is non-increasing in $q$. \end{prop}

\begin{proof} See Online Appendix \ref{sec:exist} for the constructive proof. Online Appendix \ref{sec:cross} additionally characterizes the cross-sectional properties of the equilibrium at any given date $t$. \end{proof}

\noindent \textit{Aggregate Output and Welfare}.--- Since only active experts produce, aggregate output at time $t$ is: \begin{equation} \label{eq:output} Y_t \equiv \left[  \int_{a_t}^{\infty} q dF_t(q) \right]  m^{I} h^{N-I}  \end{equation}

Welfare at time $t$ is defined as the discounted sum of the aggregate income of novices and experts from period $t$ onward: \begin{equation} \label{eq:welfare} W_t \equiv \sum_{j=t}^{\infty} \beta^{j - t}  \left[ \int_{a_j}^{\infty} S(q) dF_j(q) \right] =    \sum_{j=t}^{\infty} \beta^{j - t}  \Big(Y_j - \rho I [1-F_j(a_j) ]\Big)\end{equation}This is also equivalent to the discounted sum of aggregate production minus total machine costs. 

\section{Long-Run Equilibrium Outcomes} \label{sec:longrun}

This section characterizes the economy’s long-run behavior. I show that sustaining strictly positive output growth hinges on the initial skill distribution $F_0$ having sufficiently fat tails. When this condition holds, the asymptotic growth rate converges to a constant determined by experts’ span of control and the tail thickness of $F_0$. I then derive the asymptotic distribution of normalized expert skills, the limiting wage and income schedules, and the asymptotic baseline utility of novices.

These results set the stage for Section \ref{sec:advances}, which examines the implications of different technological advances.

\subsection{Growth Rate of Output} \label{sec:output}

For fixed technology parameters, aggregate output in \eqref{eq:output} depends solely on the skill distribution $F_t$ and the entry cutoff $a_t$. As established in Section \ref{sec:formal}, both evolve independently of equilibrium prices: the sequence $\{a_t, F_t\}_{t \ge 0}$ solves the dynamical system defined by the market-clearing condition (\ref{eq:mktclearing}) and the law of motion (\ref{eq:evolution}). Given $F_0$, the unique solution is:\begin{equation} \begin{split} \label{eq:cutF}  1 = (N-I)^{t+1}[1-F_0(a_t)], \ & \text{for $t \ge 0$} \\
    1-F_t(q) = (N-I)^t[1-F_0(q)], \ & \text{for $t \ge 1$ and $q \ge x_t = a_{t-1}$}  \end{split}\end{equation}
Since $N-I > 1$ and $F_0$ has unbounded support, the sequence $\{a_t\}_{t \ge 0}$ is strictly increasing and diverges as $t \to \infty$. Substituting this solution into \eqref{eq:output} yields:
 \begin{equation*} \begin{split} Y_t &= m^I h^{N-I} (N-I)^t[1-F_0(a_t)]\, \mathbb{E}_0[q \mid q \ge a_t] \\
    & = \frac{m^I h^{N-I}}{N-I} \, \mathbb{E}_0[q \mid q \ge a_t]  \end{split} \end{equation*}
where $\mathbb{E}_0[\cdot]$ denotes the expectation with respect to the initial distribution $F_0$.
    
Because $\mathbb{E}_0[q \mid q \ge a_t] \ge a_t$ and $a_t \to \infty$, aggregate output grows without bound ($Y_t \to \infty$). Whether the economy sustains strictly positive asymptotic growth, however, depends on the thickness of the right tail of the initial distribution. To formalize this, I classify the tail behavior of $F_0$ using regular variation \citep{bingham1989regular}:

\begin{mydef}[Tail Behavior] $\quad$  \vspace{-2mm}
\begin{enumerate}[leftmargin=*]
\item \textbf{Regularly Varying (i.e., Fat) Tails.} $1-F_0$ is regularly varying with index $-1/\theta$ (denoted $1-F_0 \in \mathrm{RV}_{-1/\theta}$) if, for all $\lambda>0$,
\[
\lim_{q \to \infty} \frac{1-F_0(\lambda q)}{1-F_0(q)} = \lambda^{-1/\theta}.
\]

\item \textbf{Rapidly Varying (i.e., Thin) Tails.} $1-F_0$ is rapidly varying (denoted $1-F_0 \in \mathrm{RV}_{-\infty}$) if, for all $\lambda>1$,
\[
\lim_{q \to \infty} \frac{1-F_0(\lambda q)}{1-F_0(q)} = 0.
\]
\end{enumerate}
\end{mydef}

\begin{rmk} \label{rmk:finite} Under regular variation, finite mean requires $0<\theta<1$. I maintain this restriction throughout. \end{rmk}

\begin{rmk} Regular variation corresponds to power-law (Pareto-type) tails, as in Pareto and Fr\'echet distributions. Rapid variation corresponds to tails that decay faster than any power law, including exponential, Weibull, and log-normal distributions. \end{rmk}

\begin{rmk} This classification is not exhaustive. Other tail behaviors that are neither regularly nor rapidly varying are possible---for example, power-like tails with persistent oscillations---but such cases need not deliver a well-defined limiting growth factor. \end{rmk}

As shown in \citet[][Appendix A.3]{embretch} and \citet{perla2014equilibrium}, these tail classes imply the following limits for conditional means:
\begin{equation} \label{eq:Erg}
\lim_{a \to \infty} \frac{\mathbb{E}_0[q \mid q \ge a]}{a}
=
\left\{ \begin{array}{cl}
\dfrac{1}{1-\theta} & \text{if $1-F_0 \in \mathrm{RV}_{-1/\theta}$} \\
1 & \text{if $1-F_0 \in \mathrm{RV}_{-\infty}$}.
\end{array} \right.
\end{equation}Combining the cutoff dynamics in \eqref{eq:cutF} with the conditional-mean limits in \eqref{eq:Erg} yields the following result:

\begin{prop}[Asymptotic Growth Rate of Output] \label{prop:growth} In equilibrium, \vspace{-2mm}
\[
\lim_{t\to\infty}\frac{a_{t+1}}{a_t}
=
\lim_{t\to\infty}\frac{Y_{t+1}}{Y_t} \equiv 1+ g
= \left\{ 
\begin{array}{cl}
(N-I)^{\theta} &  \text{if $1-F_0 \in \mathrm{RV}_{-1/\theta}$} \\
1 & \text{if $1-F_0 \in \mathrm{RV}_{-\infty}$}
\end{array} \right.
\] \end{prop}

\begin{proof} See Appendix \ref{app:growth}. \end{proof}

Although the formal proof is in the appendix, I provide a brief sketch of the mechanics behind the result here. First, note that (\ref{eq:cutF}) implies the following relationship between successive cutoffs:
\[
 \frac{1- F_0(a_{t+1})}{1- F_0(a_t)}=\frac{1}{N-I}.
\]
Define the cutoff growth factor by $\lambda_t \equiv a_{t+1}/a_t>1$. When $1-F_0$ is regularly varying with index $-1/\theta$ and $a_t\to\infty$, the tail ratio above behaves heuristically like:\[
\frac{1-F_0(a_{t+1})}{1-F_0(a_t)}
=\frac{1-F_0(\lambda_t a_t)}{1-F_0(a_t)} \approx \lambda_t^{-1/\theta}
 \ \  \text{for large } t.
\]
Since the ratio on the left is exactly $1/(N-I)$, this approximation suggests that $\lambda_t^{-1/\theta}\to 1/(N-I)$ as $t \to \infty$, and hence $a_{t+1}/a_t\to (N-I)^{\theta}= 1+g$.

To derive the asymptotic growth rate of output, decompose the ratio $Y_{t+1}/Y_t$ using $Y_t \propto  \mathbb{E}_0[q \mid q \ge a_t]$. Since $a_t \to \infty$, the asymptotic behavior of the conditional mean-to-cutoff ratio $\mathbb{E}_0[q \mid q \ge a_t]/a_t$ is governed by (\ref{eq:Erg}). Consequently: \[
\lim_{t \to \infty }\frac{Y_{t+1}}{Y_t}
=
\left( \lim_{t \to \infty}\frac{a_{t+1}}{a_t} \right)
\left( \lim_{t \to \infty} \frac{\mathbb{E}_0[q \mid q \ge a_{t+1}]}{a_{t+1}}\right)
\left( \lim_{t \to \infty} \frac{a_{t}}{\mathbb{E}_0[q \mid q \ge a_{t}]} \right)
=
1+g
\]

The rapidly varying case can be viewed heuristically as the limit $\theta \to 0$ of the regularly varying case, for which the growth factor $1+g=(N-I)^{\theta}$ converges to one. Consistent with this intuition, if $1-F_0$ is rapidly varying, then $a_{t+1}/a_t \to 1$ and $Y_{t+1}/Y_t \to 1$ as $t \to \infty$. This completes the sketch.

The economic forces underlying Proposition \ref{prop:growth} are as follows. In this economy, output grows because the most skilled experts each transmit their skill to $N-I>1$ novices. This replication expands the mass of high-skilled experts over time, pushing the active cutoff upward and raising the skill of producing experts.

Replication alone, however, is insufficient for strictly positive asymptotic growth. Sustained growth requires the cutoff $a_t$ to keep rising at a non-vanishing proportional rate, which---by the aggregate market-clearing condition \eqref{eq:mktclearing}---depends on the upper tail of $F_t$. Because skills are inherited, that upper tail is simply a rescaled version of the upper tail of the initial distribution $F_0$ (see (\ref{eq:cutF})).

If $F_0$ features a fat upper tail, multiplying the cutoff $a_t$ by any fixed factor $\lambda>1$ reduces tail mass by a corresponding constant factor, so the decline is asymptotically offset by the constant tail reweighting generated by replication. The cutoff can therefore continue to grow at a non-vanishing proportional rate. If instead the upper tail is thin, multiplying $a_t$ by a fixed factor reduces tail mass too sharply for the constant reweighting from replication to offset. The cutoff continues to rise, but its growth factor converges to one, and asymptotic growth vanishes.

These two forces---replication of high-skill experts and scarcity in the upper tail---are summarized by the asymptotic growth factor $1+g=(N-I)^{\theta}$. Here, experts' span of control, $N-I$, governs skill replication, while $\theta$ captures the thickness of the upper tail.\footnote{As the previous discussion makes clear, strictly positive asymptotic growth requires an unbounded, fat-tailed source of expertise. This assumption should not be read literally as requiring the realized initial population to contain experts of arbitrarily high skill. Rather, following the ``Plato’s realm of Forms'' interpretation discussed by \citet{buera}, it can be viewed as a reduced-form representation of an economy in which agents with bounded skills draw ideas from a fixed, unbounded latent distribution. Online Appendix \ref{sec:plato} formalizes this interpretation by showing that the same asymptotic growth logic obtains when the realized initial skill distribution is bounded, but the latent distribution of ideas has a regularly varying upper tail. In that version, the relevant tail condition applies to the latent distribution of ideas, not to the realized initial population of experts.} As $N-I \rightarrow 1$ or $\theta \rightarrow 0$, the growth factor converges to one, and asymptotic growth vanishes.\footnote{A sharper statement holds for the $N-I \to 1$ case: Online Appendix \ref{sec:Nrep} shows that if $N-I \le 1$, asymptotic growth is exactly zero.} Strictly positive asymptotic growth is therefore obtained in the regularly varying case when $N-I>1$.

Assumption \ref{assu:M1}  ensures that $N-I>1$. To focus on this sustained-growth case with a well-defined asymptotic growth factor, I additionally assume hereafter that $1-F_0 \in \mathrm{RV}_{-1/\theta}$. This, in turn, requires $\beta(1+g)<1$ to ensure that overall welfare as defined by (\ref{eq:welfare}) remains finite. Thus, for the remainder of the paper, I assume:

\begin{assu}[Sustained Growth and Bounded Welfare] \label{assu:growth} $1-F_0 \in \mathrm{RV}_{-1/\theta}$, where $\theta \in (0,1)$. Furthermore, $\beta (1+g) < 1$, where $1+g \equiv (N-I)^\theta$.\end{assu}

\subsection{Distribution of Experts' Skill} \label{sec:skills}

I now turn to the asymptotic distribution of experts' skills. Since the minimum skill level $x_t$ diverges ($x_t = a_{t-1} \to \infty$), the distribution of absolute skills does not converge. I therefore focus on the distribution of normalized skills, defined as each expert’s skill relative to the minimum skill of the current cohort, $z \equiv q/x_t$.

Formally, let $\Phi(z)$ denote the asymptotic CDF of $z$: $$1 - \Phi(z) \equiv \lim_{t \to \infty} \mathbb{P}_t\left(\frac{q}{x_t} > z\right) = \lim_{t \to \infty} \frac{1 - F_0(zx_t)}{1 - F_0(x_t)}$$where the second equality follows from (\ref{eq:cutF}) (see Appendix \ref{app:dist} for details). A direct application of regular variation then yields the following result:

\begin{prop}[Asymptotic Distribution of Experts' Skills] \label{prop:dist} Suppose $1-F_0 \in \mathrm{RV}_{-1/\theta}$. Then, the asymptotic distribution of normalized skills is a Pareto distribution with scale $1$ and shape $1/\theta$: $$ \Phi(z) = 1 - z^{-1/\theta}, \ \ \text{for } z \ge 1.$$Consequently, for large $t$, the distribution of absolute experts' skills $F_t(q)$ is approximately Pareto with a time-varying scale $x_t$ and a constant shape parameter $1/\theta$. \end{prop} 

\begin{proof} See Appendix \ref{app:dist}. \end{proof}

Intuitively, regular variation implies that the upper tail of $F_0$ is arbitrarily close to a Pareto tail for large $q$. As established in Section \ref{sec:output}, the equilibrium dynamics repeatedly truncate this initial distribution and reweight the surviving upper tail by experts' common span of control $N-I$, operations that preserve its Pareto tail for large $q$. Thus, over time, the truncation point moves farther into the rescaled upper tail of $F_0$, so the economy draws experts from regions where the distribution is arbitrarily close to Pareto. 

\subsection{Long-Run Baseline Utility, and Income and Wage Schedules} \label{sec:sch}

Finally, I characterize the asymptotic normalized baseline novice utility, $u \equiv \lim_{t \to \infty} U_t/x_t$, together with the asymptotic normalized expert income and wage schedules:$$\pi(z) \equiv \lim_{t \to \infty} \frac{\Pi_t(z x_t)}{x_t}, \qquad \omega(z) \equiv \lim_{t \to \infty} \frac{w_t(z x_t)}{x_t}$$

To start, recall from Section \ref{sec:formal} that $U_t \equiv S(a_t)/(N-I)+\beta\Pi_{t+1}(a_t)$. However, along the equilibrium path, $a_{t+1}>x_{t+1}=a_t$. Thus, a novice who inherits the marginal active skill $a_t$ at time $t$ becomes an inactive expert in period $t+1$, so $\Pi_{t+1}(a_t)=0$. Consequently, $U_t$ simplifies to: \[ U_t \equiv  \frac{S(a_t)}{N-I} + \beta \overbrace{\Pi_{t+1}(a_t)}^{=0} = \frac{S(a_t)}{N-I}\]

Normalizing by $x_t$, using $a_t/x_t\to 1+g$, and noting that the fixed machine cost becomes negligible asymptotically, the normalized baseline utility thus converges to:
\[  u \equiv \lim_{t \to \infty} \frac{U_t}{x_t} = \lim_{t \to \infty} \frac{a_t m^I h^{N-I} - \rho I}{x_t (N-I)}= \frac{m^I h^{N-I}}{N-I} (1+g) \]

\begin{figure}[t!]
\centering
\begin{subfigure}{.47\textwidth}
  \centering
    \includegraphics[scale=0.42]{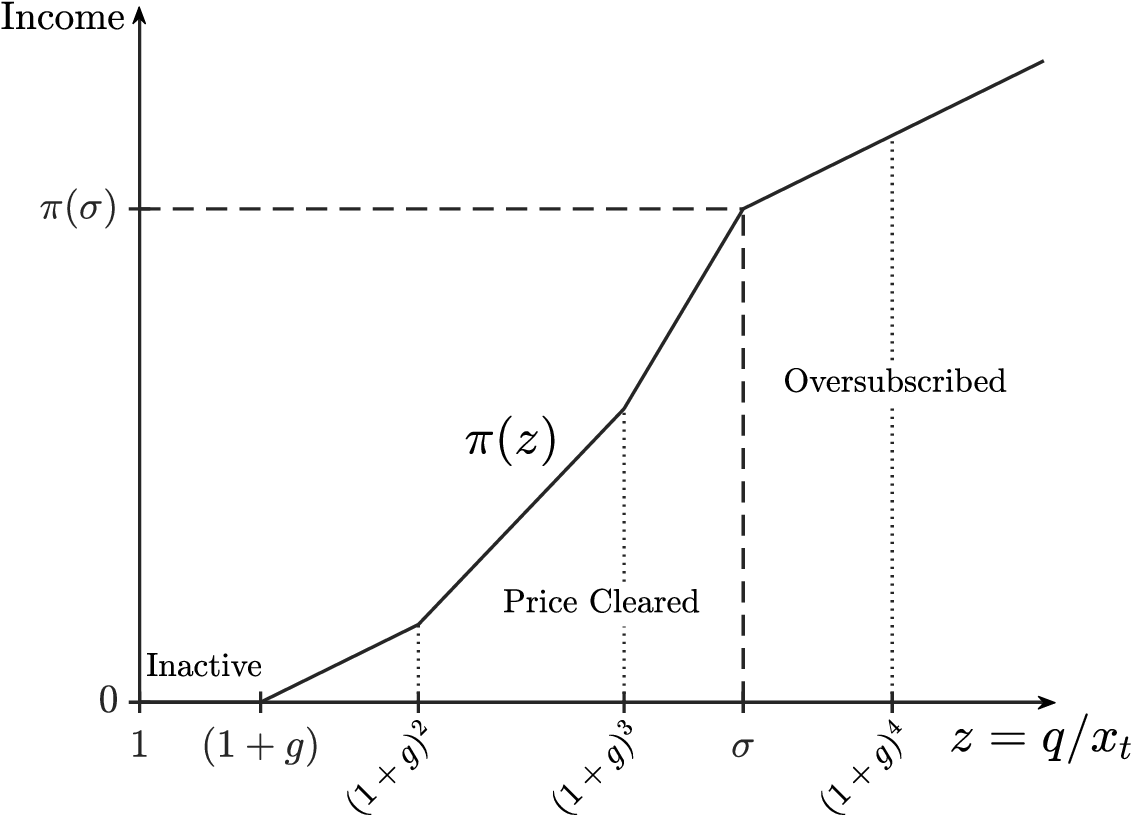}
  \caption{Asymptotic Income Schedule}
\end{subfigure} 
\begin{subfigure}{.52\textwidth}
  \centering
    \includegraphics[scale=0.42]{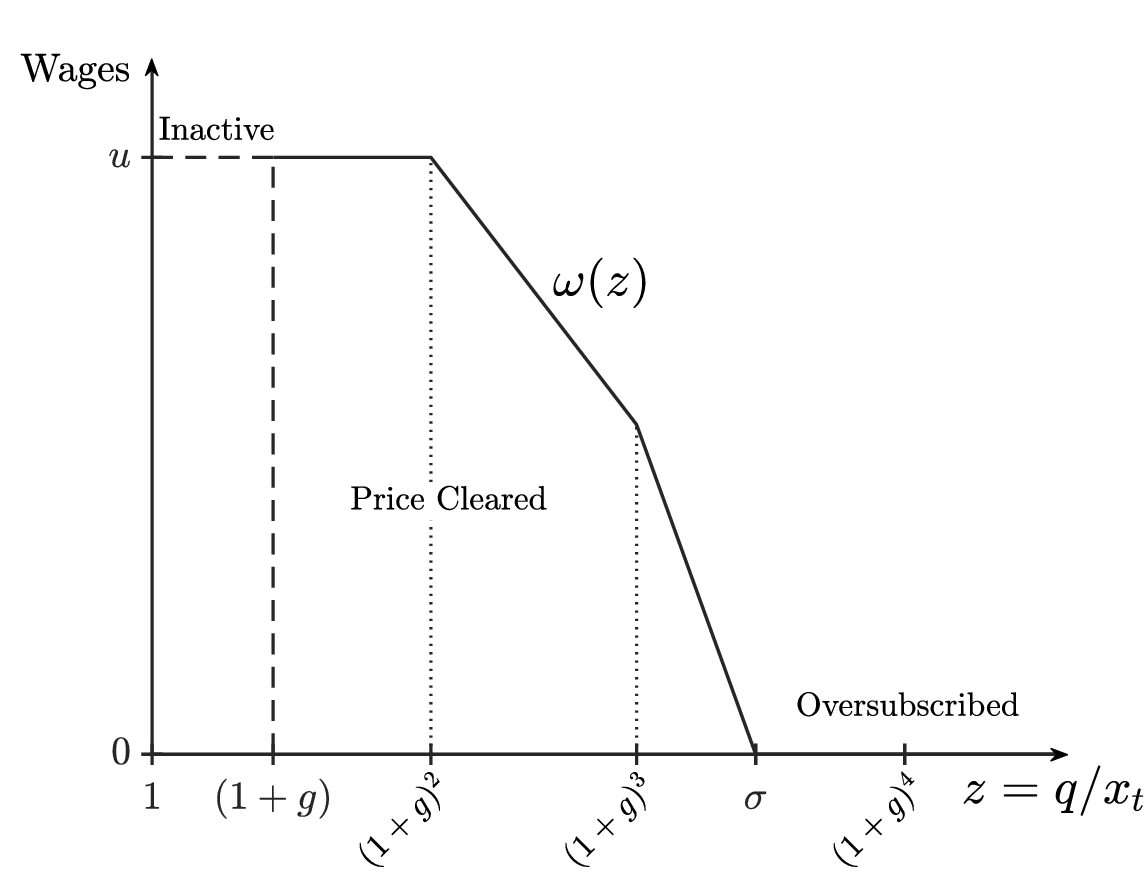}
  \caption{Asymptotic Wage Schedule}
\end{subfigure}
  \captionsetup{justification=centering}
\caption{The Construction of the Asymptotic Income and Wage Schedules \\  \justifying 
 \vspace{0.5mm}
\footnotesize \noindent \textit{Notes.} Parameter Values: $N=2.7$, $I=1$, $h=1$, $m=2$, $\beta=0.96^{10}$, and $\theta=0.5$. Panel (a) shows the asymptotic normalized income schedule $\pi(z)$, and panel (b) shows the asymptotic normalized wage schedule $\omega(z)$. The vertical dotted lines denote consecutive powers of the growth factor $(1 + g)^j$, marking the skill thresholds where the remaining active lifespans of expert dynasties increase by one period before obsolescence. These intervals create the piecewise kinks in both schedules throughout the price-cleared region $z \in [1 + g, \sigma)$.} \label{fig:asymptotics}
\end{figure}

I now turn to $\pi(z)$ and $\omega(z)$. Before deriving closed-form expressions, it is useful to describe the intuition behind their construction. To do this, consider panels (a) and (b) of Figure \ref{fig:asymptotics}, which plot $\pi(z)$ and $\omega(z)$, respectively. As shown, the expert population is partitioned into three intervals defined by the endogenous cutoffs $1+g$ and $\sigma$: 
\begin{enumerate}[noitemsep]
\item \textbf{Inactive:} Experts in $[1, 1 + g)$ who do not produce.
\item \textbf{Price-Cleared:} Experts in $[1 + g, \sigma)$ who pay strictly positive wages. 
\item \textbf{Oversubscribed:} Experts in $[\sigma, +\infty)$ for whom the zero lower bound binds.
\end{enumerate}

The construction of this partition and of the resulting schedules follows from the fact that the asymptotic normalized skill of the marginal active expert converges to $1+g$, i.e., $\lim_{t \to \infty} (a_t/x_t) = 1+g$. This implies that the absolute cutoff for becoming an active expert grows asymptotically by a factor of $1+g$ each period. Consequently, any fixed absolute skill level eventually becomes obsolete, so all future experts with that skill become inactive and earn zero income thereafter. The asymptotic schedules $\pi(z)$ and $\omega(z)$ can therefore be constructed by tracing the economic lifecycle of a given skill backward from obsolescence, subject to the zero lower bound on wages.

Figure \ref{fig:asymptotics} illustrates this process. The horizontal axis in both panels is partitioned into intervals defined by consecutive powers of the growth factor, $[(1 + g)^j, (1 + g)^{j+1})$ for $j=0,1, \dots$. All experts in $[1, 1 + g)$ are inactive. Experts with $z \in [1 + g, (1 + g)^2)$ are active, but their novices inherit a relative skill $z / (1 + g) \in [1, 1 + g)$ and become inactive in the subsequent period. Expecting zero future income, these novices must be paid the normalized baseline utility $\omega(z) = u$ upfront, generating the initial upward-sloping segment of $\pi(z)$ and the initial flat segment of $\omega(z)$.

In the next interval, $z \in [(1 + g)^2, (1 + g)^3)$, novices inherit a relative skill $z / (1 + g) \in [1 + g, (1 + g)^2)$. These novices become active experts in the following period and pay the baseline wage $u$ to their successors (who become inactive experts two periods later). This implies that current novices have a strictly positive continuation value that is strictly increasing in $z$, which current experts extract by reducing the upfront wage. Consequently, $\omega(z)$ begins to fall strictly below $u$ and decreases in $z$, generating the first downward-sloping region of $\omega(z)$. 

As $z$ crosses each successive $(1 + g)^j$ threshold, the transmitted skill survives an additional period before obsolescence. This extended lifespan increases the continuation value of transmitted knowledge, which current experts extract by further reducing the upfront wage. Iterating this logic generates a sequence of kinks in both $\pi(z)$ and $\omega(z)$ throughout the price-cleared region, $z \in [1 + g, \sigma)$. This process continues until $z = \sigma$, where the value of the transmitted knowledge exactly offsets the asymptotic novice utility $u$. At this threshold, the zero lower bound binds, and $\omega(z)=0$ for all $z \ge \sigma$. 

This oversubscribed region ($z \ge \sigma$) can rationalize the severe rationing of entry-level positions observed at elite professional services firms. For instance, McKinsey and Boston Consulting Group accept only 0.6\% and 1\% of applicants, respectively \citep{vlamis2025mbb}. Similarly, fewer than 1\% of applicants are admitted to the internship programs at JPMorgan and Goldman Sachs, which serve as the primary pipelines for full-time entry-level roles \citep{clarke2025record}.

While mechanisms outside the formal model---such as the screening of heterogeneous applicants---undoubtedly contribute to these low acceptance rates, the framework establishes that they are a natural consequence of the incomplete contracts governing the transmission of tacit knowledge. Because novices can compensate experts solely with their labor, the zero lower bound on wages caps the implicit price of elite mentorship: experts with $z > \sigma$ cannot fully extract the recursively capitalized value of the tacit knowledge they transmit, so novices strictly prefer these matches, resulting in latent ``excess demand.''\footnote{As discussed in Section \ref{sec:formal}, oversubscription is an informal interpretation, as the model characterizes only realized matches. Furthermore, as noted in Section \ref{sec:environment}, a wage of zero captures both literal unpaid labor and negligible effective compensation driven by demanding hours relative to nominal pay.} 

I now formalize the construction of $\pi(z)$ and $\omega(z)$. For ease of exposition, I do so by working directly at the asymptotic limit. In Online Appendix \ref{sec:asympSch}, I use the finite-time formulas that underlie the proof of Proposition \ref{prop:existence} to show that the normalized income and wage schedules indeed converge pointwise to the limits characterized here.

 \begin{description}[leftmargin=*, labelindent=0pt, itemindent=0pt, listparindent=\parindent]  \vspace{-1mm}
\item[Asymptotic Income and Wage Schedules.] Because the normalized skill of the marginal active expert converges to $\lim_{t \to \infty} (a_t/x_t) = 1+g$, experts with relative skill $z \in [1, 1 + g)$ are inactive. These experts earn $\pi(z) = 0$ and face a shadow wage $\omega(z) = u$, which renders production unprofitable. 

To characterize $\pi(z)$ for $z \ge 1+g$, substitute the wage schedule paid by active experts $w_t(q) = \max\{0, U_t - \beta \Pi_{t+1}(q)\}$ into the income of such experts $\Pi_t(q) = S(q) - w_t(q)(N-I)$ to obtain the recursion:$$\Pi_t(q) = \min \left\{ S(q), S(q) - (N-I)U_t + \beta(N-I) \Pi_{t+1}(q) \right\}$$Normalizing by $x_t$ and taking the limit as $t \to \infty$ yields the following functional equation (see Appendix \ref{sec:recurs} for details):$$\pi(z) = m^I h^{N-I} \min \left\{ z, z - (1+g)+ \frac{\beta(N-I)(1+g)}{m^I h^{N-I}}\pi\left(\frac{z}{1+g}\right) \right\}, \quad \text{for } z \ge 1+g$$Zero income for the marginal active expert implies continuous pasting to the inactive region, establishing the boundary condition $\pi(z) = 0$ for all $z \in [1, 1+g]$.

Appendix \ref{sec:recurs} solves this functional equation, yielding the closed-form asymptotic income schedule:$$\pi(z) = m^I h^{N-I} \left\{  \begin{array}{cl} 
      0 & \text{if }  z \in [1,1+g) \\
      \gamma(z) & \text{if }  z \in [1+g,\sigma) \\
      z & \text{if }  z \in [\sigma, +\infty)
   \end{array} \right.$$where:$$\gamma(z) = \sum_{j=0}^{n(z)-1} [\beta(N-I)]^j \left( z - (1+g)^{j+1} \right) \quad \text{with} \quad n(z) \equiv \left\lfloor \frac{\ln z}{\ln(1+g)} \right\rfloor$$and $\sigma \in (1+g,+\infty)$ is the unique solution to $\sigma = \gamma(\sigma)$. 

The term $m^I h^{N-I} \gamma(z)$ represents the \textit{value} of a dynasty of experts with skill $z$ prior to obsolescence, defined as the present discounted sum of its excess output relative to the marginal active expert in each period. The index $n(z)$ denotes the number of periods until $z$'s obsolescence, so $n(z)-1$ is the maximum integer $j$ for which the excess output condition $z \ge (1+g)^{j+1}$ holds. Absent the zero lower bound on wages, an expert with relative skill $z$ could always appropriate the full value of the dynasty by charging her novices for the skill provided, a transfer that recursively incorporates the value these novices will subsequently charge as experts to their own novices. 

However, because wages cannot be negative, an expert's current income cannot exceed the physical output generated within the period, $z m^I h^{N-I}$. Oversubscription thus emerges exactly at the skill threshold $\sigma$ where the value of the dynasty meets this limit of current production, $m^I h^{N-I} \gamma(\sigma) = m^I h^{N-I} \sigma$, yielding the condition $\sigma = \gamma(\sigma)$. From the functional equation of $\pi(z)$, this is also exactly the point at which the value of the transmitted knowledge exactly offsets the asymptotic novice utility $u$, i.e., \[  u = \beta  (1+g) \pi\left(\frac{\sigma}{1+g}\right) \]

Finally, to characterize $\omega(z)$ for $z \ge 1+g$, normalize the active expert's income $\Pi_t(q) = S(q) - w_t(q)(N-I)$ by $x_t$. Taking the limit as $t \to \infty$ and rearranging yields:$$\omega(z) = \frac{z m^I h^{N-I} - \pi(z)}{N-I}, \ \  \text{for } z \ge 1+g$$Evaluating $\omega(z)$ using the piecewise solution for $\pi$, and imposing the shadow wage $u$ for inactive experts ($z \in [1, 1+g)$), yields the complete asymptotic wage schedule:$$\omega(z) = \frac{m^I h^{N-I}}{N-I} \left\{ \begin{array}{cl} 1+g & \text{if } z \in [1,1+g) \\ z - \gamma(z) & \text{if } z \in [1+g,\sigma) \\ 0 & \text{if } z \in [\sigma, +\infty) \end{array} \right.$$ \end{description}

\section{The Impact of Different Technological Advances} \label{sec:advances}

This section studies the effects of technological change on aggregate output and welfare. I focus on two specific advances: improvements in entry-level automation and the creation of new tasks. Section \ref{sec:robust} extends the analysis to labor- and capital-augmenting technologies.

For ease of exposition, I assume the absolute skill distribution at the time of the shock $\tau$ is exactly Pareto with scale $x_\tau$ and shape $1/\theta$. This assumption admits two possible interpretations:\begin{enumerate}
\item \textbf{Exact Initial Distribution}: If the initial distribution $F_0$ is exactly Pareto with scale $x_0$ and shape $1/\theta$, equation (\ref{eq:evolution}) ensures that the skill distribution at date $\tau$ is also exactly Pareto with scale $x_\tau=x_0(N-I)^{\theta\tau}$ and shape $1/\theta$.
\item \textbf{Asymptotic Approximation}: If $F_0$ is not exactly Pareto, but $1-F_0 \in \mathrm{RV}_{-1/\theta}$, this setup can be interpreted as approximating an economy hit by the shock at an arbitrarily distant date, by which time the skill distribution is close to its Pareto limit (see Section \ref{sec:skills}). \end{enumerate}

\subsection{Improvements in Entry-Level Automation} \label{sec:entry-level}

To establish a baseline, consider the evolution of the economy from period $\tau$ onward in the absence of a shock. The distribution of expert skill then remains exactly Pareto, with its scale parameter evolving according to $x_t = x_\tau(N-I)^{\theta(t-\tau)}$ for all $t \ge \tau$. Aggregate output along this trajectory is therefore:$$Y_t = a_t \left[ \frac{m^I h^{N-I}}{(N-I)(1-\theta)} \right] = x_\tau(N-I)^{\theta(t-\tau)} \left[ \frac{m^I h^{N-I}}{(1-\theta)(N-I)^{1-\theta}} \right], \ \text{for $t \ge \tau$ }$$where the second equality follows from $a_t/x_t = a_t/a_{t-1} =1+g= (N-I)^\theta$.

Suppose an unanticipated technological shock arrives at date $\tau$ and allows an additional measure $\Delta$ of tasks to be automated, with $N-I-\Delta>1$. I impose this restriction to focus on the case in which the post-shock economy continues to grow and all novices remain employed.\footnote{As shown in Online Appendix \ref{sec:Nrep}, if $N-I-\Delta \le 1$, long-run growth collapses; if $N-I-\Delta<1$, novice unemployment also arises.} Thus, the output and welfare effects derived below do not hinge on the shock either destroying growth or generating novice unemployment. For simplicity, I also assume that machine productivity in these newly automatable tasks equals that in previously automatable tasks, $m$. This last assumption is not essential: the qualitative results require only that machines be sufficiently productive at the new automatable tasks for active experts to find it optimal to automate them.

Given Assumption \ref{assu:full}, active experts optimally assign $I+\Delta$ tasks to machines and $N-I-\Delta$ tasks to novices from date $\tau$ onward.\footnote{Note that the cutoff skill for full automation in Lemma \ref{lem:cut} is decreasing in $I$. Replacing $I$ by $I+\Delta$ therefore lowers the cutoff. Since Assumption \ref{assu:full} places the initial support above the original threshold, and subsequent supports remain above the initial one, all active experts also lie above the lower threshold associated with the enlarged automatable set.} Letting tildes denote the shocked trajectory, the absolute skill distribution at date $\tau$ is predetermined, so $\tilde{F}_\tau(q) = F_\tau(q)$. Given this initial condition and the updated span of control, the subsequent equilibrium allocations $\{\tilde{a}_t=\tilde{x}_{t+1}, \tilde{F}_t\}_{t \ge \tau}$ solve a similar dynamical system as in Section \ref{sec:output}. This yields:\[ \tilde{x}_t = x_\tau(N-I-\Delta)^{\theta(t-\tau)} \ \text{and} \ 1 - \tilde{F}_t(q) = \left( \frac{q}{\tilde{x}_t} \right)^{-1/\theta}, \  \text{for $t \ge \tau$ } \]As in the no-shock scenario, the minimum expert skill grows geometrically, but at the permanently lower factor $1+\tilde{g}=(N-I-\Delta)^{\theta}<1+g$. Moreover, the absolute skill distribution at time $t$ is also Pareto with shape $1/\theta$, but with new scale $\tilde{x}_t$. From (\ref{eq:output}), aggregate output along the new equilibrium trajectory is: \[ \tilde{Y}_t = x_\tau(N-I-\Delta)^{\theta(t-\tau)} \left[ \frac{m^{I+\Delta} h^{N-I-\Delta}}{(1-\theta)(N-I-\Delta)^{1-\theta}} \right], \  \text{for $t \ge \tau$ } \] 

The ratio of $\tilde{Y}_t$ to $Y_t$ then summarizes both the short- and long-run implications of the shock:\begin{equation} \label{eq:decomp} \frac{\tilde{Y}_t}{Y_t} = \underbrace{\overbrace{\left( \frac{m}{h} \right)^\Delta}^{\text{Intensive margin gains}} \overbrace{\left( 1 - \frac{\Delta}{N-I} \right)^{-(1-\theta)}}^{\text{Extensive margin gains}}}_{\text{Immediate adoption gains } (>1)}  \underbrace{\left( 1 - \frac{\Delta}{N-I} \right)^{\theta(t-\tau)}}_{\text{Long-run losses } (\le 1)}, \quad \text{for } t \ge \tau.\end{equation}

This equation immediately leads to the following result:

\begin{prop}[Output Effects of an Automation Shock] \label{prop:main1} Suppose an unanticipated shock at date $\tau$ expands the set of automatable tasks. Then, relative to the no-shock scenario:
\begin{enumerate}[leftmargin=*,noitemsep]
    \item \textbf{Immediate Adoption Gains}: Output is strictly higher on impact ($\tilde{Y}_\tau > Y_\tau$).
    \item \textbf{Permanent Growth Slowdown}: The growth rate of output is strictly lower for all $t \ge \tau$ ($\tilde{g} < g$).
    \item \textbf{Long-Run Reversal}: There exists a date $t^* > \tau$ such that output is strictly lower for all $t > t^*$ ($\tilde{Y}_t < Y_t$).
\end{enumerate}
\end{prop}

\begin{proof} The result follows directly from the preceding analysis. \end{proof}

  \begin{figure}[!b]
\centering
\hspace{-8mm}
\begin{subfigure}{.53\textwidth}
  \centering
    \includegraphics[scale=0.43]{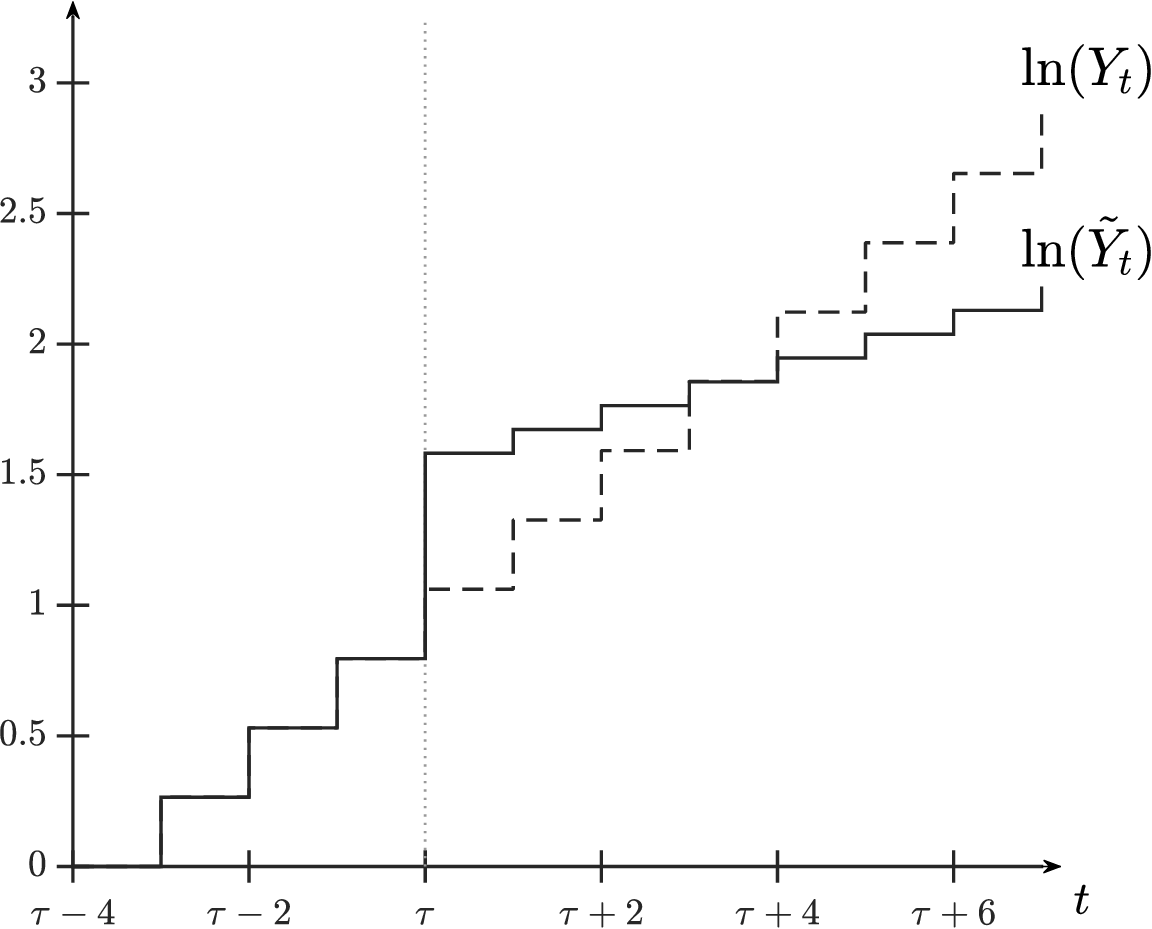}
  \caption{Log Output Trajectories}
\end{subfigure} 
\begin{subfigure}{.46\textwidth}
  \centering
    \includegraphics[scale=0.43]{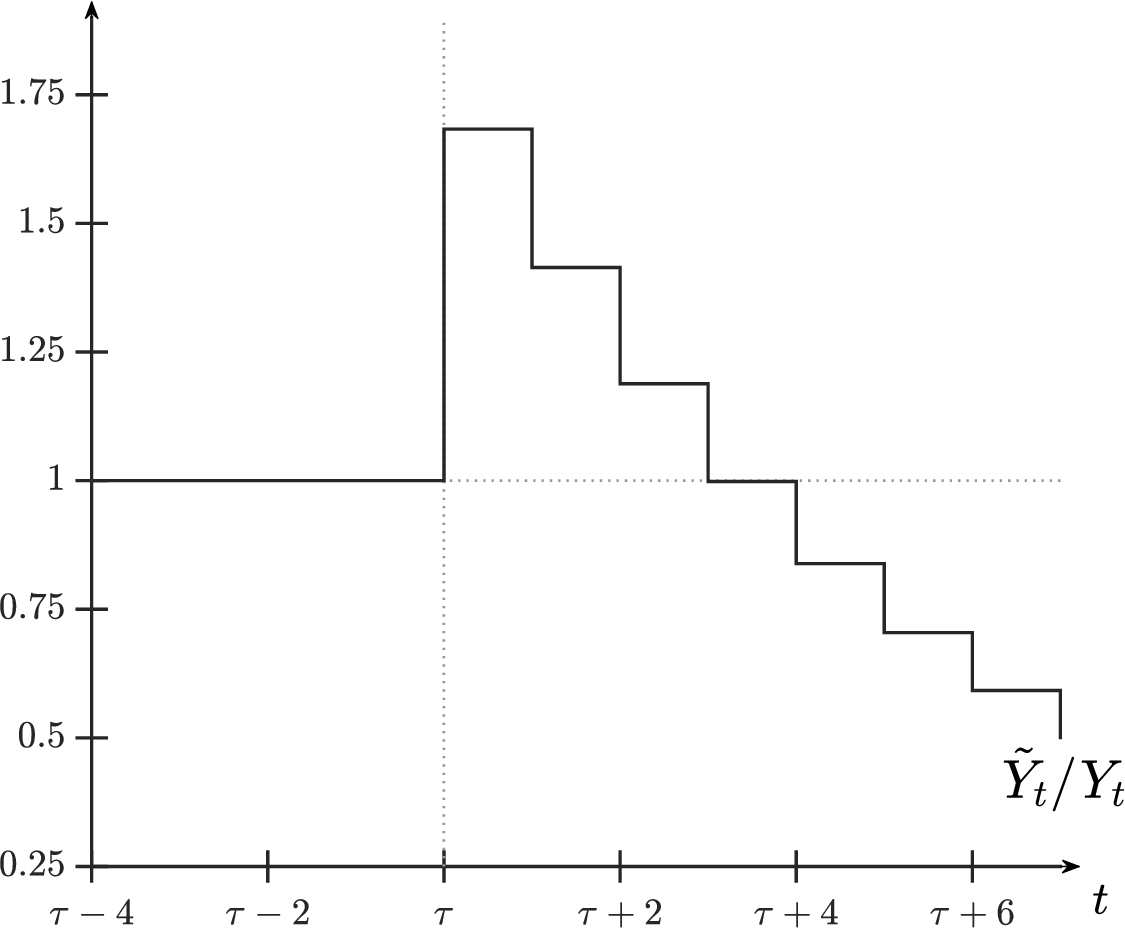}
  \caption{Relative Output}
\end{subfigure}
  \captionsetup{justification=centering}
\caption{The Short- and Long-Run Effects of an Automation Improvement on Aggregate Output \\  \justifying 
 \vspace{0.5mm}
\footnotesize \noindent \textit{Notes.} Parameter values: $N=2.7$, $I=1$, $h=1$, $m=2$, $\beta=0.96^{10}$, $\theta=0.5$, and $\Delta=0.5$. Panel (a) plots the trajectories of log aggregate output, $\ln(Y_t)$ and $\ln(\tilde{Y}_t)$, where $Y_t$ and $\tilde{Y}_t$ denote aggregate output without and with the automation shock, respectively. The underlying output paths are normalized so that output four periods prior to the shock equals 1 ($Y_{\tau-4}=\tilde{Y}_{\tau-4}=1$). Panel (b) plots the ratio of the underlying output paths, $\tilde{Y}_t / Y_t$. The vertical dotted line denotes the date of the automation shock, $t=\tau$.} \label{fig:Ytrajectories}
\end{figure}

Proposition \ref{prop:main1} is illustrated in Figure \ref{fig:Ytrajectories}. Panel (a) plots the log-level trajectories of aggregate output (normalized so that $Y_{\tau-4}=\tilde{Y}_{\tau-4}=1$). Panel (b) plots their ratio, $\tilde{Y}_t/Y_t$. On impact, the shock shifts output upward. However, as the flatter slope in Panel (a) shows, this initial expansion comes at the cost of a permanently lower growth rate. Accordingly, the output ratio in Panel (b) jumps up on impact but then declines persistently over time, eventually falling strictly below one.

Equation (\ref{eq:decomp}) clarifies the forces at work. The first two terms---both strictly greater than one and constant for all $t\ge\tau$---capture the immediate gains from adoption along two distinct margins. The first term, $(m/h)^\Delta$, reflects the intensive margin: because machines outperform novices in newly automated tasks ($m>h$), the output of all existing expert-novice teams increases on impact. The second term, $(1-\Delta/(N-I))^{-(1-\theta)}$, reflects the extensive margin: as highly skilled experts hire fewer novices, displaced labor reallocates toward less-skilled experts to clear the market. This enables a new tier of previously inactive experts to produce profitably, further increasing aggregate output.

In contrast, the third term, $(1-\Delta/(N-I))^{\theta(t-\tau)}$, captures the long-run losses. At $t=\tau$, it equals one because experts’ skills are predetermined. For $t>\tau$, however, it is strictly below one and declines over time. This decay is the dynamic counterpart of the equilibrium reallocation of labor: as highly skilled experts hire fewer novices, displaced novices are reassigned to less-skilled experts. These novices acquire lower-quality tacit knowledge and become worse mentors themselves. The best practices of top experts thus diffuse more slowly, imposing a compounding drag on future output.

A common intuition is that improvements in automation can harm skill accumulation by depriving novices of hands-on practice.\footnote{The model can also capture a version of this hands-on-experience channel, but only in an extreme form: through novice unemployment. In the model, a novice obtains hands-on experience only by being matched with an expert. If $N-I-\Delta<1$, aggregate demand for novices is smaller than the unit mass of novices, so some novices remain unmatched and therefore do not acquire expertise through joint production. In that case, the automation improvement weakens knowledge transmission along two margins: it changes the composition of mentors among novices who remain matched and reduces the number of novices who obtain hands-on experience.} While this mechanism is plausible, Proposition \ref{prop:main1} shows that it is not necessary: a deterioration in mentor composition alone is sufficient to impair growth. Anecdotal reports are consistent with such a compositional shift. Amid the recent cooling of the entry-level white-collar labor market---which, as discussed below, some have linked to nascent AI adoption---graduates appear to be increasingly sorting into smaller, lower-tier firms that previously struggled to attract such talent \citep{ellis2026,lahart2026}.

The remaining question is whether the immediate gains from automation outweigh the long-run losses in welfare terms. To evaluate this trade-off, I compute welfare at date $\tau$ with and without the shock using (\ref{eq:welfare}): \begin{align*}  &W_{\tau} = \frac{Y_{\tau}}{1-\beta(1+g)} - \frac{\rho I}{(N-I)(1-\beta)} \\
& \tilde{W}_{\tau}  = \frac{\tilde{Y}_{\tau}}{1-\beta(1+\tilde{g})} - \frac{\rho (I+\Delta)}{(N-I-\Delta)(1-\beta)}  \end{align*}Comparing these expressions yields the following result:

\begin{prop}[Welfare Effects of an Automation Shock] \label{prop:main2}
There exists a unique threshold $\beta^* \in (0, \frac{1}{1+g})$ such that the automation shock reduces welfare ($\tilde{W}_\tau < W_\tau$) if and only if $\beta > \beta^*$. \end{prop}

\begin{proof} See Appendix \ref{app:welfare}. \end{proof}

Proposition \ref{prop:main2} establishes that an improvement in automation can reduce aggregate welfare. Because experts could still use the pre-shock task allocation after the shock, laissez-faire adoption is welfare-dominated by no adoption whenever $\tilde{W}_\tau < W_\tau$. In this sense, adoption can be socially excessive relative to the no-adoption benchmark.

The possibility of such a welfare loss reflects the role of the zero lower bound on wages. In a frictionless economy, novices could borrow against future earnings and pay oversubscribed experts for access to mentorship positions. Positions whose continuation value exceeded the current output gains from automation would then be preserved. Instead, because the zero lower bound prevents these payments, oversubscribed experts fail to fully internalize the future income generated by the novice positions that automation displaces. They may therefore adopt automation even when the resulting loss in future income outweighs the current output gains.

The distortion originates with oversubscribed experts, but its consequences are transmitted through equilibrium reallocation. Price-cleared experts are not themselves distorted by the zero lower bound. However, they still matter for welfare because when oversubscribed experts automate excessively and hire fewer novices, the displaced novices are reallocated toward them. Through that channel, the distortion weakens growth and potentially reduces welfare.\footnote{These results are related to \cite{sachs2012smart} and \cite{sachs2015robots}, who show that automation can raise output on impact while reducing the welfare of future generations. In their immiserizing cases, automation lowers young workers’ wages; because the young are the economy’s savers, this reduces savings and investment, lowering future cohorts’ welfare. Here, the channel is different: contractual frictions make current labor the only means by which novices can finance mentorship; automation devalues that currency, and because the zero lower bound binds for the most productive experts, some novices are reallocated toward less productive mentors. The welfare implications also differ. In those papers, automation may fail to be Pareto-improving under laissez-faire, but the productivity gains from automation are large enough that, with appropriate intergenerational transfers, all generations could be made better off relative to the pre-automation baseline. Here, by contrast, the discounted future income losses from weaker knowledge transmission can exceed the immediate gains from adoption, so aggregate welfare can fall.}

More generally, both Propositions \ref{prop:main1} and \ref{prop:main2} are driven by the reallocation of novices away from higher-skill mentors. Section \ref{sec:robust} extends the baseline framework by allowing experts to choose a continuous scale of operation, constrained by the time required to supervise novices and machines. Once experts can adjust scale, an automation improvement may reduce the number of novices needed per project while increasing the number of projects high-skill experts oversee. If this scale effect is strong enough, novices may be reallocated toward rather than away from higher-skill mentors, thereby reversing the propositions’ conclusions.

At least in these early stages of adoption, however, the preliminary evidence on generative AI appears more consistent with displacement than with offsetting scale effects. For instance, \cite{lichtinger2025} find that GenAI-adopting firms reduce junior employment by 9\% relative to non-adopters, while senior employment remains unaffected. Likewise, \cite{canaries2025} document a 16\% relative decline in early-career employment across highly AI-exposed occupations, with no corresponding impact on senior positions. Both estimates are robust to controlling for several confounding macro-sectoral variables and trends. Additional evidence points in the same direction: entry-level white-collar vacancies have declined in occupations exposed to generative AI \citep{berger2}, employment shares have fallen for young workers in AI-exposed sectors \citep{atkinson2026young}, and similar displacement effects have been documented in the United Kingdom \citep{klein2025generative}.\footnote{Other studies find more mixed results. \cite{johnston2025labor} document employment and wage gains for younger workers following the widespread adoption of AI, though \cite{canaries2025} argue these results may reflect insufficiently granular data. \cite{de2025artificial} finds heterogeneous effects in Brazil: AI appears to displace younger workers in administrative jobs while augmenting them in production roles. \cite{humlum2026still}, in turn, document widespread chatbot adoption and task reorganization in Denmark, but little measurable effect on earnings.}

This evidence, however, does not yet track where affected junior workers are subsequently placed. It therefore suggests that some displacement of junior labor might be occurring in AI-exposed sectors, but it is not direct evidence on the reallocation mechanism emphasized here. Systematic evidence on technology-induced reallocation remains scarce, though the anecdotal reports noted above are consistent with such reallocation \citep{ellis2026,lahart2026}. Section \ref{sec:scale} therefore turns to related evidence from settings outside technological change. This evidence shows that labor-market shocks at entry---such as cyclical downturns---can redirect young workers across institutions and firms, with persistent consequences for skill development and future productivity. 

\subsection{Task-Creating Technologies} \label{sec:aug}

While preliminary evidence points to a relative decline in entry-level hiring within AI-exposed occupations, the historical record also shows that technological advances can create new labor-intensive tasks.\footnote{For instance, \cite{autor2024new} estimate that 60\% of U.S. employment in 2018 consisted of job titles introduced since 1940.} Motivated by this precedent, I now explore the effects of a task-creating technology.

Formally, consider an unanticipated technological shock at date $\tau$ that introduces a new production technology alongside the existing one. Under the new technology, only tasks in $[0, I-\Delta]$, for some $\Delta \in (0, I)$, remain automatable: on those tasks, machines and novice labor have productivities $m$ and $h$, respectively. All tasks in $(I-\Delta, N]$ are non-automatable. Among them, tasks in $(I-\Delta, N-\Delta]$ are standard tasks on which novice labor retains productivity $h$, whereas tasks in $(N-\Delta, N]$ are newly introduced tasks on which novice labor has productivity $H>m>h$.

This formulation, inspired by \cite{acemoglu2018race} and \cite{autor2024new}, captures task-creating technological change. The new technology creates an additional measure $\Delta$ of valuable labor-intensive tasks. As in those papers, I assume that the introduction of these new tasks simultaneously renders obsolete an equal measure of previously automatable tasks, so that the total measure of tasks in a production process remains fixed at $N$.

Because the baseline technology remains available, the expert's optimization problem now entails three choices: (i) whether to be active; (ii) which technology to use; and (iii) the optimal allocation of tasks between machines and novices. I then impose the following condition:

\begin{assu}[Full Adoption of New Technology] \label{assu:UAdop}
The productivity of novices on new tasks satisfies: $$ \left(\frac{H}{m}\right)^\Delta \ge 1 + \frac{\Delta}{N-I} $$
\end{assu}

This condition states that the productivity gains from newly created labor-intensive tasks are large enough. Appendix \ref{app:creating} shows that Assumption \ref{assu:UAdop} is exactly the condition for sustaining permanent full adoption. If the assumption holds, there exists an equilibrium in which all active experts adopt the task-creating technology at every date $t\geq \tau$. If it fails, no such permanent full-adoption equilibrium exists: either the marginal active expert prefers the old technology already at $\tau$, or a future marginal active expert eventually does so as absolute skills grow.

I focus on this full-adoption equilibrium for the remainder of the section. In the present environment, this equilibrium is the counterpart of the task-creation benchmark in \citet{acemoglu2018race} and \citet{autor2024new}, with adoption of the new technology made endogenous at the expert level. In this equilibrium, all active experts optimally adopt the new technology and assign $I-\Delta$ tasks to machines, $N-I$ standard tasks to novices, and $\Delta$ new tasks to novices. An active expert with skill $q$ therefore produces:$$qm^{I-\Delta}h^{N-I}H^{\Delta}$$With this new production structure and span of control $N-I+\Delta$, the post-shock equilibrium can be characterized analogously to the baseline model in Sections \ref{sec:baseline}--\ref{sec:longrun}. Consequently, aggregate output and welfare along the new trajectory are given by:\begin{align*} & \tilde{Y}_t=x_\tau(N-I+\Delta)^{\theta(t-\tau)}\left[\frac{m^{I-\Delta}h^{N-I}H^\Delta}{(1-\theta)(N-I+\Delta)^{1-\theta}}\right],\text{ for }t\ge\tau  \\
&  \tilde{W}_{\tau}  = \frac{\tilde{Y}_{\tau}}{1-\beta(1+\tilde{g})} - \frac{\rho (I-\Delta)}{(N-I+\Delta)(1-\beta)}  \end{align*}where $1+\tilde{g}\equiv(N-I+\Delta)^\theta$ is the new growth factor of output. Here I impose $\beta(1+\tilde g)<1$, so that post-shock welfare $ \tilde{W}_{\tau} $ remains finite.\footnote{If $\beta(1+\tilde g)\geq 1$, the expression for $\tilde{W}_{\tau}$ above does not apply. Under Assumption \ref{assu:growth}, however, baseline welfare remains finite while post-shock welfare diverges, so the welfare-gain conclusion of Proposition \ref{prop:main3} holds in the extended sense.} The ratio of aggregate output with and without the shock is therefore:$$\frac{\tilde{Y}_t}{Y_t} = \underbrace{\overbrace{\left(\frac{H}{m}\right)^\Delta}^{\text{Intensive margin gains}} \overbrace{\left(1+\frac{\Delta}{N-I}\right)^{-(1-\theta)}}^{\text{Extensive margin losses}} }_{\text{Immediate adoption gains } (>1)} \underbrace{\left(1+\frac{\Delta}{N-I}\right)^{\theta(t-\tau)}}_{\text{Long-run gains $(\ge 1)$}}$$

These expressions imply the following result:

\begin{prop}[Output and Welfare Effects of a Task-Creating Technology] \label{prop:main3} Suppose an unanticipated shock at date $\tau$ introduces a task-creating technology. Then, relative to the no-shock scenario:
\begin{enumerate}[leftmargin=*,noitemsep]
    \item \textbf{Immediate Adoption Gains}: Output is strictly higher on impact ($\tilde{Y}_\tau > Y_\tau$).
    \item \textbf{Growth Acceleration}: The growth rate of output is strictly higher for all $t \ge \tau$ ($\tilde{g} > g$).
    \item \textbf{Welfare Gains}: The shock is strictly welfare-increasing ($\tilde{W}_\tau > W_\tau$).
\end{enumerate}
\end{prop}

\begin{proof} Growth acceleration and welfare gains are immediate from comparing $\tilde{g}$ and $\tilde{W}_{\tau}$ with $g$ and $W_\tau$ from Section \ref{sec:entry-level}. The immediate-adoption-gains result, in turn, follows because: \[ \left(\frac{H}{m}\right)^\Delta \left(1+\frac{\Delta}{N-I}\right)^{-(1-\theta)} \ge \left(1+\frac{\Delta}{N-I}\right)^{\theta} >1 \]where the first inequality holds by Assumption \ref{assu:UAdop}. \end{proof}

\begin{figure}[t!]
\centering
\hspace{-8mm}
\begin{subfigure}{.53\textwidth}
  \centering
    \includegraphics[scale=0.43]{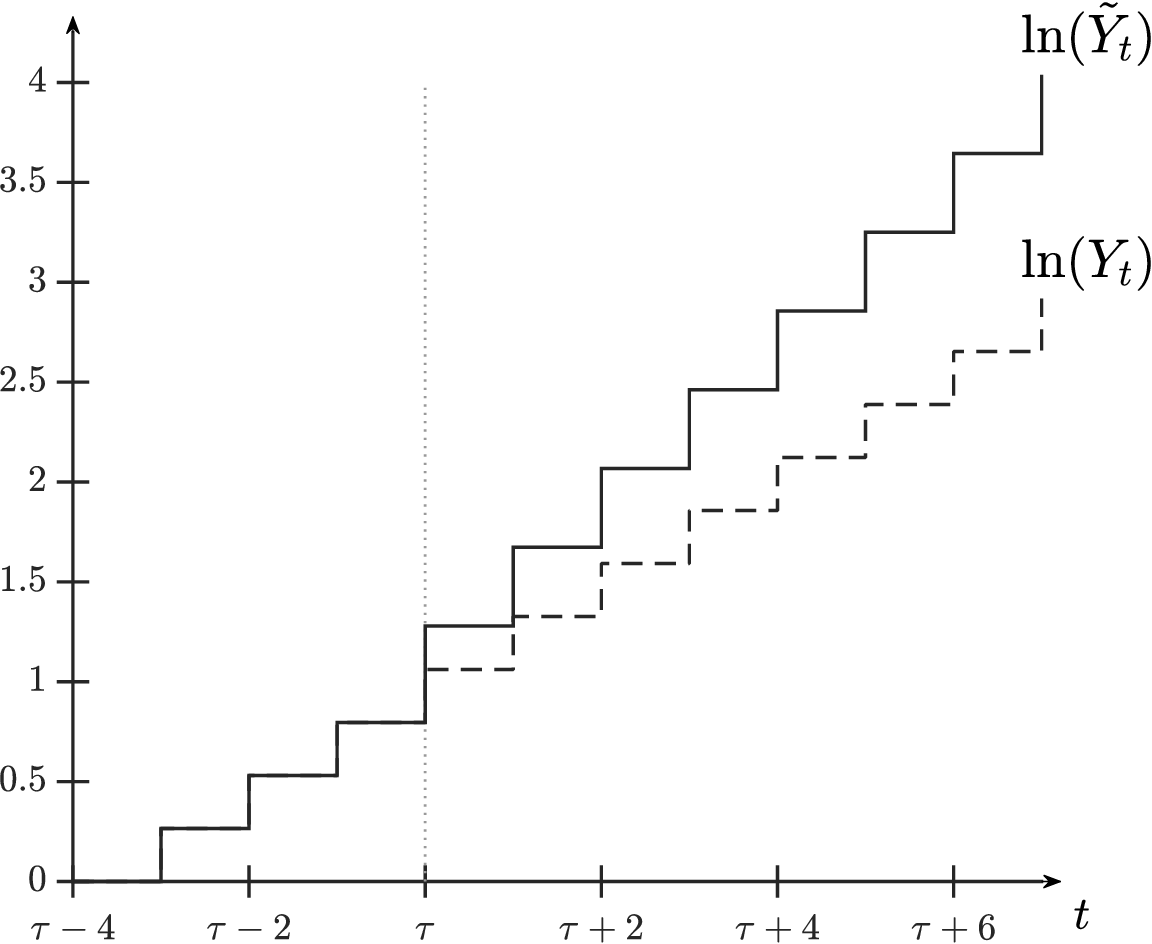}
  \caption{Log Output Trajectories}
\end{subfigure} 
\begin{subfigure}{.46\textwidth}
  \centering
    \includegraphics[scale=0.43]{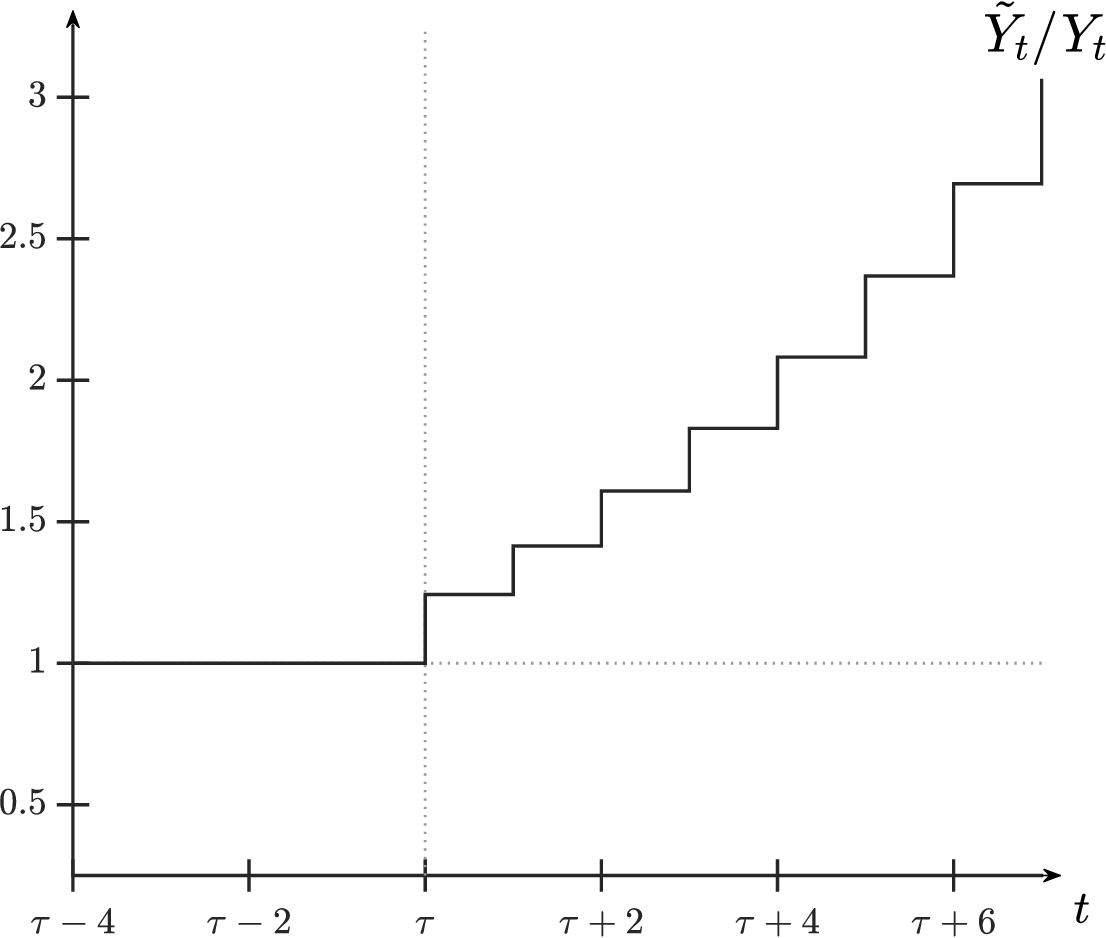}
  \caption{Relative Output}
\end{subfigure}
  \captionsetup{justification=centering}
\caption{The Short- and Long-Run Effects of Task-Creating Technology \\  \justifying 
 \vspace{0.5mm}
\footnotesize \noindent \textit{Notes.} Parameter values: $N=2.7$, $I=1$, $h=1$, $m=2$, $\beta=0.96^{10}$, $\theta=0.5$, $\Delta=0.5$, and $H=4$.  Panel (a) plots the trajectories of log aggregate output, $\ln(Y_t)$ and $\ln(\tilde{Y}_t)$, where $Y_t$ and $\tilde{Y}_t$ denote aggregate output without and with the task-creating technology, respectively. The underlying output paths are normalized so that output four periods prior to the shock equals 1 ($Y_{\tau-4}=\tilde{Y}_{\tau-4}=1$). Panel (b) plots the ratio of the underlying output paths, $\tilde{Y}_t / Y_t$. The vertical dotted line denotes the date of the task-creating shock, $t=\tau$.} \label{fig:Ytrajectories2}
\end{figure}

Figure \ref{fig:Ytrajectories2} illustrates the output effects of a task-creating technology. Unlike improvements in automation, this technology generates both immediate and long-run gains. By prompting the highest-skilled experts to hire additional novices, the introduction of a task-creating technology accelerates the diffusion of best practices. Although this expansion in span of control raises the threshold for becoming an active expert, thereby generating an extensive-margin loss, Assumption \ref{assu:UAdop} implies that the intensive-margin gains strictly dominate. Thus, aggregate output unambiguously increases on impact as well.

\section{Discussion and Extensions} \label{sec:robust}

This section examines the implications of key assumptions and introduces additional margins that shape how technological change affects the intergenerational transmission of tacit knowledge. The Online Appendix provides further extensions and robustness checks.

\subsection{Scale Effects and Other Technological Shocks}  \label{sec:scale}

\noindent \textit{Scale Effects}.--- The baseline model assumes that production is non-scalable. Due to limited time and attention, each expert oversees only the fixed measure $N$ of tasks required for a single project. To relax this assumption, let the expert choose a continuous scale of operation, $\mu$, representing the measure of concurrent projects she oversees. The total output of an expert with skill $q$ is therefore:$$\mu \cdot q m^I h^{N-I}$$

The scale of operation, $\mu$, is determined by the expert’s unit time endowment and the total supervision (or management) time required for a single project, $\Man(I,m,h)$, so that $\mu = 1/\Man(I,m,h)$. I model $\Man(I,m,h)$ as a constant-returns-to-scale CES aggregate of the supervision time required across the project’s constituent tasks, where $\eta > 0$ governs how task-level supervision requirements are aggregated. The time required to supervise an individual task is assumed to be inversely related to the productivity of the agent performing it: $1/m$ for a machine and $1/h$ for a novice. Aggregating these task-level supervision requirements across task shares and scaling by project size $N$ yields:$$\Man(I, m, h) = N \left[ \frac{I}{N} \left(\frac{1}{m}\right)^\eta + \frac{N-I}{N} \left(\frac{1}{h}\right)^\eta \right]^{\frac{1}{\eta}}$$

Because an expert manages $\mu$ projects and hires $N-I$ novices per project, her total span of control is $\ell = \mu(N-I) = (N-I)/\Man(I, m, h)$. Assuming $\ell>1$, equilibrium dynamics mirror the baseline model, except that the long-run growth factor of aggregate output becomes $1+g=\ell^\theta$. Thus, once experts can scale, the economy's growth rate depends on the total number of novices active experts supervise across projects, not only on the number of novices required within each project.

This formulation departs from standard task-based frameworks, which typically allow inputs to scale continuously within tasks (see Appendix \ref{sec:production}). Such a departure serves three purposes. First, it more faithfully captures the expert's role as an orchestrator who integrates tasks, rather than a standard factor of production. Second, it explicitly incorporates the expert’s time constraint, formalizing the idea that expert time is the fundamental bottleneck to scaling embodied expertise. Finally, it is a minimal deviation from the baseline model, allowing scale effects to be analyzed without a full recharacterization of the equilibrium.\footnote{In this specification, increases in $m$ or $h$ raise output through two channels: directly, by increasing output per project, and indirectly, by reducing project management time and allowing experts to operate at larger scale. An alternative modeling approach, closer to \citet{garicano2000hierarchies}, would take output per project to be $q$, so that total output is $\mu q$. In that case, improvements in machine or novice productivity would affect output only by relaxing the expert’s time constraint and increasing scale. The qualitative comparison between displacement and scale effects would be unchanged: for knowledge transmission, the relevant object remains the expert's total span of control, $\ell=(N-I)/\Man(I,m,h)$. I use the specification in the text because it is the minimal scalable extension of the baseline model: it introduces the expert-time constraint needed to study scale effects while leaving the original task-based production structure unchanged.}

Consider then the automation shock studied in Section \ref{sec:entry-level}. Although output continues to increase on impact, the effect on the span of control---and therefore on long-run growth---is now ambiguous. Differentiating $\ln \ell$ with respect to $I$ isolates the competing forces:$$\frac{\partial \ln \ell}{\partial I} = \underbrace{ \left(-\frac{1}{N-I} \right)}_{\text{displacement effect}} + \ \ \  \underbrace{ \left(- \frac{\partial \ln \Man(I, m, h)}{\partial I}\right)}_{\text{scale effect}}$$The first term is the displacement effect analyzed in Section \ref{sec:entry-level}: allocating tasks to machines reduces the measure of novices per project. The second term captures a new scale effect. Because machines are more productive and require less expert supervision ($m > h$), automation lowers total management time per project ($\partial \Man/\partial I < 0$), allowing experts to oversee more concurrent projects. The net impact on long-run growth depends on which of these opposing forces dominates.

When the displacement effect dominates, the central implication of Section \ref{sec:entry-level} remains: an improvement in automation lowers the scale-adjusted span of control and, by slowing knowledge diffusion, reduces growth. This occurs in two different cases. First, displacement dominates when automation is ``so-so'' \citep{AcemogluRestrepoNBER}: machines are adopted but are only marginally more productive than novices ($m \approx h$). Here, replacing a novice with a machine yields minimal time savings. Indeed, as $m \to h$, the scale effect vanishes ($\partial \Man/\partial I \to 0$), and displacement entirely drives the impact on the scale-adjusted span of control.

Second, the displacement effect dominates when management time is additive or more bottleneck-like in task-level supervision requirements ($\eta \ge 1$). The limiting case illustrates the logic: as $\eta \rightarrow \infty$, the CES aggregator converges to a maximum, or bottleneck, aggregator, so project supervision time is pinned down by the most supervision-intensive tasks---here, those performed by novices. In that limit, even if machines are substantially more productive than novices ($m \gg h$), automation frees no effective expert time while novice tasks remain. The same logic extends throughout $\eta \ge 1$: the time savings from an automation improvement are too small to offset the reduction in novices per project.

By contrast, when $\eta<1$ and machines are sufficiently productive relative to novices, the scale effect can dominate. In this case, the expert’s expanded scale of operation outweighs the displacement of novices per project. Top experts hire a larger aggregate measure of novices, accelerating the diffusion of best practices. The macroeconomic effects of the shock therefore resemble those of a task-creating technology, even without introducing new labor-intensive tasks. As the evidence reviewed in Section \ref{sec:entry-level} suggests, however, displacement rather than expert scaling appears to be the more salient margin so far in the case of generative AI.

\vspace{3mm}

\noindent \textit{Labor- and Capital-Augmenting Technologies}.--- Beyond automating and creating tasks, AI tools can increase the productivity of novices and machines at existing tasks. Consistent with labor augmentation, \cite{brynjolfsson2023generative} document that AI assistants lead to a 15\% increase in the resolution rate of customer-support agents by providing real-time response suggestions. For capital augmentation, \cite{de2025artificial} reports that AI-powered maintenance can reduce unplanned machine downtime by 50\% through analysis of real-time sensor data.

Formally, labor- and capital-augmenting technologies correspond to exogenous increases in $h$ and $m$, respectively. In the baseline model without scale effects, these improvements generate a one-time level effect on aggregate output but leave the experts' span of control---and thus long-run growth---unchanged. Scale effects alter this result. By reducing task execution time, both technologies lower total supervision time per project. Experts therefore oversee more concurrent projects, accelerating knowledge transmission and output growth.\footnote{The growth effects of these shocks are asymmetric when task supervision is highly interdependent ($\eta \to \infty$). In that case, total project supervision time is pinned down by the slowest tasks---those performed by human novices, since $m>h$. Capital-augmenting improvements therefore do not relax the supervision bottleneck and leave long-run growth unchanged. Labor-augmenting improvements, by contrast, relax that bottleneck and accelerate long-run growth.}

\vspace{3mm}

\noindent \textit{The Reallocation Margin}.--- As the previous discussion shows, new technologies can set multiple, often competing, forces in motion: they can automate tasks, create new labor-intensive ones, and augment labor and capital in existing tasks.\footnote{The electronic spreadsheet provides an example. It automated arithmetic tasks previously performed by clerical staff. At the same time, by lowering the cost of calculation, it generated new labor-intensive tasks, such as standardizing raw data and auditing complex formulas, that expanded the set of routine tasks performed by junior analysts.} Furthermore, even pure automation can accelerate the diffusion of tacit knowledge if the resulting time savings enable experts to scale their operations. Consequently, the net effect of a new technology on knowledge transmission is theoretically ambiguous.

The preceding cases nevertheless share a common structure. What matters for knowledge transmission is not the technological category per se, but how the technology reshapes the assignment of novices across experts. Technologies that move novices away from leading experts weaken the diffusion of best practices; technologies that expand novices’ access to those experts strengthen it.

Evidence outside the context of technological change supports the relevance of this reallocation margin. \citet{oyer2006initial} shows that weak academic job markets redirect new economists toward lower-ranked initial placements, with persistent effects on later research productivity. \citet{arellano2022effects} shows that workers exposed to worse labor-market conditions during the education-to-work transition acquire more formal education but end up with lower adult skills, and links this pattern to faster skill growth among young workers at larger firms. \citet{arellano2024career} provides complementary evidence from Spanish administrative data, showing that plausibly exogenous variation in access to larger first employers persistently raises earnings, with further evidence suggesting that skill development at these firms helps explain the gains. Taken together, these studies show that early-career allocation across institutions and firms can have durable consequences, partly because initial placements differ in the opportunities they provide for skill development.

For future empirical and policy work on how technological change shapes expertise formation across generations, this suggests the need to look beyond aggregate entry-level employment and examine how new technologies reallocate junior workers across firms and mentors.

\subsection{Further Considerations} \label{sec:further}

\noindent \textit{Relaxing the Strict Division of Labor}.--- The baseline framework imposes a strict division of labor: experts exclusively apply tacit knowledge as orchestrators, while novices execute routine tasks. This assumption forces the least-skilled experts in any cohort into inactivity, yielding zero second-period income. Suppose instead that experts face an endogenous occupational choice between acting as orchestrators and supplying one unit of routine labor.\footnote{This binary choice between occupations captures experts' limited time and attention.} Although this flexibility changes within-cohort allocations, the core results remain unchanged.

Experts' endogenous occupational choice induces sorting by age. Facing a zero continuation value, low-skilled experts (who would otherwise be inactive) maximize static payoffs by supplying routine labor to the lowest-skilled orchestrators, who pay the highest upfront wages to compensate for inferior mentorship. Forward-looking novices, in contrast, value skill acquisition and outbid low-skilled experts to work alongside elite orchestrators. The intergenerational transmission of the upper tail is thus preserved.

With this sorting in place, technological shocks affect knowledge transmission through the same mentor-composition margin as in the baseline model. For instance, when an automation shock reduces the number of novice positions available at elite orchestrators, the marginal novice match moves down the upper tail of the orchestrator skill distribution, just as in the baseline model without occupational choice. Some novices therefore learn from less-skilled mentors than before, worsening mentor composition and slowing growth.

Endogenous occupational choice by experts, however, changes how the remaining routine-labor market clears. In the automation case just described, displaced novices move down the orchestrator skill distribution and fill positions that low-skilled experts would otherwise have taken. The low-skilled experts displaced from those positions are, in turn, matched with lower-skill orchestrators. Market clearing is restored when some marginal low-skilled experts who would otherwise have performed routine tasks instead switch into orchestration and start hiring routine labor. This differs from the baseline model, in which the corresponding adjustment occurs through the entry of previously inactive experts into production.

\vspace{3mm}

\noindent \textit{More on Knowledge Transmission}.--- The baseline model assumes that all routine tasks provide uniform insight into the expert’s tacit knowledge. Online Appendix \ref{sec:hetero} relaxes this assumption by allowing some routine tasks to provide no meaningful interaction with the expert, and therefore no skill transmission, while the remaining tasks transmit expertise as in the baseline model. The extension shows that automating such non-learning tasks raises production efficiency without weakening knowledge transmission---and may even strengthen transmission through the scale effects discussed above.

The baseline model also assumes that skill transmission is exact: every novice matched with an expert inherits that expert’s skill. In practice, however, transmission is likely imperfect, and its effectiveness may depend on the number of novices an expert supervises---either because novices help one another or because supervising more novices dilutes the expert’s attention.\footnote{The effectiveness of transmission may also depend on communication technologies or workplace arrangements, such as remote work. Extending the model along those lines may be a fruitful direction for future research.} Online Appendix \ref{sec:imperfect_transmission} studies an extension with imperfect and span-dependent transmission. Among other things, the extension shows that, under attention dilution, improvements in entry-level automation introduce an additional countervailing force: while they reduce the number of novices supervised by each expert, they may improve the effectiveness of transmission to each remaining novice. This channel is distinct from the scale effects discussed above.

\vspace{3mm}

\noindent \textit{More on Contract Incompleteness}.--- As discussed in Section \ref{sec:environment}, the zero lower bound on wages captures, in reduced form, the contractual incompleteness inherent in the transfer of tacit knowledge: novices cannot finance access to expertise through enforceable claims on future returns. At the same time, because this constraint is imposed in reduced form rather than derived from an explicit contracting game, the analysis takes as given the contractual environment that gives rise to these frictions. Thus, new technologies can alter the wedge between private adoption incentives and the social value of knowledge transmission for a given set of contracts, but they do not alter the contracting environment itself.

The advantage of the reduced-form approach adopted here is that the results do not hinge on the finer details of the underlying contracting game. Instead, the model isolates forces that arise across many contractual environments: new technologies affect the value of expertise, the value of novice labor as a payment channel for acquiring it, and the allocation of novices across experts, with consequences for growth and welfare. In addition, the zero lower bound naturally generates the rationing of entry-level positions observed at elite professional services firms (Section \ref{sec:sch}).

The corresponding limitation is that this contractual environment does not adjust endogenously. A richer framework would allow technology to affect not only task productivity, the value of mentorship, and the return to expertise, but also the verifiability of tacit knowledge, the pledgeability of acquired human capital, and the contractual arrangements through which access to mentorship is financed.\footnote{For a model in which technology interacts with the contractual environment governing training, see \cite{GR2025AI}. In their setting, novices' inability to commit to employment post-training forces experts to transfer knowledge gradually. AI then affects apprenticeship viability through two competing channels: while automating entry-level tasks compresses the training window and limits the surplus experts can extract, a concurrent increase in the future value of expertise can offset this reduction by relaxing the novice's no-quit constraint.} Nevertheless, the results derived here are informative about such a richer environment: they indicate how technological change affects the value of contractual arrangements that mitigate these frictions.\footnote{All else equal, technologies that raise the value of mentorship or reduce the value of current labor as a payment channel increase the value of improving verifiability, pledgeability, or appropriability; technologies that reduce the importance of mentorship or raise the value of novice labor as a payment channel reduce it.}

\section{Final Remarks} \label{sec:final}

I conclude by summarizing the paper’s key insights, discussing their policy implications, and pointing to additional technological possibilities that the framework can help analyze.

\vspace{3mm}

\noindent \textit{The Main Takeaways}.--- The analysis rests on two premises. First, early-career work serves not only as an input into current production but also as a central channel for transmitting tacit knowledge across generations. Second, because tacit knowledge is embodied and non-verifiable, the contracts governing its transfer are inherently incomplete.

The first premise makes the allocation of novices across experts central to the transmission of tacit knowledge. Technological change can therefore affect future expertise not only by changing the number of entry-level positions but also by changing which experts those positions give novices access to. The second premise implies that private adoption decisions need not internalize the effects of new technologies on knowledge transmission. Technological improvements that shift novices away from the most skilled experts can thus weaken the diffusion of best practices and reduce aggregate welfare.

\vspace{3mm}

\noindent \textit{Policy Implications}.--- The incompleteness of knowledge-transfer contracts creates scope for policy intervention. At the same time, the analysis cautions against treating new technologies as a single force. This is especially important for AI as a general-purpose technology, since it is likely to take the form of a wide range of tools: some may automate entry-level tasks, while others may create new labor-intensive tasks or augment labor or capital in existing tasks. The policy challenge is therefore to harness the gains from these tools without undermining the formation of expertise in future generations.

The model offers three insights for addressing this challenge. First, policy assessment should look beyond aggregate entry-level employment: the deployment of new technologies may reduce welfare even if the total number of entry-level jobs remains unchanged, provided it reallocates novices away from the most skilled experts. Second, tools that automate entry-level tasks deserve particular scrutiny, since they are especially likely to reduce novices’ access to high-quality mentors. Third, whether this risk materializes depends on the strength of countervailing margins, such as scale effects, that may offset the decline in demand for novices among the most skilled experts.

When evidence indicates that a particular tool or technology weakens knowledge transmission enough to make laissez-faire adoption welfare-reducing, policy should aim to preserve novices’ access to high-quality learning environments. The appropriate response will depend on the institutional context, but possible instruments include taxes or limits on uses of entry-level automation tools, subsidies for junior positions that provide access to high-quality mentorship, and public support for apprenticeship- or residency-style training.

\vspace{3mm}

\noindent \textit{Additional Technological Possibilities}.--- The analysis has focused on more conventional technological improvements—such as automation, task creation, and factor augmentation—that affect the creation and execution of tasks. AI, however, may differ from earlier technological changes in its ability to encode and apply tacit knowledge at scale \citep{AIKE}. This raises the possibility that AI could democratize expertise either by \textit{democratizing performance}---narrowing the productivity gap between experts---or by \textit{democratizing learning}---facilitating novices’ acquisition of tacit knowledge through exposure to a broad range of examples and on-demand interactive guidance.\footnote{A more radical possibility is that AI renders some existing forms of tacit knowledge obsolete, shifting expertise toward capabilities many of today’s experts do not possess. Even then, much existing tacit knowledge would likely remain valuable, because human judgment would still be needed to assess when AI output is reliable.} A separate possibility is that AI could accelerate innovation by helping experts search, experiment, prototype, and recombine practices more effectively, thereby increasing the creation of new tacit ideas.

These possibilities suggest additional ways AI may reshape how tacit knowledge is created and how it is transmitted across generations, with implications for aggregate productivity. Exploring them is a natural next step, and the framework developed here provides a useful starting point for studying their consequences.

\newpage

\begin{appendix}
\counterwithin{figure}{subsection}

\begin{center} \Large \textsc{Appendix} \end{center}

\section{Glossary of Core Symbols} \label{app:glossary}
\vspace{-3mm}
\begin{table}[H]
\centering
\renewcommand{\arraystretch}{1.15} 
\resizebox{\textwidth}{!}{%
\begin{tabular}{@{} c l @{\hspace{2em}} c l @{}}
\toprule
\textbf{Symbol} & \textbf{Definition} & \textbf{Symbol} & \textbf{Definition} \\
\midrule

\multicolumn{4}{@{}l}{\textit{\textbf{Exogenous Primitives: Basic Parameters \& Initial Conditions}}} \\
\midrule
$t$ & Discrete time index. & $f_0(q)$ & Initial PDF of expert skills. \\
$\beta$ & Discount factor. & $x_0$ & Minimum of the support of $F_0$. \\
$q$ & Absolute skill level. & $\theta$ & Tail thickness parameter. \\
$F_0(q)$ & Initial CDF of expert skills. & $\tau$ & Date of technological shock.  \\
\addlinespace[1em]

\multicolumn{4}{@{}l}{\textit{\textbf{Exogenous Primitives: Production \& Technology}}} \\
\midrule
$N$ & Measure of tasks per project. & $\rho$ & Machine rental rate. \\
$I$ & Measure of automatable tasks. & $\Delta$ & Measure of newly automated or created tasks. \\
$m$ & Machine task productivity. & $H$ & Novice task productivity on new tasks. \\
$h$ & Novice task productivity. & & \\
\addlinespace[1em]

\multicolumn{4}{@{}l}{\textit{\textbf{Endogenous Variables (Finite Time)}}} \\
\midrule
$F_t(q)$ & CDF of expert skills at $t$.  &   $A_t$ & Set of active experts at $t$.  \\
$f_t(q)$ & PDF of expert skills at $t$. &  $a_t$ & Skill of the marginal active expert at $t$.   \\
$x_t$ & Minimum of the support of $F_t$. &    $\Lagr_t(E)$ & Mass of novices assigned to expert set $E$ at $t$.\\
$k$ & Measure of tasks assigned to machines. &  $U_t$ & Baseline novice utility at $t$. \\
$S(q)$ & Surplus of expert $q$ before labor costs. & $Y_t$ & Aggregate output at $t$. \\
$w_t(q)$ & Wage schedule at $t$.  & $W_t$ & Aggregate welfare at $t$. \\
$\Pi_t(q)$ & Expert income schedule at $t$. & & \\
\addlinespace[1em]

\multicolumn{4}{@{}l}{\textit{\textbf{Endogenous Variables (Normalized Asymptotics)}}} \\
\midrule
$g$ & Asymptotic growth rate of output. & $\omega(z)$ & Asymptotic normalized wage schedule. \\
$z$ & Normalized skill level. & $\sigma$ & Skill threshold where zero lower bound binds. \\
$\Phi(z)$ & Asymptotic CDF of normalized skills. & $\gamma(z)$ & Asymptotic value factor for dynasty $z$. \\
$u$ & Asymptotic normalized baseline utility. & $n(z)$ & Periods until obsolescence of dynasty $z$. \\
$\pi(z)$ & Asymptotic normalized income schedule. & & \\
\addlinespace[1em]

\multicolumn{4}{@{}l}{\textit{\textbf{Extensions (Section \ref{sec:robust})}}} \\
\midrule
$\mu$ & Measure of concurrent projects. &  $\eta$ & CES parameter for supervision-time aggregation. \\
$\Man(I, m, h)$ & Project management time. & $\ell$ & Total span of control (scale effects included). \\ 
\bottomrule
\end{tabular}%
}
\end{table}
\FloatBarrier

\section{Derivations and Omitted Proofs for Section \ref{sec:baseline}} \label{app:A}

\subsection{More on the Production Function} \label{sec:production}

In this appendix, I show that the production function of the baseline model is a special case of a more general formulation. In this general formulation, an expert with skill $q$ produces final output $y(q)$ by combining her skill with a continuum of routine tasks $s \in [0, N]$, aggregated via a Cobb–Douglas technology:$$\ln y(q) = \ln q + \int_{0}^{N} \ln  \mathcal{y}(s) \, ds$$where $\mathcal{y}(s)$ denotes the output of task $s$. Tasks can be performed by either machines or novice labor according to the following technology:$$ \mathcal{y}(s)  = \begin{cases}
\psi_h(s)\chi(s) + \psi_m(s) [1-\chi(s)] & \text{if } s \in [0,I] \\
\psi_h(s)\chi(s) & \text{if } s \in (I,N]
\end{cases}$$where $\psi_h(s)$ and $\psi_m(s)$ denote the productivity of novices and machines in task $s$, respectively, and $\chi(s)\in\{0,1\}$ indicates whether task $s$ is performed by a novice ($\chi(s)=1$) or a machine ($\chi(s)=0$).

The baseline specification corresponds to the special case where $\psi_h(s)=h$ for all $s \in [0,N]$ and $\psi_m(s)=m$ for all $s \in [0,I]$. Under these assumptions, total output depends only on the measure $k \in [0, I]$ of tasks assigned to machines. Substituting these constant productivities into the aggregator and exponentiating yields:\[ y(q) = q m^{k} h^{N-k}, \ \text{where $k \in [0, I]$}  \]
 
The fundamental difference between this specification and standard task-based models \citep[e.g.][]{acemoglu2011chapter,acemoglu2018race,AcemogluRestrepoNBER} is that in the present setting, tasks are non-scalable. Formally, this restricts $\chi(s)$ to be a binary indicator. In contrast, standard models allow for continuous input choices (the intensive margin), replacing the discrete allocation $\chi(s)$ with variable quantities of labor and capital. As discussed in Sections \ref{sec:environment} and \ref{sec:scale}, the assumption of non-scalability reflects experts' limited time and attention. 

\subsection{Proof of Lemma \ref{lem:cut}} \label{app:thresh}

This appendix shows that even with free novice labor, full automation ($k = I$) is strictly optimal for all active experts with: $$q >  \frac{\rho I}{h^N[ (m/h)^I-1]}$$An active expert chooses $k \in [0, I]$ to maximize: \[ \max_{k \in [0, I]} \left\{ q m^k h^{N-k} - \rho k - w_t(q)(N-k) \right\} \] Since $m > h$, the objective function is strictly convex in $k$, implying the optimum is a corner solution $k^* \in \{0, I\}$. Full automation ($k=I$) strictly dominates no automation ($k=0$) if and only if:$$q (m^I h^{N-I} - h^N) > I (\rho -w_t(q))$$ Since $w_t(q) \geq 0$, the condition for full automation is most restrictive when labor is free. Setting $w_t(q)=0$ and solving for $q$ in the above inequality immediately gives the desired result. \hfill $\square$

\section{Derivations and Omitted Proofs for Section \ref{sec:longrun}} \label{app:B}

Please note that Appendix \ref{app:B} uses different tail assumptions across its subsections. Appendix \ref{app:growth} works under the broader classification introduced in Section \ref{sec:longrun}, allowing $1-F_0$ to be either regularly varying or rapidly varying. From Appendix \ref{app:dist} onward, I impose Assumption \ref{assu:growth} and focus on the sustained-growth case, $1-F_0\in \mathrm{RV}_{-1/\theta}$, with $\theta\in(0,1)$.

\subsection{Proof of Proposition \ref{prop:growth}} \label{app:growth}

\begin{description}[leftmargin=*, labelindent=0pt, itemindent=0pt, listparindent=\parindent]
\item[Case 1: $1-F_0 \in \mathrm{RV}_{-1/\theta}$.] I first show that $\lim_{t \to \infty} (a_{t+1}/a_t) = (N-I)^\theta$. Define $\xi_t \equiv (N-I)^{t+1}$ and $J(q) \equiv 1/(1-F_0(q))$. According to (\ref{eq:cutF}), $J(a_t) = \xi_t$ and $J(a_{t+1}) = (N-I)\xi_t$. Because $F_0$ admits a continuous and strictly positive PDF, $J$ is strictly increasing, ensuring its inverse $J^{-1}$ is well-defined. Therefore, $a_t = J^{-1}(\xi_t)$ and $a_{t+1} = J^{-1}((N-I)\xi_t)$, yielding:$$\frac{a_{t+1}}{a_t} = \frac{J^{-1}((N-I)\xi_t)}{J^{-1}(\xi_t)}$$Since $1-F_0 \in \mathrm{RV}_{-1/\theta}$, I have $J \in \mathrm{RV}_{1/\theta}$. Because $1/\theta > 0$, an application of \citet[][Theorem 1.5.12]{bingham1989regular} yields $J^{-1} \in \mathrm{RV}_\theta$. Finally, since $\xi_t \to \infty$ as $t \to \infty$, it immediately follows that:$$\lim_{t \to \infty}\left( \frac{a_{t+1}}{a_t}\right) = \lim_{\xi_t \to \infty } \left( \frac{J^{-1}((N-I)\xi_t)}{J^{-1}(\xi_t)}\right) = (N-I)^\theta$$

I now show that $\lim_{t \to \infty} (Y_{t+1}/Y_t) = (N-I)^\theta$. Let $r(a) \equiv \mathbb{E}_0[q \mid q \ge a]/a$. As shown in the main text, $Y_{t+1}/Y_t$ can be decomposed as follows:$$\frac{Y_{t+1}}{Y_t} = \left( \frac{a_{t+1} r(a_{t+1})}{a_t r(a_t)} \right).$$By (\ref{eq:Erg}), $\lim_{a \to \infty} r(a) = (1-\theta)^{-1}$. Because the sequence of equilibrium cutoffs $\{a_t\}$ is strictly increasing and diverges to infinity as $t \to \infty$, the sequential characterization of limits directly guarantees that the sequences $\{r(a_t)\}$ and $\{r(a_{t+1})\}$ converge to the exact same limit $(1-\theta)^{-1}$ as $t \to \infty$. Applying this to the decomposed output ratio, and using the fact that $\lim_{t \to \infty} (a_{t+1}/a_t) = (N-I)^\theta$, yields the result.

\item[Case 2: $1-F_0 \in \mathrm{RV}_{-\infty}$.]  I first show that $\lim_{t \to \infty} (a_{t+1}/a_t) = 1$. Let $\lambda_t \equiv a_{t+1}/a_t$. Since the sequence of equilibrium cutoffs $\{a_t\}$ is strictly increasing, $\lambda_t > 1$ for all $t$. Suppose, by way of contradiction, that $\limsup_{t \to \infty} \lambda_t > 1$. Then, there exists a subsequence $\{t_j\}_{j \ge 0}$ and some $\epsilon > 0$ such that $\lambda_{t_j} \ge 1 + \epsilon$ for all $j$. This implies that $a_{t_j+1} \ge (1+\epsilon)a_{t_j}$. Because the survival function $1-F_0$ is strictly decreasing, it immediately follows that $1-F_0(a_{t_j+1}) \le 1-F_0((1+\epsilon)a_{t_j})$. Dividing both sides by $1-F_0(a_{t_j})$, and noting from (\ref{eq:cutF}) that the one-period forward ratio $(1-F_0(a_{t_j+1}))/(1-F_0(a_{t_j}))$ is identically equal to $1/(N-I)$, yields:$$ \frac{1}{N-I} \le \frac{1-F_0((1+\epsilon)a_{t_j})}{1-F_0(a_{t_j})}.$$By the definition of rapid variation, the right-hand side converges to zero as $j \to \infty$ because $a_{t_j} \to \infty$. However, the left-hand side is identically $1/(N-I)> 0$, yielding a contradiction. Thus, $\limsup_{t \to \infty} \lambda_t \le 1$. Since $\lambda_t > 1$, it must be that $\lim_{t \to \infty} \lambda_t = 1$.

The proof that $\lim_{t \to \infty} (Y_{t+1}/Y_t) = 1$ is analogous to the regularly varying case, except that now $\lim_{a \to \infty} r(a) = 1$ (see (\ref{eq:Erg})) and $\lim_{t \to \infty} (a_{t+1}/a_t) = 1$. \hfill $\square$
\end{description}

\subsection{Proof of Proposition \ref{prop:dist}} \label{app:dist}

By construction, $\mathbb{P}_t\left(q/x_t > z\right)=1-F_t(z x_t)$, and due to (\ref{eq:cutF}), $1-F_t(z x_t)=(N-I)^t [1-F_0(z x_t)]$. Hence, $\mathbb{P}_t\left(q/x_t > z\right)=(N-I)^t [1-F_0(z x_t)]$. Note that (\ref{eq:cutF}) also implies that $1 = (N-I)^t [1-F_0(a_{t-1})]$, and since $a_{t-1}=x_t$, it follows that $(N-I)^t= [1-F_0(x_t)]^{-1}$. Combining everything yields the relationship stated in Section \ref{sec:skills}:$$1 - \Phi(z) \equiv \lim_{t \to \infty} \mathbb{P}_t\left(\frac{q}{x_t} > z\right) = \lim_{t \to \infty} \frac{1 - F_0(zx_t)}{1 - F_0(x_t)}$$From the proof of Proposition \ref{prop:growth}, the sequence of entry cutoffs diverges, meaning the minimum skill of the active cohort $x_t = a_{t-1} \to \infty$ as $t \to \infty$. Because the initial survival function is regularly varying ($1-F_0 \in \mathrm{RV}_{-1/\theta}$), the definition of regular variation directly implies that for any constant $z \ge 1$:$$\lim_{t \to \infty}\left( \frac{1-F_0(z x_t)}{1-F_0(x_t)}\right) = \lim_{x_t \to \infty} \left( \frac{1-F_0(z x_t)}{1-F_0(x_t)}\right) = z^{-1/\theta}.$$Substituting this limit into the survival function trivially implies $\Phi(z) = 1 - z^{-1/\theta}$. \hfill $\square$

\subsection{Characterization of the Asymptotic Income Schedule} \label{sec:recurs} 

This appendix characterizes in detail the asymptotic income and wage schedules, $\pi(z)$ and $\omega(z)$.\begin{description}[leftmargin=*, labelindent=0pt, itemindent=0pt, listparindent=\parindent]
\item[1. Detailed Derivation of the Functional Equation for $\pi(z)$.] The absolute income of active experts satisfies the recursive formulation:$$\Pi_t(q) = \min \left\{ S(q), S(q) - (N-I)U_t + \beta(N-I) \Pi_{t+1}(q) \right\}, \quad \text{for } q \ge a_t$$with the boundary condition $\Pi_t(q) = 0$ for all $q < a_t$. Evaluate this expression at $q = z x_t$, divide by $x_t$, and take the limit as $t \to \infty$. From Section \ref{sec:sch}, the normalized limits are $\lim_{t \to \infty} S(zx_t)/x_t = z m^I h^{N-I}$ and $\lim_{t \to \infty} (N-I)U_t/x_t = (1+g)m^I h^{N-I}$. Furthermore, the normalized entry cutoff converges to $\lim_{t \to \infty} (a_t/x_t) = 1 + g$, and the continuation to:$$\lim_{t \to \infty} \frac{\Pi_{t+1}(z x_t)}{x_t} = \lim_{t \to \infty} \left( \frac{x_{t+1}}{x_t} \right) \frac{\Pi_{t+1}\left( z \frac{x_t}{x_{t+1}} x_{t+1} \right)}{x_{t+1}} = (1+g) \pi\left(\frac{z}{1+g}\right)$$

Substituting these limits and factoring out $m^I h^{N-I}$ yields the asymptotic functional equation:$$\pi(z) = m^I h^{N-I} \min \left\{ z, z - (1+g)+ \frac{\beta(N-I)}{m^I h^{N-I}} (1+g) \pi\left(\frac{z}{1+g}\right) \right\}, \quad \text{for } z \ge 1+g.$$Because the finite-time boundary condition dictates zero income for all inactive experts, the normalized schedule pastes continuously to the inactive region, establishing the asymptotic boundary condition $\pi(z) = 0$ for all $z \in [1, 1+g]$.  \hfill $\square$

\item[2. Uniqueness of the Income Schedule.] Global uniqueness is established by forward induction on the intervals $I_j \equiv [(1+g)^j, (1+g)^{j+1})$ for integers $j \ge 0$.

The boundary condition imposes $\pi(z) = 0$ for all $z \in I_0$. Assume $\pi(z)$ is uniquely determined on $I_{j-1}$. For any $z \in I_j$, the continuation skill $\frac{z}{1+g} \in I_{j-1}$. Because the functional equation determines $\pi(z)$ exclusively as a mapping of $\pi\left(\frac{z}{1+g}\right)$, uniqueness on $I_{j-1}$ guarantees uniqueness on $I_j$. By induction, the schedule is uniquely determined for all $z \ge 1$.  \hfill $\square$

\item[3. Characterization of the Solution.] Since global uniqueness has been established, it suffices to verify that the following piecewise schedule satisfies the functional equation globally:$$\pi(z) = m^I h^{N-I} \left\{  \begin{array}{cl} 
0 & \text{if }  z \in [1,1+g) \\
\gamma(z) & \text{if }  z \in [1+g,\sigma) \\
z & \text{if }  z \in [\sigma, +\infty)
\end{array} \right.$$where $\gamma(z)$ is the finite sum defined in Section \ref{sec:sch}:$$\gamma(z) = \sum_{j=0}^{n(z)-1} [\beta(N-I)]^j \left( z - (1+g)^{j+1} \right), \quad \text{with} \quad n(z) \equiv \left\lfloor \frac{\ln z}{\ln(1+g)} \right\rfloor$$and $\sigma \in (1+g,+\infty)$ is the unique threshold satisfying $\sigma = \gamma(\sigma)$.

To do this, the verification proceeds in two steps. First, I establish the algebraic properties of $\gamma(z)$ and prove the existence of the unique threshold $\sigma$. Second, I show that the proposed schedule indeed solves the functional equation when $z \in [1+g,\sigma)$ and $z \ge \sigma$.

For the first step, note from its closed-form expression that $\gamma(z)$ is continuous, strictly increasing for all $z\ge1+g$, and unbounded as $z \to \infty$. Moreover, $\gamma(z)$ is also the unique solution to the following recursion:$$\gamma(z) = z - (1+g) + \beta(N-I)(1+g)\gamma(z/(1+g)), \quad \text{for } z \ge 1+g$$with the boundary condition $\gamma(z) = 0$ for $z \in [1, 1+g)$. 

For $z \ge 1+g$, define $D(z) \equiv \gamma(z)-z$. Using the recursion for $\gamma(z)$, $D(z)=-(1+g)+\beta(N-I)(1+g)\gamma(z/(1+g))$. Since $\gamma(z)$ is continuous, the recursion for $\gamma(z)$ implies that $D(z)$ is continuous. Moreover, since $z/(1+g)\to\infty$ and $\gamma(z)\to\infty$ as $z\to\infty$, the same recursion implies that $D(z)\to\infty$. Hence, $D(z)$ is unbounded above. Now, when $z\in[1+g,(1+g)^{2}]$, the continuation skill $z/(1+g) \in  [1,1+g]$, so $\gamma(z/(1+g))=0$. Hence, $D(z)=-(1+g)<0$ on $[1+g,(1+g)^{2}]$. For any $z > (1+g)^{2}$, however, the continuation skill satisfies $z/(1+g) >1+g$, so $\gamma(z/(1+g))>0$. This then implies that $D(z)$ is strictly increasing for all $z\ge(1+g)^{2}$ given that $\gamma(z)$ is strictly increasing for all $z\ge1+g$. 

Because $D(z)$ is continuous, strictly negative on $[1+g,(1+g)^{2}]$, and strictly increasing and unbounded thereafter, it admits a unique root $\sigma$ in the interval $[1+g, \infty)$. The value of $\sigma$, moreover, satisfies $\sigma > (1+g)^2$. This guarantees $D(\sigma) = 0$, or equivalently, $\gamma(\sigma) = \sigma$. Consequently, $\gamma(z) < z$ for $z \in [1+g, \sigma)$, and $\gamma(z) \ge z$ for $z \ge \sigma$.

For the second step, I evaluate the right-hand side of the functional equation using the candidate $\pi(z)$ in the two active regions:\begin{enumerate}[leftmargin=*,label=(\alph*)]
\item For $z \in [1+g, \sigma)$: The continuation skill satisfies $z/(1+g) < \sigma$. The candidate therefore evaluates to $\pi(z/(1+g)) = m^I h^{N-I}\gamma(z/(1+g))$. Substituting this into the functional equation yields:$$\pi(z) = m^I h^{N-I} \min\Big\{ z, \underbrace{z - (1+g) + \beta(N-I)(1+g)\gamma(z/(1+g))}_{=\gamma(z)} \Big\}$$Thus, $\pi(z) = m^I h^{N-I} \min\{z, \gamma(z)\}$. Because $\gamma(z) < z$ in this region, the $\min$ operator strictly selects the second argument, matching the candidate $\pi(z) = m^I h^{N-I}\gamma(z)$.

\item  For $z \ge \sigma$: To match the candidate $\pi(z) = m^I h^{N-I} z$, the $\min$ operator must select the first argument, $z$. This requires:$$z - (1+g) + \beta(N-I)(1+g)\frac{\pi(z/(1+g))}{m^I h^{N-I}} \ge z$$or equivalently, $\beta(N-I)\frac{\pi(z/(1+g))}{m^I h^{N-I}} \ge 1$. Note then that $\sigma = \gamma(\sigma)$ if and only if $\beta(N-I)\gamma(\sigma/(1+g)) = 1$. Since $\sigma/(1+g) < \sigma$, the candidate schedule dictates $\frac{\pi(\sigma/(1+g))}{m^I h^{N-I}} = \gamma(\sigma/(1+g))$. Thus, the required condition binds with equality exactly at $z = \sigma$. Because the candidate schedule $\pi(z)$ is strictly increasing, it follows that $\beta(N-I)\frac{\pi(z/(1+g))}{m^I h^{N-I}} \ge 1$ for all $z \ge \sigma$. Thus, $\pi(z) = m^I h^{N-I} z$. \hfill $\square$
\end{enumerate}
\end{description}

\section{Derivations and Omitted Proofs for Section \ref{sec:advances}} \label{app:C}

\subsection{Proof of Proposition \ref{prop:main2}} \label{app:welfare}

Let $\delta(\beta)$ denote the change in total welfare as a function of $\beta \in (0, \frac{1}{1+g})$: $$\delta(\beta) \equiv  \tilde{W}_{\tau}-W_{\tau} =  \frac{\tilde{Y}_{\tau}}{1-\beta(1+\tilde{g})} - \frac{Y_{\tau}}{1-\beta(1+g)} - \frac{\rho \Delta N}{(N-I)(N-I-\Delta)(1-\beta)}$$ First, note that $\lim_{\beta \to 1/(1+g)} \delta(\beta) = -\infty$. Next, I establish that $\delta(0) > 0$, which immediately implies that $\delta(\beta)$ has at least one root in $\beta^* \in (0, \frac{1}{1+g})$.

Observe that the date-$\tau$ skill distribution is predetermined and the mass of active experts increases on impact ($\tilde{a}_\tau < a_\tau$). Thus, $\delta(0)$ can be written as:\[ \delta(0) = \int_{a_\tau}^{\infty} [\tilde{S}(q)-S(q)]dF_{\tau}(q)+\int_{\tilde{a}_\tau}^{a_\tau} \tilde{S}(q) dF_{\tau}(q)   \]where $S(q) \equiv q m^{I} h^{N-I} - \rho I$ and $\tilde{S}(q) \equiv q m^{I+\Delta} h^{N-I-\Delta} - \rho (I+\Delta)$. Per Assumption \ref{assu:full}, it follows that $\tilde{S}(q)>S(q)$ for all $q \ge x_0$. Furthermore, Assumption \ref{assu:full} implies $S(x_0) > 0$. Together, these conditions imply that $\tilde{S}(q) > S(q) \ge S(x_0) > 0$ for all $q \ge x_0$. Since both integrands are strictly positive, it immediately follows that $\delta(0) > 0$.

To establish the uniqueness of $\beta^*$, define $P(\beta) \equiv \delta(\beta)(1-\beta)\big[1-\beta(1+g)\big]\big[1-\beta(1+\tilde{g})\big]$. For all $\beta \in \left(0, \frac{1}{1+g}\right)$, the term $(1-\beta)\big[1-\beta(1+g)\big]\big[1-\beta(1+\tilde{g})\big]$ is strictly positive, so $P(\beta)$ and $\delta(\beta)$ have the same roots on that interval. Expanding $P(\beta)$ yields:$$ P(\beta) = \tilde{Y}_\tau (1-\beta)\big[1-\beta(1+g)\big] - Y_\tau (1-\beta)\big[1-\beta(1+\tilde{g})\big] - C \big[1-\beta(1+g)\big]\big[1-\beta(1+\tilde{g})\big] $$where $C \equiv \frac{\rho \Delta N}{(N-I)(N-I-\Delta)} > 0$. Because each of its three terms is the product of two linear functions of $\beta$, $P(\beta)$ is a polynomial of degree at most two. Evaluating $P(\beta)$ at the interval boundaries yields $P(0) = \delta(0) > 0$ and$$P\left(\frac{1}{1+g}\right) = -Y_\tau \left(1-\frac{1}{1+g}\right)\left( 1 - \frac{1+\tilde{g}}{1+g} \right) < 0.$$Because $P(0) > 0$ and $P\left(\frac{1}{1+g}\right) < 0$, the fact that $P(\beta)$ is a polynomial of degree at most two immediately implies that it crosses zero exactly once in $\left(0, \frac{1}{1+g}\right)$. Hence, $\beta^*$ is unique. \hfill $\square$
 
 \subsection{Equilibrium with a Task-Creating Technology} \label{app:creating}
 
 This appendix characterizes the existence of the permanent full-adoption equilibrium after the introduction of a task-creating technology. Throughout, I maintain the baseline full-automation condition in Assumption \ref{assu:full}. Under this maintained condition, I show that Assumption \ref{assu:UAdop} guarantees the existence of a permanent full-adoption equilibrium. I also show that, when Assumption \ref{assu:UAdop} fails, no such equilibrium can be sustained.
 
I begin with the following preliminary result:

\begin{lem}\label{lem:creating_full_automation} Under Assumption \ref{assu:full}, if an expert with skill $q\ge x_0$ adopts the task-creating technology, then she optimally assigns all $I-\Delta$ automatable tasks to machines. \end{lem}

\begin{proof} Consider an expert with skill $q\ge x_0$ who adopts the task-creating technology. If she assigns a measure $k\in[0,I-\Delta]$ of the remaining automatable tasks to machines, then she hires $N-k$
novices and obtains payoff:
\[
q m^k h^{N-\Delta-k}H^\Delta-\rho k-w_t(q)(N-k),
\]
where $w_t(q)\ge 0$ is the wage paid to novices.

Since $m>h$, this objective is strictly convex in $k$. Therefore, as in Lemma \ref{lem:cut}, the optimum over $k\in[0,I-\Delta]$ is attained at a corner: either $k=0$ or $k=I-\Delta$. Full automation of the
remaining automatable tasks strictly dominates no automation if and only if:
\[
q\left(m^{I-\Delta}h^{N-I}H^\Delta-h^{N-\Delta}H^\Delta\right)
>
(I-\Delta)(\rho-w_t(q)).
\]
Because $w_t(q)\ge 0$, this condition is most stringent when $w_t(q)=0$. Thus, it is enough to show that:
\[
q> \frac{\rho(I-\Delta)} {h^N\left(\frac Hm\right)^\Delta \left[\left(\frac mh\right)^I-\left(\frac mh\right)^\Delta\right]}
\]
It remains to compare this threshold with the baseline full-automation threshold. Since $H>m$ and $ (m/h)^x$ is convex in $x$:
\[ \frac{\rho(I-\Delta)} {h^N\left(\frac Hm\right)^\Delta 
\left[\left(\frac mh\right)^I-\left(\frac mh\right)^\Delta\right]} <
\frac{\rho(I-\Delta)}
{h^N
\left[\left(\frac mh\right)^I-\left(\frac mh\right)^\Delta\right]} \le
\overbrace{\frac{\rho I}
{h^N\left[\left(\frac mh\right)^I-1\right]}}^{\text{full-automation cutoff} }
\]
where the last inequality follows from the secant-slope inequality:
\[
\frac{\left(\frac mh\right)^I-1}{I}
\le
\frac{\left(\frac mh\right)^I-\left(\frac mh\right)^\Delta}{I-\Delta}.
\]
Therefore, by the maintained full-automation condition, every $q\ge x_0$ exceeds the threshold for full automation under the task-creating technology. Hence, full automation of the remaining $I-\Delta$ automatable tasks is strictly optimal. \end{proof}

\begin{lem} \label{lem:creating_existence} Suppose Assumption \ref{assu:UAdop} holds. Then there exists an equilibrium in which all active experts adopt the task-creating technology at every date $t \geq \tau$.\end{lem}

\begin{proof} By Lemma \ref{lem:creating_full_automation}, under the maintained full-automation condition, active experts assign all remaining automatable tasks to machines when they adopt the new technology. Thus, construct the candidate equilibrium by applying the baseline equilibrium characterization to the transformed technology in which active experts hire $N-I+\Delta$ novices and produce $q m^{I-\Delta}h^{N-I}H^\Delta$. Let $w_t(q)$ denote the wage schedule in this candidate equilibrium. It remains to be verified that no expert wants to revert from new technology to old technology. I proceed in three steps.

First, I show that the incentive to adopt the new technology is strictly increasing in expert skill, $q$. Second, I establish that under Assumption \ref{assu:UAdop}, the marginal active expert, $a_t$, strictly prefers the new technology for every $t \ge \tau$. Thus, by steps 1 and 2, all active experts $q \ge a_t$ permanently adopt the task-creating technology. Third, I show that no inactive expert can profitably enter using the old technology.

\begin{description}[leftmargin=*, labelindent=0pt, itemindent=0pt, listparindent=\parindent]
 \item[Step 1.] Consider an active expert with skill $q \ge a_t$ at time $t$. Given $w_t(q)$, the optimal profits under the baseline and new task-creating technologies are, respectively: \begin{align*} & \Pi_t^{old}(q) = q m^I h^{N-I} - \rho I - w_t(q)(N-I) \\
 &\Pi_t^{new}(q) = q m^{I-\Delta} h^{N-I} H^\Delta - \rho(I-\Delta) - w_t(q)(N-I+\Delta) \end{align*}The net incentive to adopt the new technology is the profit difference:$$\Pi_t^{new}(q) - \Pi_t^{old}(q) = q m^{I-\Delta} h^{N-I} (H^\Delta - m^\Delta) + \rho \Delta - w_t(q)\Delta$$Because $H > m$, the first term is strictly increasing in $q$. By the baseline equilibrium characterization applied to the transformed technology, the wage schedule $w_t(q)$ is non-increasing in $q$ (see Proposition \ref{prop:existence}), implying the cost term $-w_t(q)\Delta$ is non-decreasing in $q$. Therefore, the net incentive to adopt is strictly increasing in $q$.
  
\item[Step 2.] I now establish that under Assumption \ref{assu:UAdop}, the marginal active expert $a_t$ strictly prefers the new technology. Combined with the monotonicity established in Step 1, this guarantees that all active experts optimally adopt the task-creating technology for all $t \ge \tau$, confirming the conjectured equilibrium.

In this equilibrium, the marginal expert earns zero profit under the new technology, implying $\Pi_t^{new}(a_t) = 0$. This boundary condition pins down the wage paid by the marginal expert: \begin{equation} \label{eq:wtA} w_t(a_t) = \frac{a_t m^{I-\Delta} h^{N-I} H^\Delta - \rho(I-\Delta)}{N-I+\Delta}\end{equation}

Given $\Pi_t^{new}(a_t) = 0$, the marginal expert strictly prefers the new technology if and only if deviating to the baseline technology yields strictly negative profits, $\Pi_t^{old}(a_t) < 0$. Substituting $w_t(a_t)$ into the expression for $\Pi_t^{old}(a_t)$, multiplying by $N-I+\Delta$, and rearranging terms yields: \begin{equation} \label{eq: cond} a_t m^I h^{N-I} (N-I) \left[ 1 + \frac{\Delta}{N-I} - \left(\frac{H}{m}\right)^\Delta \right] < \rho \Delta N\end{equation}

Under Assumption \ref{assu:UAdop}, the productivity premium of new tasks satisfies:$$\left(\frac{H}{m}\right)^\Delta \ge 1 + \frac{\Delta}{N-I}$$This condition guarantees that the bracketed term on the left-hand side is non-positive. Since the right-hand side, $\rho \Delta N$, is strictly positive, the strict inequality $\Pi_t^{old}(a_t) < 0$ always holds.

\item[Step 3.] Finally, I show that no inactive expert can profitably enter using the old technology. Entry using the new technology is already ruled out by the construction of the candidate equilibrium. Consider any inactive expert $q<a_t$ who deviates by using the old technology and offering wage $w'$. Any successful deviation must deliver at least the lifetime payoff obtained by novices at the marginal active match:
\[
w' + \beta \Pi_{t+1}(q) \geq w_t(a_t)+\beta \Pi_{t+1}(a_t).
\]
Here, $\Pi_{t+1}$ denotes the continuation income schedule in the candidate full-adoption equilibrium. Since $\Pi_{t+1}$ is non-decreasing in skill and $q<a_t$, any successful deviation requires $w'\geq w_t(a_t)$. Consequently: 
\begin{equation*} \begin{split}
q m^I h^{N-I}-\rho I-w'(N-I)
& \leq q m^I h^{N-I}-\rho I-w_t(a_t)(N-I) \\
&< a_t m^I h^{N-I}-\rho I-w_t(a_t)(N-I) = \Pi_t^{old}(a_t)< \Pi_t^{new}(a_t) =0 \end{split} \end{equation*} Hence, no inactive expert can profitably enter using the old technology. This completes the verification of the full-adoption equilibrium. \end{description}\end{proof}

\begin{lem}\label{lem:creating_nonexistence} Suppose Assumption \ref{assu:UAdop} fails. Then no equilibrium exists in which all active experts adopt the task-creating technology at every date $t\geq \tau$. \end{lem}

\begin{proof} Suppose, by contradiction, that such an equilibrium exists. In any permanent full-adoption equilibrium, the allocation block is the same as in the baseline model with span of control $N-I+\Delta$, so $a_t\to\infty$ as $t\to\infty$. Moreover, the marginal active expert must weakly prefer the new technology to the old one, so the weak version of condition (\ref{eq: cond}) must hold at every date $t\geq\tau$.

When Assumption \ref{assu:UAdop} fails, the bracketed term in (\ref{eq: cond}) is strictly positive. The left-hand side of (\ref{eq: cond}) is therefore strictly positive and proportional to $a_t$. Since $a_t\to\infty$, it eventually exceeds $\rho\Delta N$. At that date, the marginal active expert strictly prefers the old technology, contradicting permanent full adoption. \end{proof}

\end{appendix}

 {\small
 \bibliographystyle{ecta}
\bibliography{KT_references}} 

@unpublished{ide2024turing,
  title={The Turing Valley: How AI Capabilities Shape Labor Income},
  author={Ide, Enrique and Talam\`{a}s, Eduard},
  note={arXiv preprint arXiv:2408.16443},
  year={2024}
}

@article{fudenberg2021working,
  title={Working to Learn},
  author={Fudenberg, Drew and Georgiadis, George and Rayo, Luis},
  journal={Journal of Economic Theory},
  volume={197},
  pages={105347},
  year={2021},
  publisher={Elsevier}
}

@article{fudenberg2019training,
  title={Training and Effort Dynamics in Apprenticeship},
  author={Fudenberg, Drew and Rayo, Luis},
  journal={American Economic Review},
  volume={109},
  number={11},
  pages={3780--3812},
  year={2019},
  publisher={American Economic Association 2014 Broadway, Suite 305, Nashville, TN 37203}
}

@article{garicano2017relational,
  title={Relational Knowledge Transfers},
  author={Garicano, Luis and Rayo, Luis},
  journal={American Economic Review},
  volume={107},
  number={9},
  pages={2695--2730},
  year={2017},
  publisher={American Economic Association 2014 Broadway, Suite 305, Nashville, TN 37203}
}

@book{beane2024skillcode,
  title={The Skill Code: How to Save Human Ability in an Age of Intelligent Machines},
  author={Beane, Matthew},
  year={2024},
  publisher={Harper Business},
  address={New York},
}

@unpublished{autor2024applying,
  title={Applying AI to Rebuild Middle Class Jobs},
  author={Autor, David},
  year={2024},
note = "NBER Working Paper No. 32140",
}

@article{moll2022uneven,
  title={Uneven Growth: Automation's Impact on Income and Wealth Inequality},
  author={Moll, Benjamin and Rachel, Lukasz and Restrepo, Pascual},
  journal={Econometrica},
  volume={90},
  number={6},
  pages={2645--2683},
  year={2022},
  publisher={Wiley Online Library}
}

@article{acemoglu2022automation,
author = {Acemoglu, Daron and Loebbing, Jonas},
title = {Automation and Polarization},
journal = {Journal of Political Economy},
volume = {134},
number = {3},
pages = {1017-1072},
year = {2026}
}

@article{garicano2000hierarchies,
  title={Hierarchies and the Organization of Knowledge in Production},
  author={Garicano, Luis},
  journal={Journal of Political Economy},
  volume={108},
  number={5},
  pages={874--904},
  year={2000},
  publisher={The University of Chicago Press}
}

@article{garicano2015knowledge,
  title={Knowledge-Based Hierarchies: Using Organizations to Understand the Economy},
  author={Garicano, Luis and Rossi-Hansberg, Esteban},
  journal={Annual Review of Economics},
  volume={7},
  number={1},
  pages={1--30},
  year={2015}
}

@article{acemoglu2018race,
  title={The Race Between Man and Machine: Implications of Technology for Growth, Factor Shares, and Employment},
  author={Acemoglu, Daron and Restrepo, Pascual},
  journal={American Economic Review},
  volume={108},
  number={6},
  pages={1488--1542},
  year={2018},
  publisher={American Economic Association 2014 Broadway, Suite 305, Nashville, TN 37203}
}

@article{zeira1998workers,
  title={Workers, Machines, and Economic Growth},
  author={Zeira, Joseph},
  journal={The Quarterly Journal of Economics},
  volume={113},
  number={4},
  pages={1091--1117},
  year={1998},
  publisher={MIT Press}
}

@article{garicano2004inequality,
  title={Inequality and the Organization of Knowledge},
  author={Garicano, Luis and Rossi-Hansberg, Esteban},
  journal={American Economic Review Papers and Proceedings},
  volume={94},
  number={2},
  pages={197--202},
  year={2004},
  publisher={American Economic Association}
}

@article{garicano2006organization,
  title={Organization and Inequality in a Knowledge Economy},
  author={Garicano, Luis and Rossi-Hansberg, Esteban},
  journal={The Quarterly Journal of Economics},
  volume={121},
  number={4},
  pages={1383--1435},
  year={2006},
  publisher={MIT Press}
}

@incollection{acemoglu2011chapter,
  title={Skills, Tasks and Technologies: Implications for Employment and Earnings},
  author={Acemoglu, Daron and Autor, David},
  booktitle={Handbook of Labor Economics},
  volume={4},
  pages={1043--1171},
  year={2011},
  publisher={Elsevier}
}

@article{acemoglu2022tasks,
author = {Acemoglu, Daron and Restrepo, Pascual},
title = {Tasks, Automation, and the Rise in U.S. Wage Inequality},
journal = {Econometrica},
volume = {90},
number = {5},
pages = {1973-2016},
year = {2022}
}

@unpublished{aghion2017artificial,
 title = "Artificial Intelligence and Economic Growth",
 author = "Aghion, Philippe and Jones, Benjamin F and Jones, Charles I",
 note = "NBER Working Paper No. 23928",
 year = "2017",
}

@article{caicedo2019learning,
Author = {Caicedo, Santiago and Lucas Jr, Robert E and Rossi-Hansberg, Esteban},
Title = {Learning, Career Paths, and the Distribution of Wages},
Journal = {American Economic Journal: Macroeconomics},
Volume = {11},
Number = {1},
Year = {2019},
Month = {January},
Pages = {49-88}
}

@incollection{AcemogluRestrepoNBER,
title = {8. Artificial Intelligence, Automation, and Work},
booktitle = {The Economics of Artificial Intelligence: An Agenda},
author = {Daron Acemoglu and Pascual Restrepo},
editor = {Ajay Agrawal and Joshua Gans and Avi Goldfarb},
publisher = {University of Chicago Press},
pages = {197--236},
year = {2019}
}

@unpublished{korinek,
 title = "Scenarios for the Transition to AGI",
 author = "Korinek, Anton and Suh, Donghyun",
 note = "NBER Working Paper No. 32255",
 year = {2024}
 }

@article{autor2003,
    author = {Autor, David H. and Levy, Frank and Murnane, Richard J.},
    title = "{The Skill Content of Recent Technological Change: An Empirical Exploration*}",
    journal = {The Quarterly Journal of Economics},
    volume = {118},
    number = {4},
    pages = {1279-1333},
    year = {2003},
    month = {11},
}

@book{becker1964human,
  title={Human Capital: A Theoretical and Empirical Analysis, with Special Reference to Education},
  author={Becker, Gary S},
  year={1964},
  publisher={University of Chicago press}
}

@unpublished{AKR,
  title={Tasks At Work: Comparative Advantage, Technology and Labor Demand},
  author={Acemoglu, Daron and Kong, Fredric and Restrepo, Pascual},
  year={2024},
note = "NBER Working Paper No. 32872",
}

@article{lucas2009ideas,
  title={Ideas and Growth},
  author={Lucas, Robert E},
  journal={Economica},
  volume={76},
  number={301},
  pages={1--19},
  year={2009}
}

@article{de2018clans,
  title={Clans, Guilds, and Markets: Apprenticeship Institutions and Growth in the Preindustrial Economy},
  author={De la Croix, David and Doepke, Matthias and Mokyr, Joel},
  journal={The Quarterly Journal of Economics},
  volume={133},
  number={1},
  pages={1--70},
  year={2018}
}

@article{jarosch2021learning,
  title={Learning From Coworkers},
  author={Jarosch, Gregor and Oberfield, Ezra and Rossi-Hansberg, Esteban},
  journal={Econometrica},
  volume={89},
  number={2},
  pages={647--676},
  year={2021}
}

@misc{amodei,
  author       = {Dario Amodei},
  title        = {Machines of Loving Grace: How AI Could Transform the World for the Better},
  note         = {{O}ctober, 2024. Available at \href{https://darioamodei.com/machines-of-loving-grace}{https://darioamodei.com/machines-of-loving-grace}  (accessed June 15, 2025)},
  year         = {2024},
}

@article{hassabis,
  author       = {Steven Levy},
  title        = {Demis Hassabis Embraces the Future of Work in the Age of AI},
  journal      = {Wired},
  year         = {2025},
 note         = {{J}une 4, 2025. Available at \href{https://www.wired.com/story/google-deepminds-ceo-demis-hassabis-thinks-ai-will-make-humans-less-selfish/}{https://www.wired.com/story/google-deepminds-ceo-demis-hassabis-thinks-ai-will-make-humans-less-selfish/} (accessed June 15, 2025)},
}

@misc{altman,
  author       = {Sam Altman},
  title        = {The Gentle Singularity},
  note         = {{J}une, 2025. Available at \href{https://blog.samaltman.com/the-gentle-singularity}{https://blog.samaltman.com/the-gentle-singularity}  (accessed June 15, 2025)},
  year         = {2025},
}

@article{brynjolfsson2023generative,
  title={Generative AI at Work},
  author={Brynjolfsson, Erik and Li, Danielle and Raymond, Lindsey},
      journal = {The Quarterly Journal of Economics},
    volume = {140},
    number = {2},
    pages = {889-942},
    year = {2025},
    month = {02}
}

@article{AIKE,
  title={Artificial Intelligence in the Knowledge Economy},
  author={Ide, Enrique and Talam\`{a}s, Eduard},
  journal = {Journal of Political Economy},
  volume = {133},
  number = {12},
  pages = {3762-3800},
  year = {2025},
}

@article{autor2025expertise,
    title = {Expertise},
    author = {Autor, David and Thompson, Neil},
  journal = {Journal of the European Economic Association},
  volume = {23},
  number = {4},
  pages = {1203-1271},
  year = {2025},
}

@book{polanyi,
  title={The Tacit Dimension},
  author={Polanyi, Michael},
  year={1966},
  publisher={Doubleday \& Co}
}

@book{foray2004economics,
  title={Economics of Knowledge},
  author={Foray, Dominique},
  year={2004},
  publisher={MIT Press}
}

@unpublished{berger2,
  title={Employer and Employee Responses to Generative AI: Early Evidence},
  author={Berger, Philip G. and Cai, Wei and Qiu, Lin and Xinyi Shen, Cindy},
  year={2024},
  note = "Available at SSRN 4874061",
}

@article{lucas2014knowledge,
  title={Knowledge Growth and the Allocation of Time},
  author={Lucas, Robert E and Moll, Benjamin},
  journal={Journal of Political Economy},
  volume={122},
  number={1},
  pages={1--51},
  year={2014},
  publisher={University of Chicago Press Chicago, IL}
}

@article{eaton1999international,
  title={International Technology Diffusion: Theory and Measurement},
  author={Eaton, Jonathan and Kortum, Samuel},
  journal={International Economic Review},
  volume={40},
  number={3},
  pages={537--570},
  year={1999},
  publisher={Wiley Online Library}
}

@article{kortum1997research,
  title={Research, Patenting, and Technological change},
  author={Kortum, Samuel S},
  journal={Econometrica},
  pages={1389--1419},
  year={1997}
}

@article{eaton2002technology,
  title={Technology, Geography, and Trade},
  author={Eaton, Jonathan and Kortum, Samuel},
  journal={Econometrica},
  volume={70},
  number={5},
  pages={1741--1779},
  year={2002},
  publisher={Wiley Online Library}
}

@article{buera2020global,
  title={The Global Diffusion of Ideas},
  author={Buera, Francisco J and Oberfield, Ezra},
  journal={Econometrica},
  volume={88},
  number={1},
  pages={83--114},
  year={2020},
  publisher={Wiley Online Library}
}

@inbook{mokyr2019,
title = "The Economics of Apprenticeship",
author = "Joel Mokyr",
year = "2019",
pages = "20--43",
booktitle = "Apprenticeship in Early Modern Europe",
publisher = "Cambridge University Press",
}

@article{buera,
   author = "Buera, Francisco J. and Lucas, Robert E.",
   title = "Idea Flows and Economic Growth", 
   journal= "Annual Review of Economics",
   year = "2018",
   volume = "10",
   pages = "315-345",
  }

@misc{WSJ,
  author       = {{The Journal Podcast}},
  title        = {AI Is Coming for Entry-Level Jobs},
  howpublished = {\textit{The Journal - WSJ Podcast}},
  note         = {{M}onday, July 7, 2025. Available at \href{https://www.wsj.com/podcasts/the-journal/ai-is-coming-for-entry-level-jobs/e9f9eb31-14ad-498d-94f3-11ce91e9c464}{https://www.wsj.com/podcasts/the-journal/ai-is-coming-for-entry-level-jobs/e9f9eb31-14ad-498d-94f3-11ce91e9c464}  (accessed July 19, 2025)},
  year         = {2025}
}

@article{beaneWSJ,
  author       = {Beane, Matthew},
  title        = {How AI Could Keep Young Workers From Getting the Skills They Need},
  journal      = {The Wall Street Journal},
  year         = {2024},
 note         = {{J}uly 26, 2024. Available at \href{https://www.wsj.com/lifestyle/careers/ai-training-young-employees-ed08cedc}{https://www.wsj.com/lifestyle/careers/ai-training-young-employees-ed08cedc} (accessed July 20, 2025)},
}

@article{GKW,
  title={The Impact of AI on Global Knowledge Work},
  author={Ide, Enrique and Talam\`{a}s, Eduard},
  journal={Journal of Monetary Economics},
volume = {157},
pages = {103876},
year = {2026}
}

@unpublished{freund2025job,
  title={Job Transformation, Specialization, and the Labor Market Effects of AI},
  author={Freund, Lukas B. and Mann, Lukas F.},
  year={2025},
  note={CESifo Working Paper No. 12072}
}

@article{perla2014equilibrium,
  title={Equilibrium Imitation and Growth},
  author={Perla, Jesse and Tonetti, Christopher},
  journal={Journal of Political Economy},
  volume={122},
  number={1},
  pages={52--76},
  year={2014},
  publisher={University of Chicago Press, Chicago, IL}
}

@article{perla2021equilibrium,
Author = {Perla, Jesse and Tonetti, Christopher and Waugh, Michael E.},
Title = {Equilibrium Technology Diffusion, Trade, and Growth},
Journal = {American Economic Review},
Volume = {111},
Number = {1},
Year = {2021},
Month = {January},
Pages = {73–128},
DOI = {10.1257/aer.20151645},
URL = {https://www.aeaweb.org/articles?id=10.1257/aer.20151645}}

@article{benhabib,
author = {Benhabib, Jess and Perla, Jesse and Tonetti, Christopher},
title = {Reconciling Models of Diffusion and Innovation: A Theory of the Productivity Distribution and Technology Frontier},
journal = {Econometrica},
volume = {89},
number = {5},
pages = {2261-2301},
year = {2021}
}

@misc{garicanoAIbecker,
  author       = {Luis Garicano},
  title        = {The AI Becker Problem},
  note         = {{J}uly 23, 2025. Available at \href{https://www.siliconcontinent.com/p/the-ai-becker-problem}{https://www.siliconcontinent.com/p/the-ai-becker-problem}  (accessed July 23, 2025)},
  year         = {2025},
}

@article{oyer2006initial,
  title={Initial Labor Market Conditions and Long-Term Outcomes for Economists},
  author={Oyer, Paul},
  journal={Journal of Economic Perspectives},
  volume={20},
  number={3},
  pages={143--160},
  year={2006},
  publisher={American Economic Association}
}

@article{arellano2024career,
  title={Career Consequences of Firm Heterogeneity for Young Workers: First Job and Firm Size},
  author={Arellano-Bover, Jaime},
  journal={Journal of Labor Economics},
  volume={42},
  number={2},
  pages={549--589},
  year={2024},
  publisher={The University of Chicago Press Chicago, IL}
}

@article{arellano2022effects,
    author = {Arellano-Bover, Jaime},
    title = {The Effect of Labor Market Conditions at Entry on Workers' Long-Term Skills},
    journal = {The Review of Economics and Statistics},
    volume = {104},
    number = {5},
    pages = {1028-1045},
    year = {2022},
    month = {09}
}

@article{beane2019shadow,
author = {Matthew Beane},
title ={Shadow Learning: Building Robotic Surgical Skill When Approved Means Fail},
journal = {Administrative Science Quarterly},
volume = {64},
number = {1},
pages = {87-123},
year = {2019}
}

@misc{bloomberg,
  author       = {{BloombergTV}},
  title        = {AI Boom, Entry-Level Bust: Why College Grads Are Struggling to Land Jobs},
  note         = {{A}ugust 9, 2025. Available at \href{https://www.bloomberg.com/news/videos/2025-08-09/ai-boom-entry-level-bust-grads-struggle-to-land-jobs-video}{https://www.bloomberg.com/news/videos/2025-08-09/ai-boom-entry-level-bust-grads-struggle-to-land-jobs-video}  (accessed August 11, 2025)},
  year         = {2025},
}

@article{roose,
  author       = {Kevin Roose},
  title        = {For Some Recent Graduates, the A.I. Job Apocalypse May Already Be Here},
  journal      = {The New York Times},
  year         = {2025},
 note         = {{M}ay 30, 2025. Available at \href{https://www.nytimes.com/2025/05/30/technology/ai-jobs-college-graduates.html}{https://www.nytimes.com/2025/05/30/technology/ai-jobs-college-graduates.html} (accessed August 11, 2025)},
}

@unpublished{canaries2025,
  title={Canaries in the Coal Mine? Six Facts about the Recent Employment Effects of Artificial Intelligence},
  author={Brynjolfsson, Erik and Chandar, Bharat and Chen, Ruyu},
  note={Working Paper},
  year={2025}
}

@unpublished{de2025artificial,
  title={Artificial Intelligence in the Office and the Factory: Evidence from Administrative Software Registry Data},
  author={de Souza, Gustavo},
  note={FRB of Chicago Working Paper No. 2025-11},
  year={2025}
}

@unpublished{johnston2025labor,
  title={The Labor Market Effects of Generative AI: A Difference-in-Differences Analysis of AI Exposure},
  author={Johnston, Andrew and Makridis, Christos},
  note={Available at SSRN 5375017},
  year={2025}
}

@unpublished{GR2025AI,
  title={Training in the Age of AI: A Theory of Apprenticeship Viability},
  author={Garicano, Luis and Rayo, Luis},
  note={Working Paper},
  year={2025}
}

@unpublished{gansG,
 title = "Growth in AI Knowledge",
 author = "Gans, Joshua S",
  note={NBER Working Paper No. 33907},
  year={2025}
}

@book{embretch,
  address = {Berlin},
  author = {Embrechts, Paul and Kl\"{o}ppelberg, Claudia and Mikosch, Thomas},
  publisher = {Springer-Verlag},
  series = {Applications of Mathematics (New York)},
  title = {Modelling Extremal Events: for Insurance and Finance},
  volume = 33,
  year = 1997
}

@book{bingham1989regular,
  title     = {Regular Variation},
  author    = {Bingham, N. H. and Goldie, C. M. and Teugels, J. L.},
  year      = {1989},
  publisher = {Cambridge University Press},
  address   = {Cambridge},
  series    = {Encyclopedia of Mathematics and its Applications},
  volume    = {27},
  edition   = {First paperback (with additions)},
  isbn      = {0-521-37943-1}
}

@book{mokyr2002gifts,
  title={The Gifts of Athena: Historical Origins of the Knowledge Economy},
  author={Mokyr, Joel},
  year={2002},
  publisher={Princeton University Press},
  address={Princeton, NJ}
}

@book{polanyi1962personal,
  title={Personal Knowledge: Towards a Post-Critical Philosophy},
  author={Polanyi, Michael},
  year={1962},
  publisher={Routledge},
  address={London},
  note={First published 1958. 2005 Taylor \& Francis e-Library edition.}
}

@book{lave1991situated,
  title={Situated Learning: Legitimate Peripheral Participation},
  author={Lave, Jean and Wenger, Etienne},
  year={1991},
  publisher={Cambridge University Press}
}

@article{brown1991organizational,
  title={Organizational Learning and Communities-of-Practice: Toward a Unified View of Working, Learning, and Innovation},
  author={Brown, John Seely and Duguid, Paul},
  journal={Organization Science},
  volume={2},
  number={1},
  pages={40--57},
  year={1991},
  publisher={Informs}
}

@article{clarke2025record,
  author       = {Clarke, Paul},
  title        = {Record Number of Graduates Apply to be Wall Street Bankers},
  journal      = {Financial News},
  year         = {2025},
 note         = {{S}eptember 1, 2025. Available at \href{https://www.fnlondon.com/articles/record-number-of-graduates-apply-to-be-wall-street-bankers-0ac689a8}{https://www.fnlondon.com/articles/record-number-of-graduates-apply-to-be-wall-street-bankers-0ac689a8} (accessed February 22, 2026)},
}

@article{vlamis2025mbb,
  author       = {Vlamis, Kelsey and Varanasi, Lakshmi and Paradis, Tim},
  title        = {MBB Explained: How Hard it is to Get Hired and What it's Like to Work for the Prestigious Strategy Consulting Firms, McKinsey, Bain, and BCG},
  journal      = {Business Insider},
  year         = {2025},
 note         = {{J}anuary 29, 2025. Available at \href{https://www.businessinsider.com/mbb-explained-mckinsey-bain-bcg-compare-big-three-strategy-consultants-2024-11}{https://www.businessinsider.com/mbb-explained-mckinsey-bain-bcg-compare-big-three-strategy-consultants-2024-11} (accessed February 22, 2026)},
}

@article{glickmanSaeedy2025,
  author       = {Glickman, Ben and Saeedy, Alexander},
  title        = {110-Hour Workweeks Drove Young Bankers at a Boutique Firm to the Brink},
  journal      = {The Wall Street Journal},
  year         = {2025},
  note         = {{A}pril 30, 2025. Available at \href{https://www.wsj.com/finance/banking/banking-culture-robert-w-baird-aeccc947}{https://www.wsj.com/finance/banking/banking-culture-robert-w-baird-aeccc947} (accessed February 25, 2026)},
}

@article{saeedy2024a,
  author       = {Saeedy, Alexander},
  title        = {How Bank of America Ignores Its Own Rules Meant to Prevent Dangerous Workloads},
  journal      = {The Wall Street Journal},
  year         = {2024},
  note         = {{A}ugust 11, 2024. Available at \href{https://www.wsj.com/finance/banking/bank-of-america-worker-death-policies-89eff5f6}{https://www.wsj.com/finance/banking/bank-of-america-worker-death-policies-89eff5f6} (accessed February 25, 2026)},
}

@article{saeedy2024b,
  author       = {Saeedy, Alexander},
  title        = {Bank of America Urges Bankers to Sound Alarm on Overwork After WSJ Investigation},
  journal      = {The Wall Street Journal},
  year         = {2024},
  note         = {{A}ugust 13, 2024. Available at \href{https://www.wsj.com/finance/bank-of-america-tells-bankers-to-report-bosses-who-pressure-them-to-underreport-hours-758cb44b}{https://www.wsj.com/finance/bank-of-america-tells-bankers-to-report-bosses-who-pressure-them-to-underreport-hours-758cb44b} (accessed February 25, 2026)},
}

@misc{atkinson2026young,
  author       = {Atkinson, Tyler and Yamco, Shane},
  title        = {Young Workers' Employment Drops in Occupations with High AI Exposure},
  year         = {2026},
  publisher    = {Federal Reserve Bank of Dallas},
  note         = {{J}anuary 6, 2026. Available at \href{https://www.dallasfed.org/research/economics/2026/0106}{https://www.dallasfed.org/research/economics/2026/0106} (accessed March 2, 2026)},
}

@unpublished{klein2025generative,
  title={Generative AI and labor market outcomes: Evidence from the United Kingdom},
  author={Klein Teeselink, Bouke},
 note = "Available at SSRN 5516798",
 year = {2025},
 }

@unpublished{lichtinger2025,
  title={Generative AI as Seniority-Biased Technological Change: Evidence from U.S. Résumé and Job Posting Data},
  author={Seyed Hosseini and Guy Lichtinger},
  note={Available at SSRN 5425555},
  year={2025}
}

@article{lahart2026,
  author       = {Lahart, Justin},
  title        = {Economists Are Studying the Slowing Job Market---and Feeling It Themselves},
  journal      = {The Wall Street Journal},
  year         = {2026},
  note         = {{J}anuary 18, 2026. Available at \href{https://www.wsj.com/economy/jobs/economists-job-market-hiring-2213807b}{https://www.wsj.com/economy/jobs/economists-job-market-hiring-2213807b} (accessed February 26, 2026)},
}

@article{ellis2026,
  author       = {Ellis, Lindsay},
  title        = {Even MBAs From Top Business Schools Are Struggling to Get Hired},
  journal      = {The Wall Street Journal},
  year         = {2026},
  note         = {{J}anuary 19, 2026. Available at \href{https://www.wsj.com/lifestyle/careers/even-mbas-from-top-business-schools-are-struggling-to-get-hired-11f4a167}{https://www.wsj.com/lifestyle/careers/even-mbas-from-top-business-schools-are-struggling-to-get-hired-11f4a167} (accessed February 26, 2026)},
}

@article{autor2024new,
  title={New Frontiers: The Origins and Content of New Work, 1940--2018},
  author={Autor, David and Chin, Caroline and Salomons, Anna and Seegmiller, Bryan},
  journal={The Quarterly Journal of Economics},
  volume={139},
  number={3},
  pages={1399--1465},
  year={2024},
  publisher={Oxford University Press}
}

@article{eeckhout2011identifying,
  title={Identifying Sorting---In Theory},
  author={Eeckhout, Jan and Kircher, Philipp},
  journal={The Review of Economic Studies},
  volume={78},
  number={3},
  pages={872--906},
  year={2011},
  publisher={Oxford University Press}
}

@article{becker1973theory,
  title={A Theory of Marriage: Part I},
  author={Becker, Gary S},
  journal={Journal of Political Economy},
  volume={81},
  number={4},
  pages={813--846},
  year={1973},
  publisher={The University of Chicago Press}
}

@unpublished{demirer2026chaining,
  title={Chaining Tasks, Redefining Work: A Theory of AI Automation},
  author={Demirer, Mert and Horton, John J and Immorlica, Nicole and Lucier, Brendan and Shahidi, Peyman},
  year={2026},
  note={NBER Working Paper 34859}
}

@unpublished{beraja2026value,
  title={The Value of Organizational Learning Technologies},
  author={Beraja, Martin and Talam{\`a}s, Eduard},
  year={2026},
  note={Working Paper}
}

@unpublished{jones2025past,
  title={Past Automation and Future AI: How Weak Links Tame the Growth Explosion},
  author={Jones, Charles I and Tonetti, Christopher},
  year={2026},
  note={Working Paper}
}

@unpublished{acemoglu2026collapse,
  title={AI, Human Cognition and Knowledge Collapse},
  author={Acemoglu, Daron and Kong, Dingwen and Ozdaglar, Asuman},
  year={2026},
  note={NBER Working Paper 34910}
}

@unpublished{humlum2026still,
  title={Still Waters, Rapid Currents: Early Labor Market Transformation under Generative AI},
  author={Humlum, Anders and Vestergaard, Emilie},
  year={2026},
  note={NBER Working Paper 33777}
}

@article{jones2024framework,
Author = {Jones, Benjamin F. and Liu, Xiaojie},
Title = {A Framework for Economic Growth with Capital-Embodied Technical Change},
Journal = {American Economic Review},
Volume = {114},
Number = {5},
Year = {2024},
Month = {May},
Pages = {1448–87},
DOI = {10.1257/aer.20221180},
URL = {https://www.aeaweb.org/articles?id=10.1257/aer.20221180}}

@unpublished{sachs2015robots,
  title={Robots: Curse or Blessing? A Basic Framework},
  author={Sachs, Jeffrey D and Benzell, Seth G and LaGarda, Guillermo},
  year={2015},
  note={NBER Working Paper 21091}
}

@unpublished{sachs2012smart,
  title={Smart Machines and Long-term Misery},
  author={Sachs, Jeffrey D and Kotlikoff, Laurence J.},
  year={2012},
  note={NBER Working Paper 18629}
}

@unpublished{afrouzi2026automation,
  title={Automation, Learning, and Career Dynamics},
  author={Afrouzi, Hassan and Blanco, Andres and Drenik, Andr{\'e}s and Hurst, Erik},
  year={2026},
  note={NBER Working Paper 35157}
}

@unpublished{friebel2026pyramids,
  title={Pyramids, Diamonds, and Oscillations: AI and the Structure of Internal Labor Markets},
  author={Friebel, Guido and Huang, Yao and Li, Jin and Shukla, Soumitra and Zhang, Andrew},
  note={Available at SSRN 6570519},
  year={2026}
}

@article{emanuel2026power,
    author = {Emanuel, Natalia and Harrington, Emma and Pallais, Amanda},
    title = {The Power of Proximity to Coworkers},
    journal = {The Quarterly Journal of Economics},
    pages = {Forthcoming},
    year = {2026}
}

@unpublished{althoff2026task,
  title={Task-Specific Technical Change and Comparative Advantage},
  author={Althoff, Lukas and Reichardt, Hugo},
  year={2026},
  note={CESifo Working Paper No. 12403}
}

\newpage

\begin{center} \large \textsc{Online Appendix:} \\ \Large  \textsc{Automation, AI, and the} \\ \textsc{Intergenerational Transmission of Knowledge} \\  \large \textsc{(Not for Publication)} \\ \textsc{Enrique Ide\begingroup
\renewcommand{\thefootnote}{\fnsymbol{footnote}}\footnote[2]{Department of Economics, IESE Business School, Carrer d'Arn\'{u}s i de Gar\'{\i} 3-7, 08034 Barcelona, Spain (eide@iese.edu).}\endgroup $\, $
  --  $\, $\today}  \end{center}

% --- ONLINE APPENDIX ---
% Reset the counter and change to Roman numerals
\setcounter{section}{0}
\renewcommand{\thesection}{\Roman{section}}
\numberwithin{equation}{section}
\renewcommand{\theequation}{\Roman{section}.\arabic{equation}}
 \renewcommand{\baselinestretch}{1.2}
\setcounter{page}{1}
\setcounter{footnote}{0}

\section{Asymptotic Analysis under Alternative Assumptions}

\subsection{No Asymptotic Growth when $N-I \le 1$} \label{sec:Nrep}

This appendix establishes that Assumption \ref{assu:M1} of the main text ($N-I>1$) is necessary for a strictly positive asymptotic growth rate of aggregate output. Throughout, I continue to maintain the full-automation condition in Assumption \ref{assu:full}. Hence, any active expert with skill at least $x_0$ optimally assigns all automatable tasks to machines and hires exactly $N-I$ novices. 

A technical complication arises when $N-I<1$. Because each cohort has unit mass, aggregate demand for novices by positive-skill producers is bounded above by $N-I$. When $N-I<1$, this maximum demand falls strictly below novice supply, leading to structural unemployment under the convention adopted here.

This rationing introduces two issues. First, it requires specifying the skill that unmatched novices acquire. Consistent with the model's premise that expertise requires joint production with a skilled expert, I assume that unmatched novices acquire zero skill. Second, unemployment renders the aggregate market-clearing condition (\ref{eq:mktclearing}) ill-defined, as an entry cutoff $a_t \ge x_t$ solving the condition may not exist. To circumvent this issue without formally amending the market-clearing condition, the subsequent analysis establishes a general upper bound on the intergenerational transmission of skill.

\begin{lem} Suppose $N-I \le 1$. The sequence of aggregate output $\{Y_t\}_{t \ge 0}$ is bounded from above, precluding strictly positive asymptotic growth. \end{lem}

\begin{proof}Let $A_t$ denote the set of experts who are active under the realized feasible allocation at time $t$. For any $q \geq 0$, the measure of active experts with skill strictly greater than $q$ is bounded above by the survival function of the cohort:
\[
\int_{A_t \cap (q,\infty)} dF_t(s) \leq 1 - F_t(q).
\]
By construction, no skills in the interval $(0,x_0)$ can arise: initial expert skills are supported on $[x_0,\infty)$, matched novices inherit their mentors' skills, and unmatched novices acquire zero skill. Therefore, for any $q\ge 0$, every expert in $A_t\cap(q,\infty)$ has skill at least $x_0$. By the maintained full-automation condition, each such expert hires exactly $N-I$ novices.

Moreover, a novice can enter period $t+1$ with skill strictly greater than $q$ only by matching with a period-$t$ expert whose skill is strictly greater than $q$. Hence,
\[
1-F_{t+1}(q)
\leq (N-I)\int_{A_t\cap(q,\infty)} dF_t(s)
\leq (N-I)[1-F_t(q)],  \ \ \text{for} \  q \geq 0.
\]
Iterating this inequality backward to $t=0$ gives:
\[
1-F_t(q) \leq (N-I)^t[1-F_0(q)], \ \ \text{for} \  q \geq 0.
\]

Next, aggregate output at time $t$ is bounded above by the output that would be obtained if all experts were active:
\[
Y_t = m^I h^{N-I} \int_{A_t} q\, dF_t(q) \leq m^I h^{N-I} \int_0^\infty q\, dF_t(q),
\]
Applying integration by parts to the right-hand side yields:
\[
\int_0^\infty q\, dF_t(q) = \int_0^\infty [1-F_t(q)]\, dq,
\]
where the upper boundary term vanishes, $\lim_{q \to \infty} q[1-F_t(q)] = 0$. This limit holds because the initial distribution $F_0$ has a finite mean (implying $\lim_{q \to \infty} q[1-F_0(q)] = 0$), and $F_t$ inherits this tail decay via the iterated bound established above.

Therefore,
\[
Y_t \leq m^I h^{N-I} \int_0^\infty [1-F_t(q)]\, dq
\leq m^I h^{N-I} (N-I)^t \int_0^\infty [1-F_0(q)]\, dq.
\]
By definition, the integral of the initial survival function from $0$ to $\infty$ is the initial finite mean, $\mathbb{E}_0[q]$. Aggregate output is thus bounded above by:$$ Y_t \le m^I h^{N-I} (N-I)^t \mathbb{E}_0[q] $$Two cases emerge: \vspace{-3mm} \begin{enumerate}[noitemsep]
\item If $N-I = 1$, aggregate output is globally bounded from above by the constant $m^I h^{N-I} \mathbb{E}_0[q]$.
\item If $N-I < 1$, the scaling factor $(N-I)^t \to 0$ as $t \to \infty$, implying $\lim_{t \to \infty} Y_t = 0$.
\end{enumerate} In both cases, the sequence $\{Y_t\}_{t \ge 0}$ is bounded from above, precluding strictly positive asymptotic growth. \end{proof}

\subsection{Relaxing Assumption \ref{assu:full}} \label{sec:full}

This appendix shows that Assumption \ref{assu:full} affects only a finite early-transition phase for the equilibrium class analyzed in the main text. I relax the requirement that the initial minimum skill lies above the full-automation threshold, while maintaining $x_0>0$. When considering possible unemployment, unmatched novices receive no wage and do not acquire expertise. I also use the natural extension of the stability condition in Definition \ref{def:within}: no expert can offer a nonnegative wage to novices, including unmatched novices, in a way that both raises her own payoff and makes those novices strictly better off.

The argument proceeds in two steps. First, I show that any equilibrium trajectory satisfying this extended stability condition employs all novices at every date. Second, under the same upper-set selection for active experts used in the main text, I show that the minimum skill level must eventually exceed the full-automation threshold.

\begin{lem} \label{lem:employ} Suppose $x_0>0$. Along any equilibrium trajectory satisfying the extended stability condition described above, all novices are employed at every date. Moreover, the minimum skill level remains strictly positive: $x_t>0$ for all $t\geq 0$. \end{lem}

\begin{proof} I first show that if $x_t>0$, then unemployment cannot occur in period $t$. Suppose, toward a contradiction, that a positive mass of novices is unemployed in period $t$. Since every producing expert chooses some $k_t(q)\in[0,I]$, each producing expert hires at least $N-I>1$ novices. If a positive mass of novices is unemployed, then the mass of employed novices is strictly less than one. Hence, the mass of producing experts must be strictly less than one as well; otherwise, aggregate novice demand would exceed the unit mass of novices. Therefore, a positive mass of experts must be inactive.

Because $x_t>0$, every inactive expert has a skill of at least $x_t$. Choose a positive mass of inactive experts small enough that each can hire $N$ unemployed novices. Let each of these experts choose no automation, $k=0$, hire those novices, and offer them a small positive wage $\varepsilon>0$. Each such expert produces $qh^N$ and pays total wages $N\varepsilon$. Since $q\geq x_t$, her profit from this
deviation is at least:
\[ x_t h^N - N\varepsilon. \]
For $\varepsilon>0$ sufficiently small, this profit is strictly positive. The unemployed novices also strictly prefer the deviation, since they receive a positive wage and acquire the skill of an expert with strictly positive skill. This contradicts the extended stability condition. Therefore, if $x_t>0$, all novices must be employed in period $t$.

The statement of the lemma now follows by induction. Since $x_0>0$, the argument above implies that all novices are employed in period $0$. Thus, every novice acquires the skill of some expert with skill at least $x_0$, so $x_1\geq x_0>0$. Repeating the same argument period by period implies that $x_t>0$ and that all novices are employed for every $t\geq 0$. \end{proof}

The previous lemma implies that novice labor is fully allocated at every date. I next use the same free-disposal selection as in the main text. Since a more skilled expert can always mimic the choices of a less skilled expert, the expert income schedule is non-decreasing in skill. Therefore, up to zero-profit production indifferences, the set of active experts can be represented as an upper set, i.e., $A_t=[a_t,\infty)$. The next lemma uses this upper-set representation to show that the economy reaches the full-automation region in finite time.

\begin{lem} \label{lem:finite-full} Suppose $x_0>0$, and consider any equilibrium trajectory satisfying the extended stability condition described above and the same upper-set selection for active experts as in the main text. Then the minimum skill level $x_t$ exceeds the full-automation threshold in finite time. \end{lem}

\begin{proof} Let: 
\[ \bar q \equiv \frac{\rho I}{h^N\left[(m/h)^I-1\right]} \]
denote the full-automation threshold. By Lemma \ref{lem:employ}, all novices are employed at every date. Let $k_t(q)\in[0,I]$ denote the realized automation choice of an active expert with skill $q$. Since $k_t(q)\leq I$, every active expert hires at least $N-I>1$ novices. Because the total mass of novices is one, and all novices are employed, it follows that:
\[ 1=\int_{A_t}\bigl(N-k_t(q)\bigr)\,dF_t(q) \geq (N-I)\int_{A_t}dF_t(q) \]
Thus,
\[ \int_{A_t}dF_t(q)\leq \frac{1}{N-I}<1 \]
If $a_t=x_t$, then the upper set $A_t=[a_t,\infty)$ would contain the entire support of $F_t$, implying $\int_{A_t}dF_t(q)=1$. This contradicts $\int_{A_t}dF_t(q)\leq 1/(N-I)<1$. Hence, $a_t>x_t$.

All novices employed in period $t$ learn from active experts with skills at least $a_t$. Therefore, the minimum skill in the next period satisfies $x_{t+1}\geq a_t>x_t$. Thus, the sequence $\{x_t\}_{t\geq 0}$ is strictly
increasing. It remains to show that this sequence diverges. For any $q\geq a_t$, the next-period upper tail satisfies: 
\[ 1-F_{t+1}(q) = \int_q^\infty \bigl(N-k_t(s)\bigr)\,dF_t(s) \geq (N-I)\bigl[1-F_t(q)\bigr] \]
Since $x_{t+1}\geq a_t$, applying the same argument at date $t+1$ gives $a_{t+1}>x_{t+1}\geq a_t$. Thus, the cutoffs are strictly increasing. Therefore, if $q\geq a_t$, then $q\geq a_s$ for every $s\leq t$, so the preceding inequality can be iterated backward. In particular, for every $q\geq a_t$,
\[ 1-F_{t+1}(q) \geq (N-I)^{t+1}\bigl[1-F_0(q)\bigr]  \]

Suppose, toward a contradiction, that $\{x_t\}_{t\geq 0}$ is bounded above by some finite $L$. Then $a_t\leq x_{t+1}\leq L$ for all $t$. Evaluating the previous inequality at $q=a_t$ gives:
\[ 1 \geq 1-F_{t+1}(a_t) \geq (N-I)^{t+1}\bigl[1-F_0(a_t)\bigr] \geq (N-I)^{t+1}\bigl[1-F_0(L)\bigr]. \]
Because $F_0$ has unbounded support, $1-F_0(L)>0$. Since $N-I>1$, the right-hand side diverges as $t\to\infty$, a contradiction. Therefore, $x_t\to\infty$.

Since the full-automation threshold is finite, there exists a finite date $T$ such that $x_T>\bar q$. From date $T$ onward, every active expert has a skill above the full-automation threshold. Lemma \ref{lem:cut} then implies that full automation is strictly optimal for every active expert. This proves the result. \end{proof}

Lemma \ref{lem:finite-full} implies that the economy reaches the full-automation region after a finite number of periods. From that date onward, the equilibrium is governed by the full-automation problem studied in the main text, with the skill distribution at that date as the relevant initial distribution.

The finite pre-threshold phase also preserves the upper-tail behavior of the initial distribution. To see this, let $T$ be the first date such that $x_T>\bar q$. Since the proof above implies $a_t\leq x_{t+1}\leq x_T$ for all $t<T$, every expert with skill $q\geq x_T$ is active in every pre-threshold period. Moreover, since $q>\bar q$, Lemma \ref{lem:cut} implies that these experts fully automate in every such period. Therefore, for every $t<T$ and every $q\geq x_T$,
\[ 1-F_{t+1}(q)=(N-I)[1-F_t(q)] \]
Applying this identity successively from $t=0$ to $t=T-1$ gives:
\[ 1-F_T(q)=(N-I)^T[1-F_0(q)], \qquad \text{for all } q\geq x_T \]
Therefore, $F_T$ has the same upper-tail class and the same tail index as $F_0$. Since removing Assumption \ref{assu:full} affects only finitely many initial periods, it does not affect the asymptotic characterization.

\section{Equilibrium Existence and Characterization} \label{sec:Oext}

Please note that Online Appendix \ref{sec:Oext} uses different assumptions across its subsections. Online Appendix \ref{sec:within} is a within-period result: it takes the current skill distribution $F_t$ and the continuation income schedule $\Pi_{t+1}$ as given, assuming that $\Pi_{t+1}$ is continuous and non-decreasing and that $F_t$ has a continuous, strictly positive density on its support $[x_t,\infty) \subseteq [x_0,\infty)$. 

Online Appendices \ref{sec:exist}-\ref{sec:cross} then apply this characterization to the finite-time dynamic equilibrium under the baseline assumptions of Section \ref{sec:baseline}, without imposing Assumption \ref{assu:growth}. Finally, Online Appendix \ref{sec:asympSch} imposes Assumption \ref{assu:growth} and focuses on the sustained-growth case, $1-F_0\in \mathrm{RV}_{-1/\theta}$ with $\theta\in(0,1)$, to characterize the asymptotic normalized income and wage schedules.

\subsection{Within-Period Competitive Matching} \label{sec:within}

This appendix characterizes the within-period competitive matching, taking the current skill distribution $F_t$ and the continuation income $\Pi_{t+1}$ as given. More precisely, I prove the following result:

\begin{lem}[Existence and Characterization of Within-Period Matching] \label{lem:static} Suppose $\Pi_{t+1}(q)$ is non-decreasing and continuous, and that $F_t$ admits a continuous, strictly positive density on its support $[x_t, \infty) \subseteq [x_0,\infty)$. An essentially unique within-period competitive matching exists at time $t$. In this equilibrium: \vspace{-3mm} \begin{enumerate}[leftmargin=*,noitemsep]
\item The aggregate market-clearing condition uniquely determines the marginal active skill $a_t$ (and the active set $A_t = [a_t, \infty)$): \begin{equation}  1=(N-I)\bigl[1-F_t(a_t)\bigr] \label{eq:Oa} \end{equation}
\item The assignment $\Lagr_t$ satisfies: \begin{equation} \label{eq:OL} \Lagr_t(E)=(N-I)\int_E dF_t(q), \ \ \text{for any measurable $E\subseteq A_t$} \end{equation}
\item The equilibrium wage schedule is:\begin{equation*} 
w_t(q) = \left\{ \begin{array}{cl} S(a_t)/(N-I) & \text{if $q \in [x_t,a_t)$} \\
 \max\{0,U_t-\beta\Pi_{t+1}(q)\}  & \text{if $q \in [a_t, \infty)$}  \end{array} \right. 
\end{equation*}where: \begin{equation*} U_t \equiv \frac{S(a_t)}{N-I} + \beta \Pi_{t+1}(a_t) \end{equation*} 
\end{enumerate}
\end{lem}

To streamline the exposition, the term ``equilibrium'' hereafter refers to a ``within-period competitive matching equilibrium.'' The proof proceeds via four claims.

\begin{cl}\label{cl:O1}  In any equilibrium, $a_t$ and $\Lagr_t$ are uniquely determined by (\ref{eq:Oa}) and (\ref{eq:OL}), respectively. Moreover, $a_t > x_t$. \end{cl}

\begin{proof} By Definition \ref{def:within} of the main text, any equilibrium assignment must satisfy (\ref{eq:Oa}) and (\ref{eq:OL}). Because $N-I>1$ and $F_t$ admits a continuous, strictly positive density on $[x_t, \infty)$, equation (\ref{eq:Oa}) yields a unique solution $a_t$. Moreover, because $N-I>1$, this solution satisfies $a_t > x_t$. \end{proof}

\begin{cl} \label{cl:O2}  In any equilibrium, an inactive expert ($q < a_t$) faces a wage $w_t(q) \ge S(q)/(N-I)$, and the marginal active expert $a_t$ pays a strictly positive wage $w_t(a_t) = S(a_t)/(N-I) > 0$.  \end{cl}

\begin{proof} By Definition \ref{def:within}, an inactive expert ($q < a_t$) cannot profitably produce, immediately implying $w_t(q) \ge S(q)/(N-I)$.

Next, suppose by way of contradiction that the marginal active expert earns strictly positive profit, $\Pi_t(a_t) \equiv c > 0$. This yields a wage $w_t(a_t) = (S(a_t) - c)/(N-I)$ and utility $U_t = w_t(a_t) + \beta\Pi_{t+1}(a_t)$ to the matched novices. Consider a deviating inactive expert $q' \in [x_t, a_t)$ offering wage $w' = U_t - \beta\Pi_{t+1}(q') + \epsilon$. For any $\epsilon > 0$, this wage strictly outbids the marginal expert, as $w' + \beta\Pi_{t+1}(q') = U_t + \epsilon > U_t$.

This deviation satisfies the non-negativity constraint on wages. Substituting $U_t$ into the wage offer yields $w' = w_t(a_t) + \beta[\Pi_{t+1}(a_t) - \Pi_{t+1}(q')] + \epsilon$. Because the equilibrium must satisfy the liquidity constraint ($w_t(a_t) \ge 0$), and because $\Pi_{t+1}$ is non-decreasing ($\Pi_{t+1}(a_t) \ge \Pi_{t+1}(q')$), it follows that $w' \ge \epsilon > 0$, ensuring the deviation is feasible.

The deviator's profit under this deviation is $S(q') - (N-I)w'$, which expands to:$$ S(q') - S(a_t) + c - \beta(N-I)[\Pi_{t+1}(a_t) - \Pi_{t+1}(q')] - \epsilon(N-I) $$By the continuity of the surplus function $S(\cdot)$ and the continuation income $\Pi_{t+1}(\cdot)$, the limit of this profit as $q' \uparrow a_t$ evaluates exactly to $c - \epsilon(N-I)$. Because $c > 0$, choosing an $\epsilon > 0$ sufficiently small and a $q'$ sufficiently close to $a_t$ guarantees a strictly positive profit. This constitutes a strictly profitable deviation, violating the stability condition of Definition \ref{def:within}. Thus, $\Pi_t(a_t) = 0$, yielding $w_t(a_t) = S(a_t)/(N-I)$.

Finally, because $a_t > x_t \ge x_0$ and $S(x_0) > 0$ (which follows from Assumption \ref{assu:full}), the strict monotonicity of $S(\cdot)$ guarantees $S(a_t) > 0$. Consequently, $w_t(a_t) > 0$. \end{proof}

As shown in Claim \ref{cl:O2}, equilibrium participation only provides a lower bound for the wages of inactive experts ($w_t(q) \ge S(q)/(N-I)$). Because the exact schedule for $q < a_t$ is indeterminate, setting $w_t(q) = S(a_t)/(N-I)$ for all $q < a_t$ is without loss of generality: these experts remain inactive, meaning they neither hire novices nor pay these wages.

\begin{cl} \label{cl:O3} In any equilibrium, the wage schedule for active experts ($q \ge a_t$) satisfies:$$w_t(q) = \max\{0, U_t - \beta \Pi_{t+1}(q) \}$$where $U_t \equiv S(a_t)/(N-I) + \beta\Pi_{t+1}(a_t)$. \end{cl}

\begin{proof} Let $V_t(q) \equiv w_t(q) + \beta\Pi_{t+1}(q)$ denote the lifetime utility delivered by active expert $q \ge a_t$. By Definition \ref{def:within}, $\Pi_t(q) = S(q) - (N-I)w_t(q)$ for all active experts. Furthermore, Claim \ref{cl:O2} establishes $w_t(a_t) = S(a_t)/(N-I)>0$, yielding $\Pi_t(a_t) = 0$ and $V_t(a_t) = U_t$. 

First, suppose, by way of contradiction, there exists an active expert $\hat{q} \ge a_t$ such that $w_t(\hat{q}) < \max\{0, U_t - \beta\Pi_{t+1}(\hat{q})\}$. Because the liquidity constraint requires $w_t(\hat{q}) \ge 0$, this strict inequality requires the maximum to be strictly positive: $U_t - \beta\Pi_{t+1}(\hat{q}) > 0$. Consequently, $w_t(\hat{q}) < U_t - \beta\Pi_{t+1}(\hat{q})$, which rearranges to $V_t(\hat{q}) < U_t$. Consider the deviation by expert $q' = a_t$ offering wage $w' = w_t(a_t) - \epsilon$. Because $w_t(a_t) > 0$ and $U_t > V_t(\hat{q})$, there exists an $\epsilon > 0$ small enough such that $w' \ge 0$ and $w' + \beta\Pi_{t+1}(q') = U_t - \epsilon > V_t(\hat{q})$. Under this deviation, $q'$ strictly increases profit: $S(q') - (N-I)w' = \Pi_t(q') + \epsilon(N-I) > \Pi_t(q')$. Furthermore, by letting the victim be $q = \hat{q}$, I obtain $w' + \beta\Pi_{t+1}(q') > w_t(q) + \beta\Pi_{t+1}(q)$. This violates the stability condition of Definition \ref{def:within}. Thus, $w_t(q) \ge \max\{0, U_t - \beta\Pi_{t+1}(q)\}$ for all $q \ge a_t$.

Second, suppose, by way of contradiction, there exists an active expert $\bar{q} \ge a_t$ such that $w_t(\bar{q}) > \max\{0, U_t - \beta\Pi_{t+1}(\bar{q})\}$. This implies both $w_t(\bar{q}) > 0$ and $w_t(\bar{q}) + \beta\Pi_{t+1}(\bar{q}) > U_t$, which simplifies to $V_t(\bar{q}) > U_t$. Consider the deviation by expert $q' = \bar{q}$ offering wage $w' = w_t(\bar{q}) - \epsilon$. Because $w_t(\bar{q}) > 0$ and $V_t(\bar{q}) > U_t$, there exists an $\epsilon > 0$ small enough such that $w' \ge 0$ and $w' + \beta\Pi_{t+1}(q') = V_t(\bar{q}) - \epsilon > U_t$. Under this deviation, $q'$ strictly increases profit: $S(q') - (N-I)w' = \Pi_t(q') + \epsilon(N-I) > \Pi_t(q')$. By letting the victim be the marginal expert $q = a_t$, I obtain $w' + \beta\Pi_{t+1}(q') > U_t = w_t(q) + \beta\Pi_{t+1}(q)$. This again violates the stability condition. Thus, $w_t(q) \le \max\{0, U_t - \beta\Pi_{t+1}(q)\}$ for all $q \ge a_t$.

Combining these bounds yields $w_t(q) = \max\{0, U_t - \beta\Pi_{t+1}(q)\}$ for all $q \ge a_t$. \end{proof}

\begin{cl} The essentially unique candidate characterized by Claims \ref{cl:O1}, \ref{cl:O2}, and \ref{cl:O3} satisfies all conditions of Definition \ref{def:within}. This establishes the existence and uniqueness of the equilibrium. \end{cl}

\begin{proof} By construction, the active set $A_t = [a_t, \infty)$ and assignment $\Lagr_t$ satisfy the feasibility conditions (\ref{eq:feas}) and (\ref{eq:mktclearing}) of Definition \ref{def:within}. 

Next, I verify expert participation. For inactive experts ($q < a_t$), producing at the shadow wage $w_t(q) = S(a_t)/(N-I)$ would yield a profit of $\Pi_t(q) = S(q) - (N-I)w_t(q) = S(q) - S(a_t)$. Because $S(\cdot)$ is strictly increasing, this profit is strictly negative for all $q < a_t$, ensuring the inactive choice ($\Pi_t(q) = 0$) is optimal. For active experts ($q \ge a_t$), the candidate wage is $w_t(q) = \max\{0, U_t - \beta\Pi_{t+1}(q)\}$. Because $\Pi_{t+1}$ is non-decreasing by assumption, $\Pi_{t+1}(q) \ge \Pi_{t+1}(a_t)$, which implies $U_t - \beta\Pi_{t+1}(q) \le U_t - \beta\Pi_{t+1}(a_t) = S(a_t)/(N-I)$. Consequently, $w_t(q) \le S(a_t)/(N-I)$. The active profit is therefore bounded below by $\Pi_t(q) = S(q) - (N-I)w_t(q) \ge S(q) - S(a_t) \ge 0$, satisfying the participation condition.

Finally, I verify stability. Let $V_t(q) \equiv w_t(q) + \beta\Pi_{t+1}(q)$ denote the lifetime utility delivered by an active expert $q \ge a_t$. The constructed schedule ensures $V_t(q) = \max\{\beta\Pi_{t+1}(q), U_t\} \ge U_t$. Suppose, by way of contradiction, a deviation exists: a poacher $q'$ offers a wage $w' \ge 0$ to strictly increase profit ($S(q') - (N-I)w' > \Pi_t(q')$) while strictly outbidding an active victim $q \ge a_t$ ($w' + \beta\Pi_{t+1}(q') > V_t(q) \ge U_t$).

This outbidding condition rearranges to $w' > U_t - \beta\Pi_{t+1}(q')$. Combined with the liquidity constraint ($w' \ge 0$), this implies $w' \ge \max\{0, U_t - \beta\Pi_{t+1}(q')\}$. \begin{itemize}[leftmargin=*]
\item If the poacher is active ($q' \ge a_t$), this inequality implies $w' \ge w_t(q')$. The poacher's profit under the deviation is thus $S(q') - (N-I)w' \le S(q') - (N-I)w_t(q') = \Pi_t(q')$, contradicting the requirement that the deviation strictly increases profit.
\item If the poacher is inactive ($q' < a_t$), monotonicity dictates $\Pi_{t+1}(q') \le \Pi_{t+1}(a_t)$. The outbidding condition therefore implies $w' > U_t - \beta\Pi_{t+1}(q') \ge U_t - \beta\Pi_{t+1}(a_t) = S(a_t)/(N-I)$. The poacher's profit becomes $S(q') - (N-I)w' < S(q') - S(a_t)$. Because $q' < a_t$ and $S(\cdot)$ is strictly increasing, this profit is strictly negative. Since the poacher earns $\Pi_t(q') = 0$ in equilibrium, the deviation is strictly unprofitable, generating a contradiction.
\end{itemize}

As verified above, this candidate satisfies all conditions of Definition \ref{def:within}, establishing existence. Furthermore, Claims \ref{cl:O1}, \ref{cl:O2}, and \ref{cl:O3} establish that any equilibrium must feature this exact assignment and wage schedule for active experts. Because this candidate is the sole possible solution, the within-period competitive matching is essentially unique. \end{proof}

\subsection{Proof of Proposition \ref{prop:existence}} \label{sec:exist}

The proof is constructive and exploits the model's decoupled structure: allocations (distributions and cutoffs) are determined independently of prices (utility, wages, and income).

\begin{description}[leftmargin=*, labelindent=0pt, itemindent=0pt, listparindent=\parindent]

    \item[Step 1: Allocations.] As explained in Section \ref{sec:output} of the main text, the sequences of marginal active experts $\{a_t\}_{t \ge 0}$ and skill distributions $\{F_t\}_{t \ge 0}$ are jointly determined by the following dynamical system: \begin{equation*} 1 =(N-I)\bigl[1-F_t(a_t)\bigr] \ \ \text{and} \ \  F_{t+1}(q) = \left\{ \begin{array}{cl} 0 & \text{if $q  \le a_t$} \\ (N-I) \left[ F_t(q)-F_t(a_t)\right]  & \text{if $q>a_t$} \end{array} \right. \end{equation*}Given $F_0$, the unique solution is given by:\begin{equation*} \begin{split}  1 = (N-I)^{t+1}[1-F_0(a_t)], \ & \text{for $t \ge 0$} \\
    1-F_t(q) = (N-I)^t[1-F_0(q)], \ & \text{for $t \ge 1$ and $q \ge x_t = a_{t-1}$}  \end{split}\end{equation*} Since $N-I > 1$ and $F_0$ has unbounded support, the sequence $\{a_t\}_{t \ge 0}$ is strictly increasing and unbounded ($a_t \to \infty$).
    
    \item[Step 2: Prices.] Given $\{a_t\}_{t \ge 0}$, the sequence $\{U_t\}_{t \ge 0}$ is determined directly from the cutoff dynamics. Since $x_{t+1}=a_t$ and $a_{t+1}>x_{t+1}$, the marginal active skill $a_t$ is inactive in period $t+1$, so $\Pi_{t+1}(a_t)=0$. Equation (\ref{eq:u}) of the main text therefore gives $U_t=S(a_t)/(N-I)$. It remains to characterize the schedules $\{\Pi_t,w_t\}_{t\ge 0}$. For every $t$ and $q \in [x_t, \infty)$, the pair $(\Pi_t(q),w_t(q))$ satisfies: \begin{equation} \begin{split} \label{eq:prices} & \Pi_t(q) = \max\{0,S(q)-(N-I)w_t(q)\}  \\
    & w_t(q) = \left\{ \begin{array}{cl} S(a_t)/(N-I) & \text{if $q \in [x_t,a_t)$} \\
 \max\{0,U_t-\beta\Pi_{t+1}(q)\}  & \text{if $q \in [a_t, \infty)$}  \end{array} \right. \end{split} \end{equation}The following lemma characterizes $\Pi_t(q)$ and $w_t(q)$ for a given $t \ge 0$ and $q \in [x_t,\infty)$. Since $t$ and $q$ are arbitrary, this lemma characterizes the schedules $\{\Pi_t,w_t\}_{t\ge 0}$.
 
\begin{lem} \label{lem:schON} Fix $t\ge 0$ and $q\in[x_t,\infty)$. Let: \[n_t(q)\equiv 0, \  \text{for } q\in [x_t,a_t] \quad \text{and} \quad n_t(q)\equiv \min\{j\ge 0: a_{t+j}\ge q\}, \ \text{for } q>a_t \]Thus, $n_t(q)$ is the first number of periods ahead at which skill $q$ is weakly below the entry cutoff. For $q\ge x_t$, define:
\[
\Gamma_t(q)\equiv \sum_{j=0}^{n_t(q)-1}[\beta(N-I)]^j\bigl[S(q)-S(a_{t+j})\bigr],
\]
with the convention that an empty sum is equal to zero. Then the unique values $\Pi_t(q)$ and $w_t(q)$ consistent with (\ref{eq:prices}) are:
\[
\Pi_t(q)=
\begin{cases}
0 &  \text{if $q \in [x_t,a_t)$}\\
\min\{S(q),\Gamma_t(q)\} & \text{if $q \in [a_t,\infty)$}
\end{cases} \qquad \text{and} \qquad w_t(q)=
\begin{cases}
\dfrac{S(a_t)}{N-I} &  \text{if $q \in [x_t,a_t)$}\\[2mm]
\dfrac{S(q)-\Pi_t(q)}{N-I} & \text{if $q \in [a_t,\infty)$}
\end{cases}
\]
\end{lem}

\begin{proof} Fix $t\ge 0$ and $q\in[x_t,\infty)$. If $q \in [x_t,a_t)$, the formula is immediate from \eqref{eq:prices}: the expert is inactive, so:
\[
\Pi_t(q)=0
\qquad\text{and}\qquad
w_t(q)=\frac{S(a_t)}{N-I}.
\]

Consider now $q \in [a_t,\infty)$. The system \eqref{eq:prices} implies that $w_t(q)=(S(q)-\Pi_t(q))/(N-I)$ and that $\Pi_t(q)$ must solve the following recursion:\begin{equation} \label{eq:recurss}
\Pi_t(q)
= S(q)-(N-I)\max\{0,U_t-\beta\Pi_{t+1}(q)\}
= \min\{S(q),\,S(q)-S(a_t)+\beta(N-I)\Pi_{t+1}(q)\},
\end{equation}where the second equality uses $U_t=S(a_t)/(N-I)$.

I now establish that this recursion has a unique solution. Define the partition of $[x_t,\infty)$ by: \[
I_0^t \equiv [x_t,a_t], \quad \text{and} \quad I_k^t \equiv (a_{t+k-1},a_{t+k}], \  \text{for } k\ge 1. \] Because the cutoff sequence $\{a_\tau\}$ is strictly increasing, $I_k^t$ is exactly the set of skills $q$ such that $n_t(q)=k$. I proceed by induction on $k$:\begin{description}[leftmargin=*, labelindent=0pt, itemindent=0pt, listparindent=\parindent]
\item[Base Case ($k=0$):] Consider any $t \ge 0$ and $q \in I_0^t = [x_t, a_t]$. \begin{itemize}[leftmargin=*]
	\item If $q \in [x_t, a_t)$, I proved above that the system uniquely requires $w_t(q) = S(a_t)/(N-I)$ and $\Pi_t(q) = 0$.
	\item If $q = a_t$, equation (\ref{eq:prices}) dictates $w_t(a_t) = \max\{0, U_t - \beta\Pi_{t+1}(a_t)\}$. As shown above, $\Pi_{t+1}(a_t) = 0$, so $w_t(a_t) = U_t = S(a_t)/(N-I)$. By equation (\ref{eq:prices}), $\Pi_t(a_t) = \max\{0, S(a_t) - S(a_t)\} = 0$.
	\end{itemize}
	Therefore, the schedules are uniquely determined for all $q$ such that $n_t(q) = 0$.

\item[Inductive Step ($k \ge 1$):] Assume that for some $k \ge 1$, $\Pi_\tau(q)$ and $w_\tau(q)$ are uniquely determined for all $\tau \ge 0$ and all $q \in I_{k-1}^\tau$. Consider any $t \ge 0$ and $q \in I_k^t=(a_{t+k-1},a_{t+k}]$. Since the cutoff sequence is strictly increasing, the same $q$ belongs to $I_{k-1}^{t+1}$. Hence $n_{t+1}(q)=k-1$, so $\Pi_{t+1}(q)$ is uniquely determined by the induction hypothesis. Because $k\ge 1$, I also have $q>a_t$, so $q$ lies in the active region. The recursion:

\[
\Pi_t(q)=\min\{S(q), S(q)-S(a_t)+\beta(N-I)\Pi_{t+1}(q)\}
\]then uniquely determines $\Pi_t(q)$.
\end{description}

By mathematical induction, $\Pi_t(q)$ and $w_t(q)$ are globally unique for all $q \ge x_t$ and all $t \ge 0$. Since global uniqueness is established, it remains only to verify that the proposed active region candidate, $\hat{\Pi}_t(q) = \min\{S(q), \Gamma_t(q)\}$, identically satisfies the recursion (\ref{eq:recurss}). 

For $q = a_t$, $n_t(a_t) = 0$, making the sum empty, so $\Gamma_t(a_t) = 0$ and $\hat{\Pi}_t(a_t) = \min\{S(a_t), 0\} = 0$. Evaluating the right-hand side (RHS) of the recursion (\ref{eq:recurss}), since $a_t < a_{t+1}$, the expert is inactive at $t+1$, meaning $\hat{\Pi}_{t+1}(a_t) = 0$. The recursion yields $\min\{S(a_t), S(a_t) - S(a_t) + 0\} = 0$, perfectly matching.

For $q>a_t$, $n_t(q)\ge 1$. Moreover, because $q>a_t$, the first date at which $q$ is weakly below the cutoff is one period closer when viewed from $t+1$, so: \begin{equation*} 
\Gamma_t(q) = \sum_{j=0}^{n_t(q)-1}[\beta(N-I)]^j [S(q)-S(a_{t+j})] = S(q)-S(a_t) + \sum_{j=1}^{n_t(q)-1}[\beta(N-I)]^j [S(q)-S(a_{t+j}) ] \end{equation*}

Using $n_t(q)=n_{t+1}(q)+1$, factoring out $\beta(N-I)$ from the second term and re-indexing with $m=j-1$ yields:\begin{equation*} \begin{split}
\Gamma_t(q)&=S(q)-S(a_t)+\beta(N-I)\sum_{m=0}^{n_{t+1}(q)-1}[\beta(N-I)]^m [ S(q)-S(a_{t+1+m})] \\
&= S(q)-S(a_t)+\beta(N-I)\Gamma_{t+1}(q). \end{split} \end{equation*}

Now substitute the candidate $\hat{\Pi}_{t+1}(q)=\min\{S(q),\Gamma_{t+1}(q)\}$ into the RHS of the recursion (\ref{eq:recurss}):
\[ \mathrm{RHS}=\min\{S(q), S(q)-S(a_t)+\beta(N-I)\min\{S(q),\Gamma_{t+1}(q)\}\}. \] I now distinguish the two possible cases for the inner minimum:\begin{description}[leftmargin=*, labelindent=0pt, itemindent=0pt, listparindent=\parindent]
\item[Case A: $\Gamma_{t+1}(q) \le S(q)$.] Then $\mathrm{RHS}=\min\{S(q), S(q)-S(a_t)+\beta(N-I)\Gamma_{t+1}(q)\} =\min\{S(q),\Gamma_t(q)\}$. This matches the candidate exactly.

\item[Case B: $\Gamma_{t+1}(q) > S(q)$.] Then $\mathrm{RHS}=\min\{S(q), S(q) - S(a_t) + \beta(N-I) S(q)\}$. I must prove that this evaluates exactly to $S(q)$, which requires proving that $S(q) - S(a_t) + \beta(N-I) S(q) \ge S(q)$, or equivalently, $\beta(N-I) S(q) \ge S(a_t)$.

Suppose, by way of contradiction, that $\beta(N-I) S(q) < S(a_t)$. Because $q > a_t$, $S(q) > S(a_t) > 0$. Thus $\beta(N-I) S(q) < S(a_t) < S(q)$, meaning $\beta(N-I) < 1$. Because the sequence of equilibrium cutoffs $\{a_\tau\}$ is strictly increasing, for all $j \ge 0$:$$ S(a_{t+1+j}) \ge S(a_{t+1}) > S(a_t) > \beta(N-I) S(q) $$Hence, every term in the sum for $\Gamma_{t+1}(q)$ is strictly bounded above: $S(q) - S(a_{t+1+j}) < S(q) - \beta(N-I) S(q) = S(q)[1-\beta(N-I)]$. Substituting this strict bound into the geometric expression for $\Gamma_{t+1}(q)$ gives:\begin{multline*} \Gamma_{t+1}(q) = \sum_{j=0}^{n_{t+1}(q)-1} [\beta(N-I)]^j [S(q) - S(a_{t+1+j})] < \sum_{j=0}^{n_{t+1}(q)-1} [\beta(N-I)]^j S(q)[1-\beta(N-I)]  \\ = S(q)(1 - [\beta(N-I)]^{n_{t+1}(q)}) \le S(q) \end{multline*}This yields $\Gamma_{t+1}(q)<S(q)$, contradicting the premise of Case B. Therefore, whenever $\Gamma_{t+1}(q)>S(q)$ it must be that $\beta(N-I)S(q)\ge S(a_t)$.

Hence $S(q)-S(a_t)+\beta(N-I)S(q)\ge S(q)$, so the RHS equals $S(q)$. Moreover,
\[
\Gamma_t(q)=S(q)-S(a_t)+\beta(N-I)\Gamma_{t+1}(q) > S(q)-S(a_t)+\beta(N-I)S(q)\ge S(q) \]
and therefore the candidate also equals $\min\{S(q),\Gamma_t(q)\}=S(q)$. \end{description}

This confirms that the closed-form solutions in the statement of the lemma satisfy \eqref{eq:recurss}. Since uniqueness was already established, the proof is complete.
 \end{proof}
    
\item[Step 3: Verification.] By construction, wages are bounded below by zero, satisfying the liquidity constraint. It therefore remains to verify that $\Pi_t(q)$ is continuous and non-decreasing. I first establish continuity. On the inactive region $[x_t,a_t)$, the schedules are given by $\Pi_t(q)=0$ and $w_t(q)=S(a_t)/(N-I)$, so both are continuous there. On the active region $[a_t,\infty$), Lemma \ref{lem:schON} gives:
\[
\Pi_t(q)=\min\{S(q),\Gamma_t(q)\}
\qquad\text{and}\qquad
w_t(q)=\frac{S(q)-\Pi_t(q)}{N-I}.
\]
Fix $k\ge 1$. On the interval $(a_{t+k-1},a_{t+k}]$, the index $n_t(q)$ is constant, so $\Gamma_t(q)$ is continuous there. Moreover, when $q$ crosses the cutoff $a_{t+k}$, the only additional term in the sum defining $\Gamma_t(q)$ is $[\beta(N-I)]^k[S(q)-S(a_{t+k})]$, which is equal to zero at $q=a_{t+k}$. Hence $\Gamma_t(q)$ is continuous at every cutoff. Since $\Gamma_t(a_t)=0$, the active-region formula for $\Pi_t(q)$ pastes continuously with $\Pi_t(q)=0$ on the inactive region. It follows that $\Pi_t(q)$ is continuous on $[x_t,\infty)$. Moreover, since $w_t(q)=S(a_t)/(N-I)$ on $[x_t,a_t)$ and:
\[
w_t(q)=\frac{S(q)-\Pi_t(q)}{N-I}
\]
on $[a_t,\infty)$, with $\Pi_t(a_t)=0$, the wage schedule is also continuous.

I now establish monotonicity. Since $\Pi_t(q)=0$ on $[x_t,a_t)$, it suffices to show that $\Pi_t(q)$ is non-decreasing on $[a_t,\infty)$. On $[a_t,a_{t+1})$, one has $\Pi_{t+1}(q)=0$, so $w_t(q)=U_t$ is constant. For $q\ge a_{t+1}$, Lemma \ref{lem:schON} yields $\Pi_{t+1}(q)=\min{S(q),\Gamma_{t+1}(q)}$. Because $S(q)$ is strictly increasing and, by the same argument used above, $\Gamma_{t+1}(q)$ is strictly increasing on $[a_{t+1},\infty)$, it follows that $\Pi_{t+1}(q)$ is non-decreasing on $[a_{t+1},\infty)$. Hence $w_t(q)=\max\{0,U_t-\beta\Pi_{t+1}(q)\}$ is non-increasing on $[a_t,\infty)$. Since $\Pi_t(q)=S(q)-(N-I)w_t(q)$ for $q\ge a_t$ and $S(q)$ is strictly increasing, $\Pi_t(q)$ is non-decreasing on $[a_t,\infty)$. Together with $\Pi_t(q)=0$ on $[x_t,a_t)$ and $\Pi_t(a_t)=0$, this proves that the income schedule is globally continuous and non-decreasing. This completes the proof of Proposition \ref{prop:existence}. \end{description}

\subsection{Cross-Sectional Properties of the Equilibrium at Any Given Date} \label{sec:cross}

This section characterizes the cross-sectional distribution of wages, income, and novice allocations across all heterogeneous experts at a fixed date $t$.

As established in Section \ref{sec:formal}, wages are given by $w_t(q) = \max\{0, U_t - \beta \Pi_{t+1}(q)\}$. Let $o_t$ denote the oversubscription cutoff, defined as the unique skill level where the non-negativity constraint on wages exactly binds: $$U_t = \beta \Pi_{t+1}(o_t)$$I now show that $o_t$ exists and is unique. Note that the marginal expert at time $t$ becomes inactive at $t+1$ (since $a_{t+1}>a_t$), implying $\Pi_{t+1}(a_t)=0$. Thus, $U_t-\beta\Pi_{t+1}(a_t)=U_t>0$. Moreover, by Online Appendix \ref{sec:exist}, $\Pi_{t+1}(q)$ is continuous and non-decreasing, with $\Pi_{t+1}(q)=0$ for all $q \in [a_t,a_{t+1})$. Hence any solution to $U_t=\beta\Pi_{t+1}(q)$ must satisfy $q>a_{t+1}$. Finally, because $S(q)\to\infty$ as $q\to\infty$, one also has $\Pi_{t+1}(q)\to\infty$. The Intermediate Value Theorem therefore guarantees the existence of a finite cutoff $o_t \in (a_{t+1},\infty)$. Since $\Pi_{t+1}(q)$ is strictly increasing on $[a_{t+1},\infty)$, this cutoff is unique.

This threshold, combined with the entry cutoff $a_t$, partitions the set of experts into three groups:\vspace{-2mm}  \begin{enumerate}
\item \textbf{Inactive $q \in [x_t, a_t)$:} Do not produce, as $S(q) < (N-I)U_t$. Thus, $\Pi_t(q) = 0$.
\item \textbf{Price-Cleared $q \in [a_t, o_t)$:} Active experts whose strictly positive wages adjust to perfectly clear the market, yielding the following wage and income schedules:$$w_t(q) = \frac{S(q)-\Gamma_t(q)}{N-I} = U_t - \beta \Gamma_{t+1}(q) > 0 \qquad \text{and} \qquad \Pi_t(q)=\Gamma_t(q)$$
The equality $w_t(q) =U_t - \beta \Gamma_{t+1}(q)$ uses the fact that $q<o_t$ implies $w_t(q)>0$, so $\Pi_t(q)<S(q)$ and Lemma \ref{lem:schON} gives $\Pi_t(q)=\Gamma_t(q)$; comparing the corresponding recursions then yields $\Pi_{t+1}(q)=\Gamma_{t+1}(q)$.

Novices are indifferent among working for any expert in this interval, as matching with any one of them delivers the baseline lifetime expected utility $U_t$.
\item \textbf{Oversubscribed $[o_t, +\infty$):} Active experts who pay $w_t(q)=0$ and obtain $\Pi_t(q)=S(q)$. Novices matching with them obtain $\beta \Pi_{t+1}(q) \ge U_t$, with equality at $q=o_t$ and strict inequality for $q>o_t$. Thus, positions strictly above the cutoff are strictly preferred. \end{enumerate}

\begin{figure}[t!]
\centering
\begin{subfigure}{.5\textwidth}
  \centering
   \includegraphics[scale=0.45]{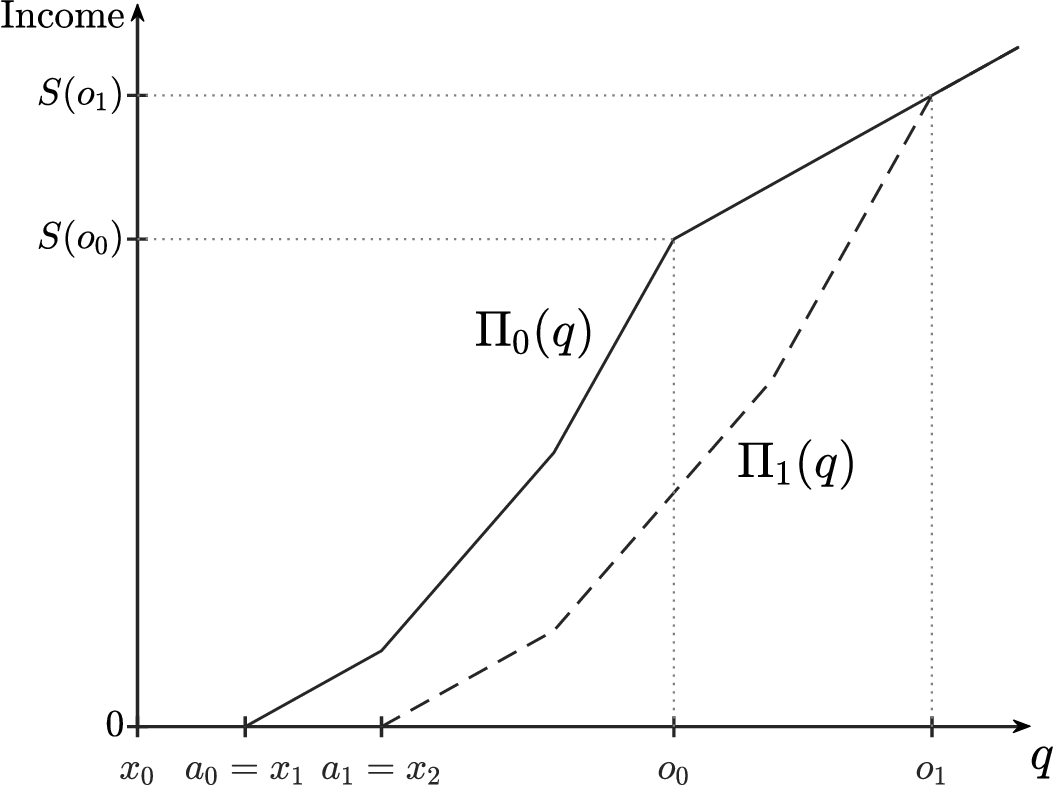}
  \caption{Income Schedule} 
\end{subfigure}%
\begin{subfigure}{.5\textwidth}
  \centering
    \includegraphics[scale=0.45]{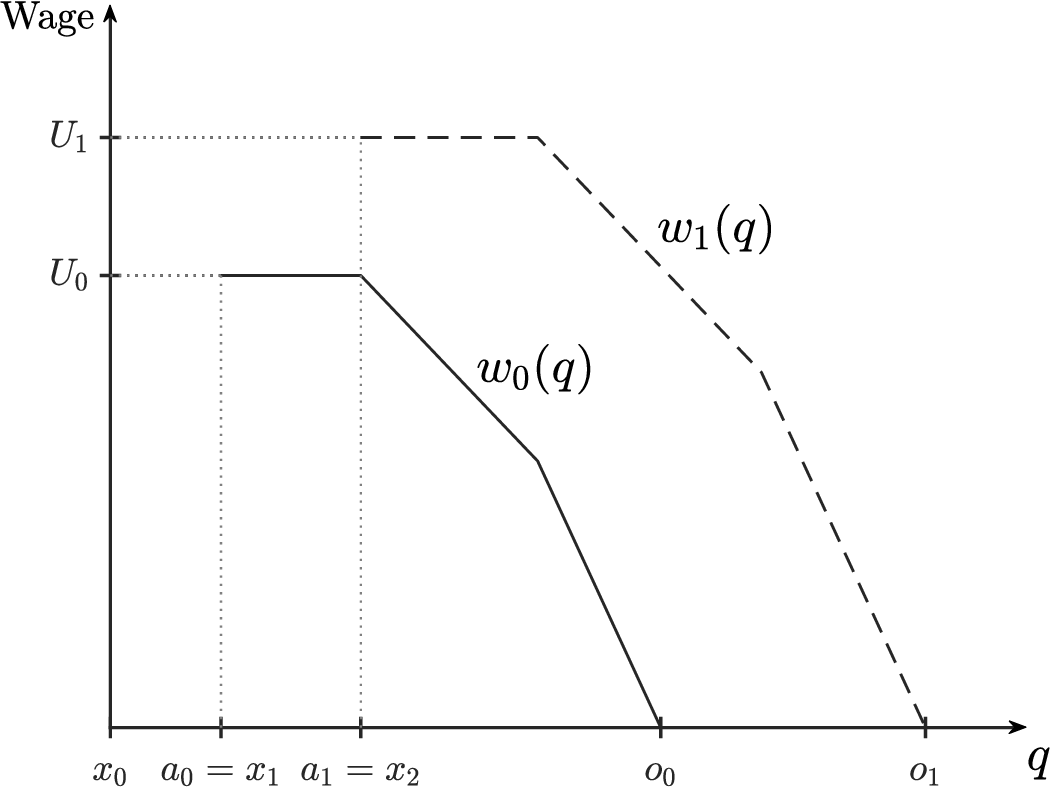}
  \caption{Wage Schedule}
\end{subfigure}
  \captionsetup{justification=centering}
\caption{Cross-Sectional Properties of the Equilibrium at Any Given Date\\  \justifying 
 \vspace{0.5mm}
\footnotesize \noindent \textit{Notes.} The initial skill distribution is Pareto: $F_0(q)=1-(q/x_0)^{-1/\theta}$ with $x_0=3$ and $\theta=0.5$. Remaining parameters: $N=2.7$, $I=1$, $h=1$, $m=2$, $\rho=1$, and $\beta=0.96^{10}$. Panel (a) plots the equilibrium income schedules $\Pi_{t}(q)$, and Panel (b) plots the equilibrium wage schedules $w_{t}(q)$, for periods $t=0$ (solid lines) and $t=1$ (dashed lines).} \label{fig:static_eq}
\end{figure}

By construction, the piecewise schedules $w_t(q)$ and $\Pi_t(q)$ are continuous at both thresholds. At the entry cutoff $a_t$, the marginal skill becomes obsolete, yielding $\Gamma_t(a_t) = 0$. The price-cleared schedules therefore evaluate to $w_t(a_t) = U_t$ and $\Pi_t(a_t) = 0$, matching the inactive region. At the oversubscription cutoff $o_t$, the defining condition $U_t=\beta\Pi_{t+1}(o_t)$ implies:
$$ \Pi_t(o_t)=\min\{S(o_t),S(o_t)-(N-I)U_t+\beta(N-I)\Pi_{t+1}(o_t)\}=S(o_t) $$Hence $w_t(o_t)=0$, so the price-cleared and oversubscribed schedules meet continuously at $q=o_t$.

Finally, these schedules are monotonic in skill. The wage schedule $w_t(q)$ is globally non-increasing: it is constant at $U_t$ on $[x_t,a_{t+1})$, strictly decreasing on $[a_{t+1},o_t)$, and equal to zero on $[o_t,\infty)$. Conversely, $\Pi_t(q)$ is globally non-decreasing: it is zero on $[x_t,a_t)$ and strictly increasing on $[a_t,\infty)$, with $\Pi_t(q)=\Gamma_t(q)$ on $[a_t,o_t)$ and $\Pi_t(q)=S(q)$ on $[o_t,\infty)$.

Figure \ref{fig:static_eq} illustrates the equilibrium income and wage schedules, along with the resulting cross-sectional partition of experts, for two periods: $t = 0$ (solid lines) and $t = 1$ (dashed lines).

\subsection{Existence of the Asymptotic Income and Wage Schedules} \label{sec:asympSch}

This appendix proves that the limits defining the normalized income and wage schedules in Section \ref{sec:sch} of the main text exist when $1-F_0\in \mathrm{RV}_{-1/\theta}$ with $\theta\in(0,1)$. Specifically, I prove the following lemma: 

\begin{lem}[Existence of the Asymptotic Income and Wage Schedules] Suppose Assumption \ref{assu:growth} holds. Then, for every $z \ge 1$, the limits: $$\pi(z) \equiv \lim_{t \to \infty} \frac{\Pi_t(z x_t)}{x_t}, \qquad \omega(z) \equiv \lim_{t \to \infty} \frac{w_t(z x_t)}{x_t}$$exist and are given as in Section \ref{sec:sch} of the main text. \end{lem}

\begin{proof} To prove this lemma, I evaluate the pointwise limits of the normalized finite-time formulas derived in Online Appendix \ref{sec:exist} as $t \to \infty$. The proof proceeds in five steps. 

\begin{description}[leftmargin=*, labelindent=0pt, itemindent=0pt, listparindent=\parindent]

\item[Step 1: Asymptotics of cutoffs, surpluses, and baseline utility.] By Proposition \ref{prop:growth}, under Assumption \ref{assu:growth}, $a_{t+1}/a_t \to 1+g$. Since $x_t=a_{t-1}$, for any fixed integer $j\ge 0$:
\[
\frac{a_{t+j}}{x_t}
=
\frac{a_{t+j}}{a_{t+j-1}}
\frac{a_{t+j-1}}{a_{t+j-2}}
\cdots
\frac{a_t}{a_{t-1}}
\to (1+g)^{j+1}.
\]
Moreover, because $S(q)=qm^Ih^{N-I}-\rho I$, for every fixed $z\ge 1$ and $j\ge 0$:
\[
\frac{S(zx_t)}{x_t}\to z\,m^Ih^{N-I}
\qquad \text{and} \qquad
\frac{S(a_{t+j})}{x_t}\to (1+g)^{j+1}m^Ih^{N-I}.
\]
Finally, since $x_{t+1}=a_t$ and $a_{t+1}>x_{t+1}$, the marginal active skill $a_t$ is inactive in period $t+1$, so $\Pi_{t+1}(a_t)=0$. Equation (\ref{eq:u}) of the main text therefore gives $U_t=S(a_t)/(N-I)$, so:
\[
\frac{U_t}{x_t}\to \frac{m^Ih^{N-I}}{N-I}(1+g)\equiv u.
\]

 \item[Step 2: The inactive region.]  Fix $z\in[1,1+g)$. Since $a_t/x_t\to 1+g$, for all sufficiently large $t$ we have $zx_t<a_t$. Hence the expert with skill $zx_t$ is inactive, so $\Pi_t(zx_t)=0$ and $w_t(zx_t)=S(a_t)/(N-I)$. Therefore:
\[
\frac{\Pi_t(zx_t)}{x_t}\to 0
\qquad \text{and} \qquad
\frac{w_t(zx_t)}{x_t}\to u.
\]
Thus $\pi(z)=0$ and $\omega(z)=u$ for all $z\in[1,1+g)$.
    
 \item[Step 3. The normalized dynasty value.] Fix $z>1+g$. Since $a_t/x_t\to 1+g<z$, the expert with skill $zx_t$ is active for all sufficiently large $t$. Online Appendix \ref{sec:exist} shows that for active experts:
\[
\Pi_t(q)=\min\{S(q),\Gamma_t(q)\},
\qquad
\Gamma_t(q)=\sum_{j=0}^{n_t(q)-1}[\beta(N-I)]^j\bigl[S(q)-S(a_{t+j})\bigr],
\]
where $n_t(q)$ is the remaining active lifespan of skill $q$ at date $t$. 

Let $n(z)\equiv \lfloor \ln z/\ln(1+g)\rfloor$, as in Section \ref{sec:sch}. If $z$ is not an exact integer power of $1+g$, then $(1+g)^{n(z)}<z<(1+g)^{n(z)+1}$. Using the cutoff limit from Step 1, it follows that for all sufficiently large $t$, $a_{t+n(z)-1}<zx_t<a_{t+n(z)}$, so $n_t(zx_t)=n(z)$. Hence:
\[
\frac{\Gamma_t(zx_t)}{x_t}
=
\sum_{j=0}^{n(z)-1}[\beta(N-I)]^j
\frac{S(zx_t)-S(a_{t+j})}{x_t}
\to
m^Ih^{N-I}\sum_{j=0}^{n(z)-1}[\beta(N-I)]^j\bigl[z-(1+g)^{j+1}\bigr].
\]
The term on the right is exactly $m^Ih^{N-I}\gamma(z)$.

Now consider the boundary case $z=(1+g)^k$ for some integer $k\ge 2$. Then, for all sufficiently large $t$, I have $n_t(zx_t)\in\{k-1,k\}$. If $n_t(zx_t)=k$, the extra term relative to the case $n_t(zx_t)=k-1$ is:
\[
[\beta(N-I)]^{k-1}\frac{S(zx_t)-S(a_{t+k-1})}{x_t},
\]
whose limit is:
\[
[\beta(N-I)]^{k-1}m^Ih^{N-I}\bigl[z-(1+g)^k\bigr]=0.
\]
Therefore, the same limit obtains in both cases. Thus, for every $z>1+g$:
\[
\frac{\Gamma_t(zx_t)}{x_t}\to m^Ih^{N-I}\gamma(z).
\]

\item[Step 4. Pointwise convergence of the normalized income schedule.] First, consider the boundary point $z=1+g$. If $(1+g)x_t<a_t$, then the expert is inactive and $\Pi_t((1+g)x_t)=0$. If instead $(1+g)x_t\ge a_t$, then the expert is active and has a remaining active lifespan equal to one, so:
\[
\Gamma_t((1+g)x_t)=S((1+g)x_t)-S(a_t).
\]
By Step 1, $\Gamma_t((1+g)x_t)/x_t\to 0$, and therefore also $\Pi_t((1+g)x_t)/x_t\to 0$. Hence $\pi(1+g)=0$.

Now let $z>1+g$. Since $zx_t$ is eventually active:
\[
\frac{\Pi_t(zx_t)}{x_t}
=
\min\left\{\frac{S(zx_t)}{x_t},\frac{\Gamma_t(zx_t)}{x_t}\right\}
\to
m^Ih^{N-I}\min\{z,\gamma(z)\}.
\]
Appendix \ref{sec:recurs} shows that there exists a unique $\sigma\in(1+g,\infty)$ such that $\gamma(\sigma)=\sigma$, with $\gamma(z)<z$ for $z\in[1+g,\sigma)$ and $\gamma(z)\ge z$ for $z\ge \sigma$. Therefore, the limit above coincides with the piecewise expression for $\pi(z)$ reported in Section \ref{sec:sch}. This proves pointwise convergence of the normalized income for every $z\ge 1$.

\item[Step 5. Pointwise convergence of the normalized wage schedule.] The result for $z\in[1,1+g)$ was established in Step 2. At the boundary $z=1+g$, if $(1+g)x_t<a_t$, then the expert is inactive and $w_t((1+g)x_t)=S(a_t)/(N-I)$, so $w_t((1+g)x_t)/x_t\to u$. If instead $(1+g)x_t\ge a_t$, then the expert is active and:
\[
w_t((1+g)x_t)=\frac{S((1+g)x_t)-\Pi_t((1+g)x_t)}{N-I}.
\]
By Step 1 and the fact that $\Pi_t((1+g)x_t)/x_t\to 0$, this again implies $w_t((1+g)x_t)/x_t\to u$. Hence $\omega(1+g)=u$.

Finally, let $z>1+g$. Since $zx_t$ is eventually active:
\[
\frac{w_t(zx_t)}{x_t}
=
\frac{1}{N-I}\left(\frac{S(zx_t)}{x_t}-\frac{\Pi_t(zx_t)}{x_t}\right)
\to
\frac{z\,m^Ih^{N-I}-\pi(z)}{N-I}.
\]
Substituting the limit for $\pi(z)$ yields exactly the piecewise expression for $\omega(z)$ characterized in Section \ref{sec:sch}. This proves pointwise convergence of the normalized wage for every $z\ge 1$. This completes the proof. 
\end{description} \end{proof}

\section{Plato's Realm of Forms} \label{sec:plato}

The baseline model assumes the initial skill distribution $F_0$ has an unbounded upper tail and that joint production is the sole mechanism for skill acquisition. I relax both assumptions by introducing exogenous expert innovation. This shows that the core results do not require an initial distribution with unbounded support, nor joint production as the sole source of skill growth.

I modify the baseline environment in two ways. First, $F_0$ now has bounded support. Second, prior to matching and production in period $t$, each expert draws $\nu \in \mathbb N_+$ independent ideas from an unbounded latent distribution $\Omega(q)$---representing ``Plato’s realm of Forms'' ({\color{MidnightBlue} Buera and Lucas}, {\color{MidnightBlue}2018}). An expert’s effective skill is the maximum of her inherited skill and these new draws.
 
I assume that $\Omega$ has finite mean, unbounded support, and admits a continuous and strictly positive probability density function. As in the main text, I consider two potential tail classes for the latent distribution: $1-\Omega \in \mathrm{RV}_{-1/\theta}$, with $\theta\in(0,1)$, and $1-\Omega \in \mathrm{RV}_{-\infty}$. Within these classes, sustained positive asymptotic growth obtains in the regularly varying case, whereas asymptotic growth vanishes in the rapidly varying case.

Let $F_t^{*}(q)$ denote the CDF of effective skills at time $t$. By independence, an expert's effective skill is the maximum of her inherited skill and her new innovation draws, yielding $F_t^{*}(q) = F_t(q)\Omega(q)^\nu$. The law of motion for the inherited skill of the $t+1$ cohort for $q \ge a_t$ is then:
$$F_{t+1}(q) = (N-I)[F_t^{*}(q) - F_t^{*}(a_t)]$$and the aggregate market-clearing condition requires $1 = (N-I)[1-F_t^{*}(a_t)]$.

Combining these three conditions yields the following recursion for any $q \ge a_t=x_{t+1}$:$$F_{t+1}^{*}(q)= \big( 1 - (N-I)[1-F_t^{*}(q)] \big)\Omega(q)^\nu.$$Iterating this recursion for points $q$ that remain in the upper tail throughout the iteration---that is, for $q\ge a_s$ for every $s=0,\ldots,t-1$---and using the initial condition $F_0^*(q)=F_0(q)\Omega(q)^\nu$ yields:\footnote{The closed-form expression is displayed for points $q$ in the upper-tail region over which the recursion has been iterated and such that $(N-I)\Omega(q)^\nu\neq 1$. At points where $(N-I)\Omega(q)^\nu=1$, the same recursion remains well defined and has a linear rather than geometric solution. This knife-edge is immaterial for the asymptotic limits: the argument below establishes that the relevant cutoffs diverge, so the analysis eventually concerns arbitrarily large $q$. Since $\Omega(q)\to 1$ as $q\to\infty$ and $N-I>1$, all sufficiently large $q$ satisfy $(N-I)\Omega(q)^\nu>1$.}$$F_t^{*}(q) = (N-I)^t \Omega(q)^{\nu(t+1)} \left(F_0(q) + \frac{N-I-1}{1-(N-I)\Omega(q)^\nu}\right)- \frac{ \Omega(q)^\nu (N-I-1)}{1-(N-I)\Omega(q)^\nu}.$$Because the latent distribution $\Omega$ has unbounded support, $\Omega(q) < 1$ for all finite $q$. By induction, $F_t^{*}(q) < 1$ for all finite $q$. Because $F_t^{*}$ retains an unbounded upper tail and $N-I > 1$, the aggregate market-clearing condition implies that the sequence of equilibrium cutoffs $\{a_t\}_{t \ge 0}$ is strictly increasing. Moreover, as in the baseline model, the cutoffs cannot converge to a finite limit: if they stayed below some finite $q$, repeated replication by active experts would eventually make the effective mass above $q$ exceed the $1/(N-I)$ mass required for market clearing. Hence $a_t\to\infty$. 

The following lemma characterizes the asymptotic behavior of $\{a_t\}_{t \ge 0}$:

\begin{lem} \label{lem:innoAsymCut} 
The sequence of cutoffs $\{a_t\}_{t \ge 0}$ satisfies: 
\[ 
\lim_{t \to \infty} \frac{1-\Omega(a_{t+1})}{1-\Omega(a_t)} = \frac{1}{N-I}.
\] 
\end{lem}

\begin{proof}Because $F_0$ has bounded support, there exists a finite $L$ such that $F_0(q) = 1$ for all $q > L$. Since $a_t \to \infty$, $F_0(a_t) = 1$ for sufficiently large $t$. Evaluating $F_t^{*}(q)$ at $q = a_t$, defining $b_t \equiv \Omega(a_t)^\nu$, and rearranging yields the simplified survival function:
$$1 - F_t^{*}(a_t) = (1 - b_t) \left[ \frac{1 - (N - I)^{t+1} b_t^{t+1}}{1 - b_t(N - I)} \right].$$
The market-clearing condition requires $1 - F_t^{*}(a_t) = \frac{1}{N-I}$. Equating these expressions, multiplying by $(N-I)(1 - b_t(N-I))$, and rearranging yields:
$$1 - b_t = \left(\frac{N-I-1}{N-I}\right) \frac{1}{[(N-I)b_t]^{t+1}}.$$
Thus, the ratio of this expression for consecutive periods $t+1$ and $t$ is:
$$\frac{1 - b_{t+1}}{1 - b_t} = \frac{1}{N-I} \left( \frac{b_t^{t+1}}{b_{t+1}^{t+2}}\right).$$

To evaluate the limit as $t \to \infty$, note that $a_t \to \infty$ implies $b_t \to 1$. Because $N-I > 1$, the term $1-b_t$ converges to zero exponentially, ensuring $b_t^{t+1} \to 1$ and $b_{t+1}^{t+2} \to 1$. Thus:
$$\lim_{t \to \infty} \frac{1 - b_{t+1}}{1 - b_t} = \frac{1}{N-I}.$$
To map this limit back to the latent survival function $1-\Omega(q)$, observe that $1 - b_t = 1 - \Omega(a_t)^\nu$. Since $\nu\in\mathbb{N}_+$, and $a_t \to \infty$ as $t \to \infty$, then:
\[
\lim_{t \to \infty}\frac{1-b_t}{1-\Omega(a_t)} =\lim_{t \to \infty}\frac{1- \Omega(a_t)^\nu}{1-\Omega(a_t)} = \lim_{t \to \infty} \sum_{j=0}^{\nu-1}\Omega(a_t)^j = \nu
\] Because this limit is a strictly positive constant, it follows that:$$\lim_{t \to \infty} \frac{1 - b_{t+1}}{1 - b_t} = \lim_{t \to \infty} \frac{1-\Omega(a_{t+1})}{1-\Omega(a_t)}.$$Equating this to the limit above yields the result. \end{proof}

The following lemma utilizes this fundamental limit to establish the macroeconomic implications of the extension, directly mirroring Propositions \ref{prop:growth} and \ref{prop:dist} from the baseline model.

\begin{lem} In equilibrium,\[
\lim_{t\to\infty}\frac{a_{t+1}}{a_t}
=
\lim_{t\to\infty}\frac{Y_{t+1}}{Y_t} \equiv 1+ g
= \left\{ 
\begin{array}{cl}
(N-I)^{\theta} &  \text{if $1-\Omega \in \mathrm{RV}_{-1/\theta}$} \\
1 & \text{if $1-\Omega \in \mathrm{RV}_{-\infty}$}
\end{array} \right.
\]Furthermore, if $1-\Omega \in \mathrm{RV}_{-1/\theta}$, then the asymptotic CDF of normalized effective skill $z \equiv q/x_t$, $\Phi(z)$, is Pareto with scale $1$ and shape $1/\theta$, i.e., $$ \Phi(z) = 1 - z^{-1/\theta}, \ \ \text{for } z \ge 1.$$ \end{lem}

\begin{proof} The proof proceeds in three steps.

\begin{description}[leftmargin=*, labelindent=0pt, itemindent=0pt, listparindent=\parindent]

    \item[Step 1. Cutoffs.] Lemma \ref{lem:innoAsymCut} establishes that $\lim_{t \to \infty} (1-\Omega(a_{t+1}))/(1-\Omega(a_t)) = 1/(N-I)$. From here, the proof proceeds exactly as that of Proposition \ref{prop:growth} in Appendix \ref{app:growth}, substituting the latent survival function $1-\Omega$ for the initial survival function $1-F_0$.

    \item[Step 2. Asymptotic Distribution when $1-\Omega \in \mathrm{RV}_{-1/\theta}$.] I now characterize the asymptotic distribution of normalized effective skills, $z \equiv q/x_t$ when $1-\Omega \in \mathrm{RV}_{-1/\theta}$ because I use this in the proof on output. 
    
Since $F_0$ has bounded support and the lower bound $x_t=a_{t-1}$ of the cohort-$t$ inherited-skill support diverges, there exists a finite time $T$ such that $F_0(q)=1$ for all $q \ge x_t$ whenever $t > T$. Substituting this into the closed-form solution for $F_t^{*}(q)$ yields the exact upper tail for $q \ge x_t$: 
$$1 - F_t^{*}(q) = \frac{(1-\Omega(q)^\nu) \left[ (N-I)^{t+1}\Omega(q)^{\nu(t+1)} - 1 \right]}{(N-I)\Omega(q)^\nu - 1}$$

By definition, the asymptotic survival function of normalized effective skills, $1-\Phi(z)$, is the limit of the ratio $\frac{1-F_t^{*}(zx_t)}{1-F_t^{*}(x_t)}$ as $t \to \infty$. Substituting the exact upper tail, this ratio decomposes multiplicatively: 
\begin{equation} \label{eq:ratioO}
\frac{1-F_t^{*}(z x_t)}{1-F_t^{*}(x_t)} = \underbrace{ \left( \frac{1-\Omega(z x_t)^\nu}{1-\Omega(x_t)^\nu} \right) }_{\text{Term 1}} \underbrace{ \left( \frac{(N-I)\Omega(x_t)^\nu - 1}{(N-I)\Omega(z x_t)^\nu - 1} \right) }_{\text{Term 2}} \underbrace{ \left( \frac{\Omega(z x_t)^{\nu(t+1)} - (N-I)^{-(t+1)}}{\Omega(x_t)^{\nu(t+1)} - (N-I)^{-(t+1)}} \right) }_{\text{Term 3}} 
\end{equation}

\noindent Evaluating the limit of each term as $t \to \infty$: 
\begin{itemize}[leftmargin=*]
    \item As $x_t \to \infty$, the latent CDF approaches one ($\Omega(x_t) \to 1$ and $\Omega(zx_t) \to 1$). Consequently, Term 2 converges to 1.
    \item The market-clearing condition dictates that $1-\Omega(x_t)^\nu$ decays exponentially at a rate proportional to $(N-I)^{-t}$. This exponential decay strictly dominates the linear growth of $t+1$, yielding $(t+1)(1-\Omega(x_t)^\nu) \to 0$. This ensures the exponents $\Omega(x_t)^{\nu(t+1)}$ and $\Omega(zx_t)^{\nu(t+1)}$ both converge to 1. Since $(N-I)^{-(t+1)} \to 0$, Term 3 converges to 1.
    \item Since $\nu\in\mathbb{N}_+$, $\lim_{q \to \infty} (1-\Omega(q)^\nu)/(1-\Omega(q)) = \nu$. The limit of Term 1 is therefore $\lim_{t \to \infty} (1-\Omega(zx_t))/(1-\Omega(x_t))$. By regular variation ($1-\Omega \in \mathrm{RV}_{-1/\theta}$), this limit exactly equals $z^{-1/\theta}$.
\end{itemize}

\noindent Multiplying these limits yields the Pareto law: $$1 - \Phi(z) = \lim_{t \to \infty} \frac{1-F_t^{*}(zx_t)}{1-F_t^{*}(x_t)} = z^{-1/\theta}, \quad \text{for } z \ge 1$$
    
    \item[Step 3. Aggregate Output.]  Aggregate output at time $t$ is:  $$Y_t = m^I h^{N-I} \int_{a_t}^{\infty} q dF_t^{*}(q)$$ Integrating by parts yields $$\int_{a_t}^{\infty} q dF_t^{*}(q) = a_t[1-F_t^{*}(a_t)] + \int_{a_t}^{\infty} [1-F_t^{*}(q)]dq$$where the upper boundary term vanishes, $\lim_{q\rightarrow\infty} q[1-F_t^{*}(q)] = 0$. This follows because $F_0$ has bounded support, implying that for sufficiently large $q$, the upper tail of $F_t^{*}(q)$ is asymptotically proportional to the latent survival function $1-\Omega(q)$ (as derived in Step 2). The finite mean of the latent distribution $\Omega$ (guaranteed by $1-\Omega \in \mathrm{RV}_{-1/\theta}$ with $\theta \in (0,1)$, or $1-\Omega \in \mathrm{RV}_{-\infty}$) ensures this tail decay.
    
 The aggregate market-clearing condition gives $1-F_t^{*}(a_t) = \frac{1}{N-I}$. Substituting this equality into the expression for output and applying the change of variables $q = za_t$ yields:
$$Y_t = \frac{m^I h^{N-I}}{N-I} a_t \left( 1 + \int_1^{\infty} \frac{1-F_t^{*}(za_t)}{1-F_t^{*}(a_t)} dz \right)$$
 
 Evaluating the limit of the integral as $t \to \infty$ requires analyzing the pointwise limit of the integrand $(1-F_t^*(za_t))/(1-F_t^*(a_t))$. I apply the same decomposition used in (\ref{eq:ratioO}), now with $a_t$ in place of $x_t$. The only term requiring separate verification is the time-dependent exponential term. The cutoff equation in Lemma \ref{lem:innoAsymCut} implies that $(t+1)(1-\Omega(a_t)^\nu)\to 0$. Since $z\geq 1$, the same is true for $(t+1)(1-\Omega(za_t)^\nu)$. Hence $\Omega(a_t)^{\nu(t+1)} \to 1$ and $\Omega(za_t)^{\nu(t+1)} \to 1$. Since $(N-I)^{-(t+1)}\to 0$, the third term in the $a_t$-based decomposition converges to one. The growth factor of output is now characterized first when $1-\Omega\in \mathrm{RV}_{-1/\theta}$, and then when $1-\Omega\in \mathrm{RV}_{-\infty}$.
 
    \begin{description}[leftmargin=*, labelindent=0pt, itemindent=0pt, listparindent=\parindent]
      \item[Case 1: $1-\Omega \in \mathrm{RV}_{-1/\theta}$.]  Applying the $a_t$-based decomposition described above, the first term converges to $z^{-1/\theta}$ by regular variation, while the second and third terms converge to one. Therefore,
\[ \lim_{t\to\infty}\frac{1-F_t^*(za_t)}{1-F_t^*(a_t)} = z^{-1/\theta}. \]Using the $a_{t}$-based decomposition above, note that for all $z \geq 1$ and sufficiently large $t$, the second factor is uniformly bounded by one (because $\Omega$ is non-decreasing and $(N-I)\Omega(a_t)^\nu>1$ for large $t$), while the third factor is bounded by a constant (because its numerator is bounded above by one and its denominator converges to one). Furthermore, because $1-\Omega \in \mathrm{RV}_{-1 / \theta}$ implies $1-\Omega^\nu \in \mathrm{RV}_{-1 / \theta}$, Potter's bounds ({\color{MidnightBlue} Bingham et al.}, {\color{MidnightBlue}1989}, Theorem 1.5.6) guarantee that for any $\epsilon>0$ and sufficiently large $t$, the full integrand $\left(1-F_{t}^{*}\left(z a_{t}\right)\right) /\left(1-F_{t}^{*}\left(a_{t}\right)\right)$ is bounded from above by a function proportional to $z^{-1 / \theta+\epsilon}$. 

As established in Remark \ref{rmk:finite}, a finite mean requires $\theta \in (0,1)$. I can therefore choose $\epsilon$ small enough such that $-1/\theta + \epsilon < -1$, ensuring this upper bound is Lebesgue integrable on $[1, \infty)$. With an integrable dominating function established, the Dominated Convergence Theorem applies, yielding:
    $$\lim_{t \to \infty} \int_1^{\infty} \frac{1-F_t^{*}(za_t)}{1-F_t^{*}(a_t)} dz = \int_1^{\infty} z^{-1/\theta} dz = \frac{1}{1/\theta - 1} = \frac{\theta}{1-\theta}$$Substituting this limit into the expression for aggregate output yields:
    $$\lim_{t \to \infty} \frac{Y_t}{a_t} = \frac{m^I h^{N-I}}{N-I} \left( 1 + \frac{\theta}{1-\theta} \right) = \frac{m^I h^{N-I}}{(N-I)(1-\theta)}$$Because aggregate output asymptotically scales linearly with the entry cutoff, its growth factor matches that of the cutoffs derived in Step 1:
    $$\lim_{t \to \infty} \frac{Y_{t+1}}{Y_t} = \lim_{t \to \infty} \left( \frac{Y_{t+1}/a_{t+1}}{Y_t/a_t} \right) \left( \frac{a_{t+1}}{a_t} \right) = (N-I)^\theta$$
      
      \item[Case 2: $1-\Omega \in \mathrm{RV}_{-\infty}$.] The same $a_t$-based decomposition applies. The second and third terms converge to one by the argument above, while the first term converges to zero for every $z>1$ because $1-\Omega\in \mathrm{RV}_{-\infty}$. Hence,
\[
\lim_{t\to\infty}\frac{1-F_t^*(za_t)}{1-F_t^*(a_t)} = 0, \ \  z>1. \]Applying the Dominated Convergence Theorem requires an integrable dominating function. As in the regularly varying case, the second factor in the decomposition is bounded by one for all sufficiently large $t$, and the third factor is uniformly bounded because its denominator converges to one. Moreover, since $1-\Omega\in \mathrm{RV}_{-\infty}$ implies that $1-\Omega^\nu \in \mathrm{RV}_{-\infty}$, the definition of rapid variation implies that, for any $\alpha>1$, the first factor is bounded by a constant multiple of $z^{-\alpha}$, uniformly for large $t$, away from an arbitrarily small neighborhood of $z=1$. In that neighborhood, the full survival-ratio integrand is bounded by one. These bounds provide the required integrable domination, and the Dominated Convergence Theorem implies that the integral converges to $0$, yielding:$$\lim_{t \to \infty} \frac{Y_t}{a_t} = \frac{m^I h^{N-I}}{N-I}$$Consequently, the asymptotic growth factor of aggregate output is 1:$$\lim_{t \to \infty} \frac{Y_{t+1}}{Y_t} = \lim_{t \to \infty} \left( \frac{Y_{t+1}/a_{t+1}}{Y_t/a_t} \right) \left( \frac{a_{t+1}}{a_t} \right) = 1$$
     \end{description}
      \end{description}
\end{proof}

\section{Innovation-Driven Growth} \label{sec:nu}

In the ``Plato's Realm of Forms'' specification of Online Appendix \ref{sec:plato}, the fixed latent knowledge pool $\Omega$ plays an essential role: its upper tail determines the asymptotic shape of the effective skill distribution and, when fat-tailed, allows positive asymptotic growth despite bounded initial skills. However, because $\Omega$ itself does not evolve over time, new draws from this fixed pool do not generate an additional source of asymptotic growth. As the active skill threshold diverges, the marginal contribution of these fresh draws becomes asymptotically negligible, and long-run growth is governed by the diffusion of skills whose tail behavior is inherited from the fixed upper tail of $\Omega$.

For innovation to contribute independently to asymptotic growth, the latent pool of ideas must evolve in tandem with experts' skills, allowing successive cohorts to draw from an improving frontier by ``standing on the shoulders of giants.'' One way to operationalize this is to assume that, while the initial pool $\Omega_0(q)$ is exogenous (and $1-\Omega_0 \in \mathrm{RV}_{-1/\theta}$), the pool available to subsequent cohorts, $\Omega_t$, matches the distribution of effective practices used by the active experts of the preceding generation. This appendix formalizes this extension.

Unlike in the baseline model, however, this specification makes the law of motion for the skill distribution nonlinear in the distribution itself. For that reason, I do not attempt here to establish the existence of an asymptotic balanced growth path, nor convergence to it from arbitrary initial conditions. Instead, I characterize the properties of any such path, if it exists.

\subsection{Asymptotic Balanced Growth Path} \label{sec:BGP}

Let $F_t(q)$ denote the distribution of inherited skill of cohort $t$, and let $\Omega_t(q)$ denote the latent distribution of ideas available to that cohort. An expert’s effective skill is the maximum of her inherited skill and $\nu\in\mathbb N_+$ independent draws from the latent pool, yielding the effective skill distribution $F_t^{*}(q)=F_t(q)\Omega_t(q)^\nu$. 

Let $a_t$ denote the minimum effective skill among active experts. Market clearing requires $1=(N-I)[1-F_t^{*}(a_t)]$. As mentioned above, although the initial pool of latent ideas $\Omega_0$ is exogenous, the latent pool available to cohort $t+1$ coincides with the cross-sectional distribution of effective skills among active experts in period $t \ge 0$:$$\Omega_{t+1}(q)=\frac{F_t^{*}(q)-F_t^{*}(a_t)}{1-F_t^{*}(a_t)},\ \ q\ge a_t.$$Substituting the market-clearing condition into the denominator yields: $$\Omega_{t+1}(q)=(N-I)[F_t^{*}(q)-F_t^{*}(a_t)]=F_{t+1}(q) \  \text{for $q\ge a_t$}$$ That is, $\Omega_{t+1}(q)$ coincides with the inherited skill distribution of the next cohort. Relabeling time indices yields $\Omega_t(q)=F_t(q)$ for all $t \ge 1$.

The effective skill distribution therefore simplifies to $F_t^{*}(q)=F_t(q)^{1+\nu}$. Let $x_{t+1}\equiv a_t$ denote the lower bound of the support of $F_{t+1}$. Substituting the simplified effective skill distribution into the equation for $F_{t+1}(q)$ provides the law of motion:$$F_{t+1}(q)=1-(N-I)[1-F_t(q)^{1+\nu}],\ \ q\ge a_t=x_{t+1}.$$For each cohort $t$, define the normalized inherited-skill distribution $\Phi_t(z)\equiv F_t(x_t z)$ for $z\ge 1$.

\begin{mydef}[Asymptotic BGP] An asymptotic balanced growth path consists of a scalar $1+g>1$ and a nondegenerate distribution function $\Phi$ on $[1,\infty)$ such that: \begin{enumerate}[noitemsep]
\item The lower bound of the support of the normalized inherited-skill distribution is $1$: $\Phi(1)=0$
\item The lower bound of the support of the unnormalized inherited-skill distribution, $x_t$, satisfies:$$\lim_{t\to\infty}\frac{x_{t+1}}{x_t}=1+g>1.$$
\item The limiting normalized distribution $\Phi$ is continuous on $[1,\infty)$, and the normalized inherited-skill distribution converges pointwise to $\Phi$:
$$ \Phi_t(z)\to \Phi(z), \qquad \text{for all } z \ge 1 $$
\item The first moment of the normalized inherited-skill distribution converges to a finite positive constant: \[
\lim_{t\to\infty}\int_1^\infty z \,d\Phi_t(z) = \int_1^\infty z\,d\Phi(z) \in (0,\infty).
\]
\end{enumerate} \end{mydef}

The subsequent analysis characterizes the properties of any asymptotic balanced growth path satisfying this definition. I do not show here that such a path exists, nor that the economy converges to it from arbitrary initial conditions.

The following lemma derives the functional equation satisfied by any asymptotic balanced growth path.

\begin{lem} \label{lem:recOV2} Suppose the economy admits an asymptotic balanced growth path with growth factor $1+g>1$ and limiting normalized distribution $\Phi$. Then, for every $z \ge 1$,$$1-\Phi(z)=(N-I)[1-\Phi((1+g)z)^{1+\nu}]$$Moreover,$$\Phi(1+g)=\left(1-\frac{1}{N-I}\right)^{1/(1+\nu)}.$$ \end{lem} 

\begin{proof} By definition, the normalized distribution is $\Phi_{t+1}(z)=F_{t+1}(x_{t+1}z)$. Applying the law of motion for the inherited skill distribution yields $\Phi_{t+1}(z)=1-(N-I)[1-F_t(x_{t+1}z)^{1+\nu}]$. Because $x_{t+1}=(x_{t+1}/x_t)x_t$, this expression can be rewritten in terms of the period-$t$ normalized distribution as: \[ \Phi_{t+1}(z)=1-(N-I)\left[1-\Phi_t\left(\frac{x_{t+1}}{x_t}z\right)^{1+\nu}\right] \]Taking the limit as $t\to\infty$, and using $\Phi_t\to\Phi$ together with $x_{t+1}/x_t\to 1+g$, yields $\Phi(z)=1-(N-I)[1-\Phi((1+g)z)^{1+\nu}]$. Rearranging gives:$$1-\Phi(z)=(N-I)[1-\Phi((1+g)z)^{1+\nu}].$$Since $\Phi(1)=0$, evaluating at $z=1$ yields $1=(N-I)[1-\Phi(1+g)^{1+\nu}]$, which implies:$$\Phi(1+g)=\left(1-\frac{1}{N-I}\right)^{1/(1+\nu)}.$$  \end{proof}

To determine the growth factor $1+g$, I impose the same tail condition used in the baseline analysis. Specifically, the limiting normalized distribution $\Phi$ has a regularly varying upper tail with index $-1/\theta$, where $\theta\in(0,1)$:$$\lim_{z\to\infty}\frac{1-\Phi(\lambda z)}{1-\Phi(z)}=\lambda^{-1/\theta}\qquad\text{for every }\lambda>0.$$

The existence of a constant balanced-growth factor does not, by itself, imply this assumption. Because the stationary recursion restricts the distribution's tail solely across the discrete dilation $z\mapsto (1+g)z$, it admits log-periodic deviations from a pure power law. Imposing regular variation explicitly precludes such oscillations.

The next lemma establishes that this tail restriction uniquely determines the balanced-growth factor.

\begin{lem} \label{lem:recOV3} Suppose the economy admits an asymptotic balanced growth path satisfying $1-\Phi \in \mathrm{RV}_{-1/\theta}$. Then: $$1+g=[(N-I)(1+\nu)]^\theta$$ \end{lem} 

\begin{proof} The stationary recursion from Lemma \ref{lem:recOV2} dictates $1-\Phi(z)=(N-I)[1-\Phi((1+g)z)^{1+\nu}]$. Dividing by the survival function $1-\Phi(z)$ and decomposing the right-hand side multiplicatively yields:$$1=(N-I)\left[\frac{1-\Phi((1+g)z)^{1+\nu}}{1-\Phi((1+g)z)}\right]\left[\frac{1-\Phi((1+g)z)}{1-\Phi(z)}\right].$$Applying the limit $z\to\infty$ on both sides: \begin{equation} \label{eq:Lrec} 1=(N-I)\left[ \lim_{z\to\infty}\frac{1-\Phi((1+g)z)^{1+\nu}}{1-\Phi((1+g)z)}\right]\left[ \lim_{z \to \infty}\frac{1-\Phi((1+g)z)}{1-\Phi(z)}\right].\end{equation}Because $\Phi((1+g)z)\to 1$ as $z\to\infty$, the first bracketed term converges to $1+\nu$. The assumption of regular variation ensures the second term converges to $(1+g)^{-1/\theta}$. Therefore, (\ref{eq:Lrec}) can be written as $1=(N-I)(1+\nu)(1+g)^{-1/\theta}$. Rearranging yields: $$1+g=[(N-I)(1+\nu)]^\theta.$$ \end{proof}

Finally, I derive the growth rate of aggregate output along the asymptotic balanced growth path:

\begin{lem} Suppose the economy admits an asymptotic balanced growth path satisfying $1-\Phi \in \mathrm{RV}_{-1/\theta}$. Then, the asymptotic growth factor of aggregate output satisfies:$$\lim_{t\to\infty}\frac{Y_{t+1}}{Y_t}=[(N-I)(1+\nu)]^\theta.$$\end{lem}

\begin{proof} By definition, aggregate output in period $t$ is $Y_t=m^I h^{N-I}\int_{a_t}^{\infty} q \,dF_t^{*}(q)$. Differentiating the law of motion for inherited skill yields the measure $dF_t^{*}(q)=\frac{1}{N-I} \,dF_{t+1}(q)$ on the support $[a_t,\infty)$. Substituting this differential yields:
$$Y_t=\frac{m^I h^{N-I}}{N-I}\int_{a_t}^{\infty} q dF_{t+1}(q).$$
Since $a_t=x_{t+1}$, this can be written as:
$$Y_t=\frac{m^I h^{N-I}}{N-I}\int_{x_{t+1}}^{\infty} q dF_{t+1}(q)=\frac{m^I h^{N-I}}{N-I}x_{t+1}\int_1^{\infty} z d\Phi_{t+1}(z).$$By the definition of an asymptotic balanced growth path, the normalized first moment converges to a finite positive constant. Therefore, aggregate output is asymptotically proportional to $x_{t+1}$, and:
$$\lim_{t\to\infty}\frac{Y_{t+1}}{Y_t}=\lim_{t\to\infty}\frac{x_{t+2}}{x_{t+1}}=1+g.$$
Applying Lemma \ref{lem:recOV3} yields:
$$\lim_{t\to\infty}\frac{Y_{t+1}}{Y_t}=[(N-I)(1+\nu)]^\theta.$$ \end{proof}

The characterization also clarifies how technologies that affect innovation capacity can change long-run growth. Writing $\ell$ for the span of control, the growth factor along any asymptotic balanced growth path characterized above can be expressed as:
\[
1+g=[\ell(1+\nu)]^\theta.
\]
Consider a technological shock that changes the span of control from $\ell$ to $\widetilde{\ell}$ and the innovation parameter from $\nu$ to $\widetilde{\nu}$. If the shocked economy admits an asymptotic balanced growth path satisfying the same regularity conditions and tail parameter $\theta$, its growth factor is:
\[
1+\widetilde g=[\widetilde{\ell}(1+\widetilde{\nu})]^\theta.
\]
Therefore, the shock raises long-run growth if and only if:
\[
\widetilde{\ell}(1+\widetilde{\nu})>\ell(1+\nu).
\]
This condition illustrates how an innovation-augmenting technology can offset a reduction in the span of control. In particular, an improvement in entry-level automation may reduce the span of control, $\widetilde{\ell}<\ell$, and still raise long-run growth if it increases innovation capacity sufficiently.\footnote{This contrasts with Online Appendix \ref{sec:plato}. There, the latent distribution $\Omega$ is fixed over time. Consequently, increasing the number of innovation draws from $\nu$ to $\widetilde{\nu}$ can affect transition dynamics, but not the asymptotic growth factor, which remains $1+g=(N-I)^\theta$. In the present appendix, by contrast, the latent pool evolves with the economy and innovation contributes to asymptotic growth, so changes in $\nu$ can affect the long-run growth factor.} This observation complements the main text, which isolates the diffusion channel: how technological change reshapes the transmission of tacit knowledge through workplace assignment.

\section{Heterogeneity in Task Learning Value} \label{sec:hetero}

This appendix relaxes the baseline assumption that all routine tasks provide uniform access to the expert’s tacit knowledge. I allow novice-performed tasks to differ in their learning value: some tasks remain productive inputs but generate no meaningful interaction with the expert and therefore do not transmit expertise. Exploiting the block-recursive structure of the equilibrium, I characterize the implications of this heterogeneity for aggregate output and long-run growth, without solving for the associated wage and income schedules.\footnote{A full characterization of these schedules would require additional assumptions about task assignment and contracting. In particular, one would need to specify whether novices know before matching which tasks they will perform, whether task assignments are contractible, and whether experts can credibly commit to assigning novices to learning-relevant rather than non-learning tasks. These assumptions affect the price block of the equilibrium but are not needed for the allocation and growth results derived here.} The analysis turns on the distinction between the total span of control, which determines novice-market clearing, and the learning-relevant span of control, which determines the intergenerational transmission of expertise.

\subsection{Setup}

Fix a technology $(I,m,h)$, and let $\Man \equiv \Man(I,m,h)$ denote the supervision time required for one project. As in Section \ref{sec:scale} of the main text, an active expert operates $1/\Man$ projects. Assuming full adoption of automation technologies, tasks in $[0,I]$ are performed by machines, while tasks in $[I,N]$ are performed by novices.

I now allow novice-performed tasks to differ in their learning value. Specifically, tasks in $[I,I+\zeta]$, with $0\leq \zeta < N-I$, are non-learning tasks: they are productive inputs, but they do not provide meaningful access to the expert's tacit knowledge. Hence, novices assigned to them acquire zero skill. Tasks in $[I+\zeta,N]$, by contrast, transmit expertise as in the baseline model. Thus, each project requires $N-I$ novice tasks, but only $N-I-\zeta$ of them are learning-relevant. This ordering of tasks captures the idea that low-learning routine tasks may be automated first.

Since each active expert operates $1/\Man$ projects, two distinct spans of control arise. The production span of control---which measures total novice demand per active expert---is:
\[
\ell_{\mathrm P}\equiv \frac{N-I}{\Man}
\]
while the learning span of control---which measures the mass of novices to whom an active expert transmits her skill---is:
\[
\ell_{\mathrm L}\equiv \frac{N-I-\zeta}{\Man},
\]
The baseline model corresponds to the special case $\zeta=0$ and $\Man=1$, in which $\ell_{\mathrm P}=\ell_{\mathrm L}=N-I$. 

Throughout this appendix, I assume that $\ell_{\mathrm L}>1$, so each active expert transmits her skill to more than one novice. This condition is analogous to Assumption \ref{assu:M1} of the main text, and it is necessary for positive asymptotic growth. I also maintain the baseline regularity assumptions on the initial skill distribution: $F_0$ has unbounded support, finite mean, and a continuous, strictly positive density on $[x_0,\infty)$. To focus on the sustained-growth case, I further assume that $1-F_0\in \mathrm{RV}_{-1/\theta}$, with $\theta\in(0,1)$.

Let $a_t$ denote the marginal active expert at time $t$. Since all novice positions absorb novice labor, aggregate market clearing is governed by the production span of control:
\begin{equation} \label{eq:mktCH}
1=\ell_{\mathrm P}[1-F_t(a_t)].
\end{equation}
By contrast, the law of motion for the learning-generated upper tail is governed by the learning span of control. For $q\geq a_t$,
\begin{equation} \label{eq:distH}
1-F_{t+1}(q)=\ell_{\mathrm L}[1-F_t(q)].
\end{equation}
Equivalently, for $q\geq a_t$,
\[
F_{t+1}(q)
=
\frac{\zeta}{N-I}
+
\ell_{\mathrm L}[F_t(q)-F_t(a_t)].
\]
The first term is the mass of novices assigned to non-learning tasks. It is independent of $\Man$, because scale effects multiply both learning and non-learning positions. The second term is the learning-generated upper tail: it is obtained by truncating the current skill distribution at the active cutoff and reweighting the surviving tail by the learning span of control. Thus, task-learning heterogeneity separates the span of control that clears the novice labor market from the span of control that replicates expertise. 

Finally, since each active expert operates $1/\Man$ projects, aggregate output at time $t$ is:
\begin{equation} \label{eq:outputH} Y_t
\equiv
\frac{m^Ih^{N-I}}{\Man} \int_{a_t}^{\infty} q\,dF_t(q).
\end{equation}

The next subsection uses these three objects---the market-clearing condition, the law of motion for the learning-generated upper tail, and the aggregate-output equation---to characterize long-run growth and the asymptotic distribution of skills.

\subsection{Long-Run Growth and the Asymptotic Distribution of Skills}

I first solve the allocation block defined by the market-clearing condition (\ref{eq:mktCH}) and the law of motion for the learning-generated upper tail (\ref{eq:distH}). Given $F_0$, the solution is: \begin{equation*} \begin{split}   1 =\ell_{\mathrm P}\ell_{\mathrm L}^t[1-F_0(a_t)], \ & \text{for $t \ge 0$} \\
    1-F_t(q)=\ell_{\mathrm L}^t[1-F_0(q)], \ & \text{for $t \ge 1$ and $q \ge a_{t-1}$}  \end{split}\end{equation*}The first equation determines the active cutoff at date $t$. The second characterizes the learning-generated upper tail of the skill distribution. Unlike in the baseline model, this is not the full distribution: a fixed mass of novices is assigned to non-learning tasks and remains outside the upper tail. Since $\ell_{\mathrm L}>1$ and $F_0$ has unbounded support, the sequence of cutoffs $\{a_t\}_{t\geq 0}$ is strictly increasing and diverges to infinity. 
    
    Substituting the upper-tail solution into the aggregate-output equation  (\ref{eq:outputH}) gives:
\[
Y_t = \frac{m^Ih^{N-I}}{\Man} \ell_{\mathrm L}^t[1-F_0(a_t)]  \mathbb{E}_0[q \mid q \ge a_t].
\]
Using the cutoff equation, this becomes: \[  Y_t = \frac{m^I h^{N-I}}{N-I} \, \mathbb{E}_0[q \mid q \ge a_t]  \]
    
I now characterize the asymptotic growth rate. From the cutoff equation,
\[
\frac{1-F_0(a_{t+1})}{1-F_0(a_t)}
=
\frac{1}{\ell_{\mathrm L}}.
\]Since $1-F_0\in \mathrm{RV}_{-1/\theta}$, the same argument as in Section \ref{sec:output}, with $\ell_{\mathrm L}$ replacing $N-I$, yields: \[
\lim_{t\to\infty}\frac{a_{t+1}}{a_t}
=
\lim_{t\to\infty}\frac{Y_{t+1}}{Y_t} \equiv 1+ g = \ell_{\mathrm L}^\theta = \left(\frac{N-I-\zeta}{\Man}\right)^\theta. \]Thus, long-run growth is governed by the learning span of control, not by the production span of control. Non-learning tasks matter for production and market clearing, but only learning-relevant tasks replicate expertise across generations.

I now turn to the asymptotic distribution of skills. For $t\geq 1$, the support of $F_t$ consists of two components. Novices assigned to non-learning tasks acquire no skill, resulting in a lower component at zero. Novices assigned to learning-relevant tasks inherit the skill of their mentors, generating an upper component with support $[a_{t-1},\infty)$. Thus,
\[
\operatorname{supp}(F_t)=\{0\}\cup [a_{t-1},\infty).
\]
I denote by $x_t$ the lower bound of the learning-generated upper tail, so $x_t\equiv a_{t-1}$. Unlike in the baseline model, $x_t$ is not the lower bound of the full support of $F_t$; it is the minimum positive skill in cohort $t$.

Evaluating the upper-tail solution at $q=x_t=a_{t-1}$ gives
\[
1-F_t(x_t)
=
\ell_{\mathrm L}^t[1-F_0(a_{t-1})].
\]
Using the cutoff equation at date $t-1$, this becomes
\[
1-F_t(x_t)
=
\frac{\ell_{\mathrm L}}{\ell_{\mathrm P}}
=
\frac{N-I-\zeta}{N-I}.
\]
Thus, the learning-generated upper tail has mass $(N-I-\zeta)/(N-I)$, while the remaining mass $\zeta/(N-I)$ belongs to the lower non-learning component. Because $\ell_{\mathrm L}>1$, this upper-tail mass exceeds $1/\ell_{\mathrm P}$, the mass of active experts required by market clearing. Hence, the active cutoff remains inside the learning-generated upper tail.

The next step is to characterize the shape of the learning-generated component. Since the full distribution contains the lower component, the relevant analog of Proposition \ref{prop:dist} of the main text is not the unconditional distribution of $q/x_t$, but the conditional distribution of normalized skills among experts in the learning-generated upper tail. For $z\geq 1$,
\[
 \mathbb{P}_t\left(
\frac{q}{x_t}>z
\,\middle|\,
q\geq x_t
\right)
=
\frac{1-F_t(zx_t)}{1-F_t(x_t)}.
\]
Using the upper-tail solution, the learning-span terms cancel:
\[
\frac{1-F_t(zx_t)}{1-F_t(x_t)}
=
\frac{1-F_0(zx_t)}{1-F_0(x_t)}.
\]
Since $x_t\to\infty$ and $1-F_0\in \mathrm{RV}_{-1/\theta}$, it follows that:
\[
 \mathbb{P}_t\left(
\frac{q}{x_t}>z
\,\middle|\,
q\geq x_t
\right)
\to
z^{-1/\theta}.
\]
Thus, conditional on belonging to the learning-generated upper tail, normalized skills converge to a Pareto distribution with scale 1 and shape $1/\theta$:
\[
\Phi^{\mathrm L}(z)=1-z^{-1/\theta},
\ \text{for} \  z\geq 1.
\]

The full normalized distribution, $\Phi^{\mathrm{full}}(z) = \lim_{t \to \infty} F_t(z x_t)$, is obtained by combining this conditional limit with the mass decomposition above. The lower non-learning component has mass $\zeta/(N-I)$. Because these novices acquire zero skill, this component remains at zero under normalization by $x_t$. The learning-generated upper tail has mass $(N-I-\zeta)/(N-I)$, and conditional on belonging to this tail, its normalized survival function converges to $z^{-1/\theta}$. Hence, for $z\geq 1$, its contribution to the unconditional survival function is $[(N-I-\zeta)/(N-I)]z^{-1/\theta}$. Therefore,
\[
\Phi^{\mathrm{full}}(z) 
=
\begin{cases}
\qquad \qquad 0, & \text{if $z<0$},\\[4pt]
\qquad \ \ \ \ \dfrac{\zeta}{N-I}, & \text{if $0\leq z<1$} ,\\[8pt]
1-\left(\dfrac{N-I-\zeta}{N-I}\right)z^{-1/\theta}, & \text{if $z\geq 1$}.
\end{cases}
\]
The analog of Proposition \ref{prop:dist} is therefore not Pareto convergence of the full skill distribution, but Pareto convergence of the learning-generated upper tail, conditional on belonging to that tail.

\subsection{The Output Effects of an Improvement in Entry-Level Automation}

I now study the effects of an unanticipated improvement in automation occurring at a date $\tau$, paralleling the analysis in Section \ref{sec:entry-level} of the main text. The objective is to examine how task-learning heterogeneity modifies the output and growth effects of such a technological shock.

As in Section \ref{sec:entry-level}, I impose a Pareto restriction at the shock date to obtain transparent closed-form expressions. Here, however, the restriction applies only to the learning-generated upper tail, since the full distribution includes the lower non-learning component. Specifically, I assume that the distribution of expert skill at time $\tau$ satisfies: \begin{equation}  \label{eq:assuD}
1-F_\tau(q) = \frac{N-I-\zeta}{N-I} \left(\frac{x_\tau}{q}\right)^{1/\theta}, \ \text{for} \  q\geq x_\tau. \end{equation}
As in the main text, this assumption admits two possible interpretations:\begin{enumerate}
\item \textbf{Exact Initial Distribution}: If the initial distribution $F_0$ is exactly Pareto with scale $x_0$ and shape $1/\theta$, the pre-shock equilibrium path implies that, by any date $\tau \ge 1$, the learning-generated upper tail at time $\tau$ satisfies (\ref{eq:assuD}). 
\item \textbf{Asymptotic Approximation}: If $F_0$ is not exactly Pareto, but $1-F_0 \in \mathrm{RV}_{-1/\theta}$, this setup can be interpreted as approximating an economy hit by the shock at an arbitrarily distant date, by which time the learning-generated upper tail is close to its conditional Pareto limit given by (\ref{eq:assuD}). \end{enumerate}

To establish a baseline, consider the evolution of the economy from period $\tau$ onward in the absence of a shock. Under (\ref{eq:assuD}), the no-shock active cutoff satisfies:
\[  a_t = x_\tau \ell_{\mathrm L}^{\theta(t-\tau+1)}, \ \text{for} \  t\geq \tau. \]
Aggregate output along this trajectory is therefore:
\[ Y_t = a_t \left[ \frac{m^Ih^{N-I}}{(N-I)(1-\theta)}  \right] = x_\tau \ell_{\mathrm L}^{\theta(t-\tau+1)} \left[ \frac{m^Ih^{N-I}}{(N-I)(1-\theta)} \right], \ \text{for} \  t\geq \tau. \]

Now suppose an unanticipated automation shock arrives at date $\tau$ and expands the automatable interval from $[0,I]$ to $[0,I+\Delta]$, where $0<\Delta<N-I$. Let $ \widetilde{\Man}\equiv \Man(I+\Delta,m,h)$ denote the post-shock supervision time required for one project. As in the main text, assume that $ \widetilde{\Man} \le \Man$ because machines are more productive than novices. The post-shock production and learning spans of control are, respectively:
\[ \widetilde{\ell}_{\mathrm P} \equiv \frac{N-I-\Delta}{\widetilde{\Man}} \quad \text{and} \quad \widetilde{\ell}_{\mathrm L} \equiv \frac{N-I-\max\{\zeta,\Delta\}}{\widetilde{\Man}}. \]
The term $\max\{\zeta,\Delta\}$ reflects the fact that automation removes non-learning tasks first. If $\Delta\leq\zeta$, the shock automates only tasks that did not transmit expertise. If $\Delta>\zeta$, it also automates a measure $\Delta-\zeta$ of learning-relevant tasks.

As in Section \ref{sec:entry-level}, I focus on the case where all novices remain employed after the shock. At date $\tau$, the mass of positive-skill experts is $\ell_{\mathrm L}/\ell_{\mathrm P}$. Since zero-skill experts cannot produce, only these positive-skill experts can absorb novice labor. If all such experts were active after the shock, aggregate novice demand would be $\widetilde{\ell}_{\mathrm P}(\ell_{\mathrm L}/\ell_{\mathrm P})$. I therefore assume:
\[
\widetilde{\ell}_{\mathrm P}\frac{\ell_{\mathrm L}}{\ell_{\mathrm P}}>1.
\]
This condition ensures that all novices remain employed after the shock. It also implies $\widetilde{\ell}_{\mathrm L}>1$, so the shocked economy remains in the sustained-growth case.

Letting tildes denote the shocked trajectory, the skill distribution at date $\tau$ is predetermined. The new active cutoff therefore solves $\widetilde{\ell}_{\mathrm P}[1-F_\tau(\widetilde a_\tau)] = 1$. Using (\ref{eq:assuD}), this gives:
\[ \widetilde a_\tau = x_\tau \left( \widetilde{\ell}_{\mathrm P}\frac{\ell_{\mathrm L}}{\ell_{\mathrm P}} \right)^\theta. \]
From date $\tau$ onward, the learning-generated upper tail evolves according to the post-shock learning span $\widetilde{\ell}_{\mathrm L}$. Thus, for $t\geq \tau$,
\[ \widetilde a_t = \widetilde a_\tau \widetilde{\ell}_{\mathrm L}^{\theta(t-\tau)}. \]

Aggregate output along the shocked trajectory is therefore:
\[
\widetilde Y_t
=
\widetilde a_t
\left[
\frac{m^{I+\Delta}h^{N-I-\Delta}}{(N-I-\Delta)(1-\theta)}
\right] = 
x_\tau
\left(
\widetilde{\ell}_{\mathrm P}\frac{\ell_{\mathrm L}}{\ell_P}
\right)^\theta
\widetilde{\ell}_{\mathrm L}^{\theta(t-\tau)}
\left[
\frac{m^{I+\Delta}h^{N-I-\Delta}}{(N-I-\Delta)(1-\theta)}
\right], \ \text{for} \ t\geq \tau.
\]

Dividing the shocked path by the no-shock path gives:
\[
\frac{\widetilde Y_t}{Y_t}
=
\left(\frac{m}{h}\right)^\Delta
\left(\frac{N-I}{N-I-\Delta}\right)
\left(\frac{\widetilde{\ell}_{\mathrm P}}{\ell_P}\right)^\theta
\left(\frac{\widetilde{\ell}_{\mathrm L}}{\ell_{\mathrm L}}\right)^{\theta(t-\tau)}.
\]
Equivalently, substituting the definitions of $\ell_P,\ell_{\mathrm L},\widetilde{\ell}_{\mathrm P}$, and $\widetilde{\ell}_{\mathrm L}$,
\begin{equation} \label{eq:outputRH}
\frac{\widetilde Y_t}{Y_t}
=
\underbrace{\left(\frac{\Man}{\widetilde{\Man}}\right)^\theta
\left(\frac{m}{h}\right)^\Delta
\left(1-\frac{\Delta}{N-I}\right)^{-(1-\theta)} }_{\text{Immediate adoption gains}}
\underbrace{
\left[
\frac{N-I-\max\{\zeta,\Delta\}}{N-I-\zeta}
\frac{\Man}{\widetilde{\Man}}
\right]^{\theta(t-\tau)}
}_{\text{Long-run gains or losses}}.
\end{equation} Note that this expression nests the baseline result in Section \ref{sec:entry-level}. If all novice-performed tasks are learning-relevant ($\zeta=0$) and there are no scale effects ($\widetilde \Man=\Man=1$), then the expression collapses to equation (\ref{eq:decomp}) of the main text. 

To understand the effects of this shock in this more general setting, consider first the case in which automation removes only non-learning tasks, $\Delta\leq\zeta$. Then $\max\{\zeta,\Delta\}=\zeta$, and equation \eqref{eq:outputRH} becomes:
\[
\frac{\widetilde Y_t}{Y_t}
=
\underbrace{
\left(\frac{\Man}{\widetilde \Man}\right)^\theta
\left(\frac{m}{h}\right)^\Delta
\left(1-\frac{\Delta}{N-I}\right)^{-(1-\theta)}
}_{\text{Immediate adoption gains}}
\underbrace{
\left(\frac{\Man}{\widetilde \Man}\right)^{\theta(t-\tau)}
}_{\substack{\text{Long-run gains} \\ \text{(scale effects)}}}.
\]
In this case, automation raises output on impact without directly weakening knowledge transmission. If $\widetilde\Man=\Man$, the second term equals one, so the shock leaves the long-run growth rate unchanged. If $\widetilde\Man<\Man$, the second term grows over time: automation increases long-run growth by expanding experts' scale of operation while leaving the measure of learning-relevant tasks per project unchanged.

The result potentially changes when automation reaches learning-relevant tasks. If $\Delta>\zeta$, then $\max\{\zeta,\Delta\}=\Delta$, and the long-run term in \eqref{eq:outputRH} becomes:
\[
\left[
\frac{N-I-\Delta}{N-I-\zeta}
\frac{\Man}{\widetilde\Man}
\right]^{\theta(t-\tau)}.
\]
Equivalently, the post-shock growth factor exceeds the pre-shock growth factor if and only if
\[
\widetilde{\ell}_{\mathrm L}>\ell_{\mathrm L}
\quad\Longleftrightarrow\quad
\frac{N-I-\Delta}{N-I-\zeta}
\frac{\Man}{\widetilde\Man}>1.
\]
Absent scale effects, $\widetilde\Man=\Man$, this inequality fails: once automation reaches learning-relevant tasks, it reduces the learning span and lowers long-run growth. With scale effects, the sign of the growth effect depends on whether the reduction in supervision time offsets the loss of learning-relevant positions.

This extension therefore separates the effect of automation on production from its effect on knowledge transmission. When automation is confined to non-learning routine tasks, it raises output on impact without directly reducing the positions through which novices acquire expertise; if it also lowers supervision time, it can increase the learning span and raise long-run growth. The force toward lower growth arises only when automation reaches learning-relevant tasks. At that point, automation reduces the measure of positions through which expertise is transmitted, slowing knowledge diffusion unless scale effects offset the reduction.

Thus, the effect of automation on growth can be non-monotonic in the extent of automation: early automation of low-learning tasks may raise output, and possibly growth, while later automation can slow growth once it encroaches on tasks that transmit expertise.

\section{Imperfect and Span-Dependent Skill Transmission} \label{sec:imperfect_transmission} 

This appendix relaxes the assumption that novices exactly inherit their mentor's skill. It also allows the effectiveness of transmission to depend on the number of novices supervised by a given expert. 

To simplify, I return to the baseline environment without task-learning heterogeneity or scale effects: all novice-performed tasks provide uniform access to the expert's tacit knowledge, and each active expert hires $N-I$ novices. Moreover, as in Online Appendix \ref{sec:hetero}, I exploit the block-recursive structure of the equilibrium to study the implications for aggregate output and growth, without characterizing the associated wage and income schedules.

Let $\ell \equiv N-I$ denote the mass of novices hired by each active expert. I assume that knowledge transmission is imperfect. When an expert supervises $\ell$ novices, only a fraction $\varphi(\ell)\in[0,1]$ of them acquire the expert's skill; the remaining fraction acquire no skill. The function $\varphi(\ell)$ may depend on the number of novices supervised. If $\varphi$ is decreasing, supervising more novices dilutes the expert's attention and weakens transmission. If $\varphi$ is increasing, a larger group of novices may instead create learning synergies.

I interpret imperfect transmission as a deterministic transmission technology at the level of the continuum. Thus, among the novices assigned to an expert with skill $q$, a fraction $\varphi(\ell)$ enters the next period with skill $q$, while the remaining fraction enters with skill zero. This formulation avoids the need to construct independent transmission shocks across a continuum of novices.

I assume that $\ell\varphi(\ell)>1$. This condition is analogous to Assumption \ref{assu:M1} of the main text, and it is necessary for positive asymptotic growth. It also implies that the mass of positive-skill experts is large enough to absorb all novice labor, implying no novice unemployment. 

Let $a_t$ denote the marginal active expert at time $t$. Since each active expert still hires $\ell=N-I$ novices, aggregate market clearing is unchanged:
\[
1=\ell[1-F_t(a_t)].
\]

The law of motion for the distribution of experts' skills changes. Since only a fraction $\varphi(\ell)$ of the novices matched with active experts acquire their mentor's skill, the next-period skill distribution satisfies
\[
F_{t+1}(q)
=
\begin{cases}
0, & q<0,\\[4pt]
1-\varphi(\ell), & 0\leq q<a_t,\\[4pt]
1-\varphi(\ell)+\ell\varphi(\ell)[F_t(q)-F_t(a_t)], & q\geq a_t.
\end{cases}
\]
Equivalently, for $q\geq a_t$,
\[
1-F_{t+1}(q)
=
\ell\varphi(\ell)[1-F_t(q)].
\]Thus, as in Online Appendix \ref{sec:hetero}, imperfect transmission separates the span of control relevant for novice-market clearing from the span relevant for knowledge transmission. The ``production span'' is $\ell_{\mathrm P}=\ell$, while the ``learning span'' is $\ell_{\mathrm L}=\ell\varphi(\ell)$. In Online Appendix \ref{sec:hetero}, the two spans differ because some novice tasks do not transmit expertise. Here, the spans differ because transmission is imperfect, even though all novice positions are symmetric.

The derivations then proceed as in Online Appendix \ref{sec:hetero}. Hence, the long-run growth factor is:
\begin{equation} \label{eq:growthI}
1+g=[\ell\varphi(\ell)]^\theta.
\end{equation}
Exact inheritance corresponds to the special case $\varphi(\ell)=1$, in which the growth factor reduces to $\ell^\theta=(N-I)^\theta$. Imperfect transmission lowers the effective learning span whenever $\varphi(\ell)<1$, but leaves the structure of the growth result unchanged.

As in Online Appendix \ref{sec:hetero}, the full normalized skill distribution is not Pareto, because a mass $1-\varphi(\ell)$ of each cohort acquires zero skill. Let $x_t\equiv a_{t-1}$ denote the lower bound of the learning-generated upper tail. Conditional on belonging to this upper tail, normalized skills converge to the same Pareto law as in the baseline model:
\[
1-\Phi^{\mathrm L}(z) = \lim_{t \to \infty}  \mathbb{P}_t\left(\frac{q}{x_t}>z \,\middle|\, q\geq x_t\right) = z^{-1/\theta}, \ \text{for} \ z\geq 1.
\]
Thus, imperfect transmission changes the mass of the learning-generated upper tail and the growth factor, but not the asymptotic shape of that tail.

Equation (\ref{eq:growthI}) clarifies how imperfect inheritance affects the results of the main text. Consider, for instance, an improvement in entry-level automation that expands the set of automatable tasks by $\Delta$.  Without scale effects, the improvement reduces the raw span of control from $\ell=N-I$ to $\widetilde\ell=N-I-\Delta$. Under exact inheritance, this reduction directly lowers the rate of knowledge transmission. With imperfect and span-dependent transmission, however, the relevant object is the effective learning span, $\ell\varphi(\ell)$. 

Provided the shocked economy remains in the full-employment growth case both at impact and asymptotically, the post-shock growth factor exceeds the pre-shock growth factor if and only if:
\[
\widetilde{\ell}\varphi(\widetilde{\ell})
>
\ell\varphi(\ell).
\]

If $\varphi$ is increasing, so that supervising more novices creates learning synergies, automation is unambiguously detrimental for long-run knowledge transmission. Since automation reduces the raw span of control, $\widetilde{\ell}<\ell$, it also lowers transmission effectiveness, $\varphi(\widetilde{\ell})\leq \varphi(\ell)$. Hence, $\widetilde{\ell}\varphi(\widetilde{\ell})<\ell\varphi(\ell)$, so the post-shock growth factor is necessarily lower.

In contrast, if $\varphi$ is decreasing, so supervising more novices dilutes the expert's attention, the automation shock has two opposing effects. The displacement effect lowers the number of novices supervised by each expert. However, the transmission-quality effect increases transmission effectiveness because each expert supervises fewer novices. Thus, in this case, improvements in entry-level automation can increase long-run growth even without scale effects, provided the transmission-quality effect dominates the displacement effect. 

Thus, once transmission is imperfect and span-dependent, the raw span of control is no longer in one-to-one correspondence with knowledge transmission. The relevant object is the effective learning span. Exact inheritance is therefore not essential to the mechanism, but it makes the mapping between span of control and knowledge transmission exact.

\section*{References}

{\small

\textsc{Bingham, N. H., C. M. Goldie, and J.L. Teugels} (1989): \textit{Regular Variation}, vol. 27 of \textit{Encyclopedia of Mathematics and its Applications}, Cambridge: Cambridge University Press, first paperback  (with additions) ed.

\vspace{3pt}

\noindent \textsc{Buera, F. J. and R. E. Lucas} (2018): ``Idea Flows and Economic Growth,'' \textit{Annual Review of Economics}, 10, 315–345

}

\end{document}